\documentclass[12pt,french]{article}
\usepackage[french]{babel}
\usepackage{color,epsfig}

\definecolor{violet}{rgb}{0.4,0,0.4}
\definecolor{vert}{rgb}{0,0.5,0.0}
\definecolor{navy}{rgb}{0.0,0.0,0.6}
\definecolor{orange}{rgb}{0.8,0.2,0.0}
\definecolor{bleu}{rgb}{0.3,0.0,0.8}

\def\thf{ }
\def\fff{ }

\def\pp{\ \bot}

\def\eqdef{\fff\ \vbox{\hbox{$_{_{\rm def}}$} \hbox{$=$} }\ \thf }

\def\ug{\underline g}
\def\ov{\overline}
\def\olg{\ov{\ov g}}

\def\onab{\ov{\ov \nabla}}
\def\ug{\underline g}

\def\vv{_{\ \vert}}

\def\af{\fff\vbox{\hbox{$_{_{\vv}}$} \hbox{$\overline {\ f} $} }\thf }
\def\ag{\fff\vbox{\hbox{$_{_{\vv}}$} \hbox{$\overline {\ g } $} }\thf }
\def\aG{\fff\vbox{\hbox{$_{_{\vv}}$} \hbox{$\overline {G } $} }\thf }
\def\anab{\fff\vbox{\hbox{$_{_{\vv}}$} \hbox{$\overline \nabla $} }\thf }
\def\alam{\fff\vbox{\hbox{$_{_{\vv}}$} \hbox{$\overline{\ \lambda} $} }\thf }
\def\acu{\fff\vbox{\hbox{$_{_{\vv}}$} \hbox{$\overline{\ u} $} }\thf }
\def\axi{\fff\vbox{\hbox{$_{_{\vv}}$} \hbox{$\overline{\ \xi } $} }\thf }
\def\alJ{\fff\vbox{\hbox{$_{_{\vv}}$} \hbox{$\overline{\underline J} $} }\thf }
\def\alT{\fff\vbox{\hbox{$_{_{\vv}}$} \hbox{$\overline{\underline T} $} }\thf }

\def\eqdef{\fff\ \vbox{\hbox{$_{_{def}}$} \hbox{$=$} }\ \thf }
\def\dag{\dagger}
\def\el{\ell}

\def\r{r}
\def\si{\varphi   }

\def\bfPhi{ {\mit\Phi} }
\def\bfDelta{ {\mit\Delta} }
\def\upbPhi{\fff\vbox{\hbox{$_{_{\pp}}$} \hbox{$\ \bfPhi$} }\thf }
\def\upbJ{\fff\vbox{\hbox{$_{_{\pp}}$} \hbox{$\ {\bf J}$} }\thf }

\begin{document}

\bf

MECHANICS AND EQUILIBRIUM GEOMETRY OF

 BLACK HOLES, MEMBRANES, AND STRINGS.

\vskip 1.2 cm

B. Carter.
\vskip 0.8 cm\rm

Dept. of Relativistic Astrophysics and Cosmology,

C.N.R.S.,  Observatoire de Paris,

92 Meudon, France.

\vskip 1.2 cm

{\bf Abstract.\ \ }\it  This course is designed to give a mathematically coherent
introduction to the classical thory of black holes and also of strings and
membranes (which are like the horizon of a black hole in being examples of
physical systems based on a dynamically evolving world sheet) giving particular
attention given to the study of the geometry of their equilibrium states.
\rm
\vskip 1.2 cm

\parindent = 0 cm
{\bf  Preface.}
 \medskip\parindent=1.2 cm

The purpose of this course is to provide a mathematically coherent introduction
to the classical  theory of black holes and also to the related and more
recently developed topic of the classical theory of relativistic strings and
membranes for which many of the same techniques are required. The stategy of the
course will be to concentrate on general results rather than special examples,
and to distinguish as clearly as possible what has been completely proved from
what has only been partly established or merely conjectured so as to give some
idea of the main open problems for future research. The discussion is developed
on the basis of a chain of key results for which it has been possible to provide
reasonably complete and self contained mathematical proofs without resort to
disproportionate technical complication.  The level of previous knowledged
required corresponds to what is obtainable from the relevant sections of a
textbook such as that of Misner, Thorne and Wheeler$^{[1]}$ (whose notation
will be used as far as possible) or, in a less encyclopaedaic but more
conveniently accessible (and up to date) form, that of Wald$^{[2]}$. There are
already several textbooks specifically devoted to various aspects of black hole
theory $^{[3][4][5][6]}$; attention is particularly to be drawn that of
Hawking and Ellis$^{[7]}$ for advanced mathematical background reading, and to
that of Novikov and Frolov$^{[8]}$ for an exceptionally comprehensive survey
of the published litterature including more than 600 references. 

The organisation of the course is as follows.

Section 1 provides a brief astrophysical introduction consisting essentially of
a simple explanation$^{[9]}$ of the orders of magnitude that are relevant to
the conventional idea of the formation of ``ordinary" black holes  by stellar
collapse (no such simple and clear picture being available for the more exotic
phenomenon of the giant black holes that are commonly believed to be lie at the
heart of active galactive nuclei).

After this physical introduction, the main part of the course is more
essentially mathematical in nature following more or less the same lines as my
previous reviews$^{[10][11][12]}$ though with the omission, except for the
necessary references, of certain parts in order to make way for the inclusion of
new results. Section 2 presents some of the main results of the theory of
exactly spherical gravitational collapse, which is the only case for which a
precise dynamical analysis is available. Section 3 gives a brief account of what
little is known about dynamical formation of black holes in more realistic
situations where spherical symmetry is broken by effects such as rotation.
Section 4 deals with the theory of stationary rotating black hole equilibrium
states in the general case for which externally orbiting matter rings may be
present. Section 5 deals more specifically with the uniqueness theorem that is
available when no external sources are present. Section 6 concludes the course
on black hole theory by describing some of the rather miraculous special
properties of the ensuing Kerr Newman metrics, whose stability is one of the
most important topics that (for lack of time and space) has not been included in
this course: for the most complete result, going a long way towards confirming
that these equilibrium solutions can indeed be considered to be stable, the
interested reader is referred to the recent work of Whiting$^{[13]}$

Section 7 moves on to present a covariant formulation$^{[14]}$
 of the basic mechanical
principles of classical brane theory meaning the subject that includes the
theories of point particles, strings, membranes and continua as special cases.
Section 8 deals more specificly with the theory of spacially isotropic branes, a
category that includes all classical string models and in particular those
representing ``superconducting cosmic strings".

Finally in a purely mathematical appendix, some of the most important tensorial
quantities (which are useful for black hole theory and indispensible for brane
theory) characterising the different kinds of curvature of an imbedding  are
presented in a readily utilisable form$^{[15]}$ that is not
yet readily available elsewhere.

\medskip
\bigskip\parindent=0cm
{\bf 1.  The astrophysical context of Black Hole formation.}
\medskip\parindent=2cm

The study of black holes in general, and of black hole equilibrium states in
particular, arises as a natural offshoot of the study of stellar equilibrium
states whose theoretical foundations were established by workers such as
  Eddington and Chandrasekhar in the years following the elucidation of the basic
principles of quantum mechanics.  In terms of the fundamental Plank type unit
system that will be used throughout this course (in which the speed of light
$c$, Newton's gravitational constant $G$, the Dirac - Plank constant
$\hbar$ and the Boltzman constant $k$ are all simulltaneously set equal to unity)
the dominant physical mechanisms governing the situation can be
described$^{[9][16]}$ in crude order of magnitude terms (give or take a power of
ten here or there) in terms of just three particularly important dimensionless
parameters, namely the masses $m_e$ and $m_p$ of the electron and the proton,
and the magitude $e$ of their electric charge, which are expressible as the
moderately small ``fine structure" coupling contant $e^2\simeq{1/ 137}$ the
considerably smaller mass ratio ${m_e/ m_p}\simeq{1/ 1800} $ and the
extremely small gravitational coupling constant $m_p{^2}\approx 10^{-39}$.

In the low temperature limit, the equilibrium states of small, medium, and even
moderately large bodies, on scales ranging from single molecules through sand
grains up to entire planets, are characterisable in crude order of magnitude by
a typical density $\rho$ given by
$$\rho \approx e^6 m_e{^3} m_p  \eqno(1.1)$$
which works out (by no means accidentally) to be very roughly of the order of
unity in ``ordinary" units, $gm/cm^3$ (which have of course  been deliberately
normalised to give such a result). Taking account of the fact that in all such
states the mean mass per baryon is given to a very good (within one per cent)
accuracy by $m_p$, so that $\rho\simeq m_p n$ where $n$ is the baryon number
density, the relation (1.1) expresses the condition that the mean separation
$\lambda\simeq n^{-1/3}$ between baryons will be of the same order as that
between the (within a factor of two equally numerous) electrons, and therefore of
the order of the Bohr radius, $\lambda\simeq 1/e^2 m_e$, which is the result
that is obtainable from the consideration that the equilibrium is determined by
the balance between Fermi (exclusion principle) repulsion between electrons and
electrostatic attraction between negatively charged electrons and positively
charged ions. 

Although applicable to bodies on scales ranging from that of a hydrogen atom to 
that of the earth, the formula (1.1) loses its validity for bodies so large
that the long range cumulative effect of the (individually very weak) 
gravitational attraction forces becomes stronger than the effect of the
electrostatic attraction forces (which of course only act locally because of 
the long range cancellation resulting from overall electric neutrality).
For a body of mass $M$, mean density $\rho$ and hence characteristic mean 
radius $R\simeq (M/\rho)^{1/3}$ resistance to collapse under the influence
of gravitational self attraction requires a mean central pressure $P$ given 
according to the well known ``virial theorem" by
$$ P\approx M^{2/3} \rho^{4/3}\ , \eqno(1.2)$$
which expresses a balance between the typical radial pressure gradient, of order
$P/R$, and the gravitational force density, of order $\rho M/R^2$.

The pressure contribution resulting from the application of the Fermi exclusion
principle to the electrons is of the order of the corresponding kinetic energy
density, and therefore will be given - in the non relativistic limit - roughly 
by 
$$ P\approx {1\over m_e}\left( {\rho\over m_p}\right)^{5/3}\eqno(1.3)$$
in view of the fact that the mean momentum per electron will just be the inverse
$\lambda^{-1}$ of the corresponding De Broglie wavelength, which will itself be
of the same order of magnitude as the mean separation,  $\lambda\simeq
n^{-1/3}$, where $n\simeq\rho/m_p$.  So long as the mass $M$ is small compared
with a critical value given roughly by $M\approx e^3/m_p{^2}$, the virial
pressure requirement (1.2) is small compared with the Fermi energy density (1.3)
at the ``ordinary" matter density (1.1) which means that the gravitational
compression effect will be unimportant, but but beyond this critical mass (which
is of the order of that of the giant planet Jupiter) the long range
gravitational attraction will dominate over the short range electrostatic
binding so that the corresponding equilibrium states will be of white dwarf
type, with the central pressure determined by direct equation of (1.2) and (1.3)
which means that the characteristic mean central density $\rho$ will be given as
a function of the mass $M$ by an order of magnitude relation of the form
$$\rho\approx m_e{^3} m_p{^5} M^2 \ . \eqno(1.4)$$

The range of validity of the relation (1.4) is of course limited to that of
the non relativistic degenerate electron gas pressure formula (1.3) from which
it is derived. When the relevant DeBroglie wavelength $\lambda\simeq
n^{-1/3}$ becomes short compared with the Compton wavelength $\lambda\simeq
m_e$, the kinetic energy per electron is no longer given by 
$1/\lambda^2 m_e$ but just by $1/\lambda$, so that the non relativistic
formula (1.3) must then be replaced by the corresponding relativistic
degenerate gas pressure formula
$$P\approx \left({\rho\over m_p}\right)^{4/3}\ . \eqno(1.5)$$
The (by now generally accepted) recognition 
 that the theory of black holes must be taken seriously as something whose
implications are directly relevant and testable in observational astrophysics
derives from the startling (1930) discovery by Chandrasekhar$^{[17]}$ 
that substitution
of (1.5) instead of (1.4) in the virial equilibrium condition (1.2) does not
just give a modified version of the functional relation (1.4)
 for the equilibrium density $\rho$ as a function of the mass $M$, but
instead gives an absolute cut off at a critical mass
$$M\approx {1\over m_p{^2}} \eqno(1.6)$$
above which no ordinary cold equilibrium state is possible at all!

The existence of this upper mass limit does not of course mean that there are no
cold equilibrium states beyond the critical density $\rho\simeq m_e{^3}m_p$ at
which the white dwarf range (1.4) reaches the Chandrasekar limit (1.6), since it
is also possible to have high density states in which the electrons are combined
with protons to form neutrons for which the relevant analogue of the non
relativistic degenerate gas pressure formula (1.3) is
$$ P\approx {1\over m_p}\left( {\rho\over m_p}\right)^{5/3} \ . \eqno(1.7)$$
However the resulting range of neutron star equilibrium states,
with density given, by substitution of (1.7) in (1.2), as
$$ \rho\approx m_p{^8} M^2, \eqno(1.8)$$
will be cut off by an upper mass limit that is still given$^{[18][19]}$ 
by the same crude order of magitude formula (1.6) as before, because the
relativistic degenerate gas pressure has the same form (1.5) for neutrons as for
electrons. More exact calculations (whose results are still subject to a
considerable uncertainty due to the imprecision of our present understanding of
the detailed physical properties of neutron star matter) indicate that the upper
mass limit for neutron stars is somewhat larger (though only by a modest factor
not much in excess of two), than that for white dwarfs:  this conclusion is of
great astrophysical importance, and would appear to have been observationally
confirmed by the discovery of pulsars (since if the exact upper mass limit for
neutron stars had turned out to be smaller  than that for white dwarfs then the
formation of neutron stars by gravitational collapse would have been rendered
virtually impossible).

As the astronomical community belatedly recognised (after more than thirty years
of general indifference or incredulity) Chandrasekhar's discovery$^{[17]}$ made
it absolutely necessary to take the possibility of runaway gravitational
collapse - and ensuing formation of massive or ultramassive black holes - very
seriously as a phenomenon of potentially crucial relevance to many directly
observable phenomena. This contrasts  with the situation that still applies to
speculation on the subject of microscopic black holes (for which quantum
phenomena such as Hawking radiation$^{[20]}$ are significant) whose relevance to
anything actually observable remains subject to reasonable doubt. Nevertheless
the existence of several categories of observational ``black hole candidates"
(of which the most famous prototype example is the galactic X-ray source Cygnus
X-1) does not yet amount to a firm confirmation of that black holes with the
properties described in the following sections of this course really do exist.
The most numerically numerous (and perhaps ultimately most atrophysically
important) category of observationally detected ``candidates" is that of nuclei
of ``active" galaxies, but such (ultramassive) objects are all too fuzzy and far
away to have been of any use so far from the point of view of verification of
the basic physical theory. As far as the more conveniently tractable candidates
within our galaxy are concerned, the awkward fact to be faced is that sixty
years after our colleague Chandra's precocious and revolutionary theoretical
discovery, and more than twenty years after the collapse of psychological
resistance to the notion of a ``black hole" following the coining of the term
itself (by John Wheeler) and the (approximately simultaneous and no less
psychologically significant) experimental discovery  (by Jocelyne Bell and Tony
Hewish) of the pulsars  whose identification as neutrons stars has long been
unquestionable, there are still disappointingly few observationally discovered
objects that can plausibly be interpreted as ``ordinary" (moderate sized) black
holes.

The relative scarcity of black holes in the mass range immediately above
the Chandrasekhar limit (which is about one and a half times the mass of the
sun) might at first seem surprising in view of the fact that this particular
mass range is precisely that of the most numerous subclass of the ordinary stars
that are visible at night to the naked eye. However from a theoretical point of
view this apparent paradox can easily be understood as follows.

\begin{figure}
\centering
\epsfig{figure=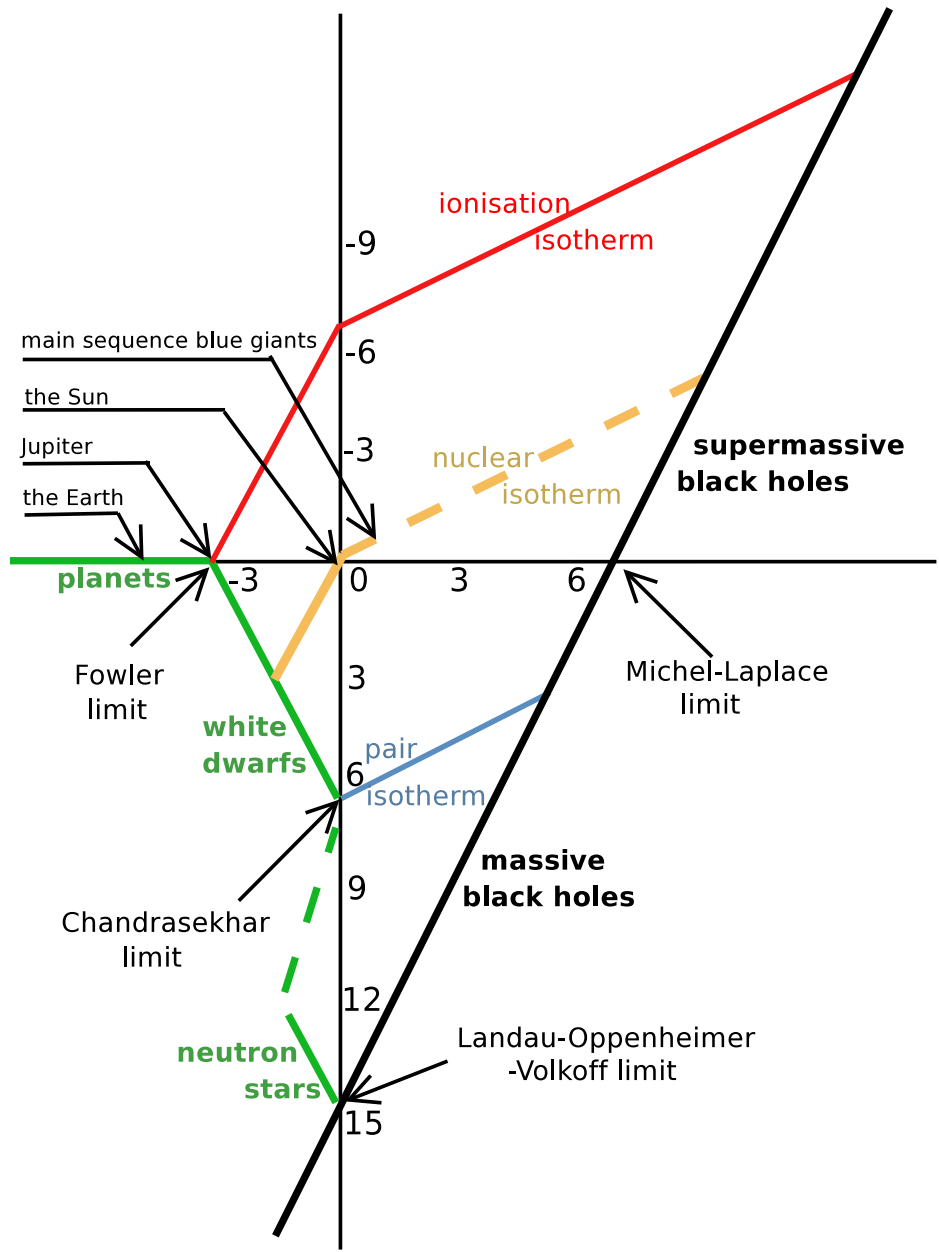, width=12.4 cm}
\caption{\color{navy}
\label{MassDensity}
{\it Logarithmic plot of characteristic orders of magnitude
for density $\rho$ against mass $M$ relative to the standard values given
by (1.1) and (1.6), which  roughly characterise  the Sun.}     
}
\end{figure}

The non-existence of any material ``ground state"  configurations, i.e. minimum
energy (cold, static, absolutely stable)  equilibrium states, above a critical
mass value given in order of magnitude by (1.6) does not of course exclude the
existence of  more massive ``excited" (and therefore in principle ultimately
unstable) equilibrium states whose support depends on having more than minimal
energy in thermal or other forms (such as that involved in differential
rotation) which in the long run are subject to dissipation and loss by radiation
but which in practice may be preserved over astrophysically or even
cosmologically long time scales. The most important examples of such excited
states are of course ordinary hot stars (including most notably those of the
main, i.e. hydrogen burning, sequence) which are characterisable by a mean
central temperature $\Theta$ say, in terms of which the pressure $P$ will be
given by the sum of a radiation contribution $P\simeq \Theta^4$ and of a
non-relativistic gas contribution $P\simeq n\Theta$ where the relevant number
density of non relativistic particles will be of the same order as the baryon
number density $n\simeq\rho/m_p$. Evidently the radiation contribution will be
dominant for $\Theta^3>>n$ while the non relativistic gas contribution will be
dominant for $\Theta^3<<n$. On substituting these formulae in the virial
equilibrium condition (1.2) it can seen that the only criterion for radiation
dominance is the mass $M$ of the star, the critical value (which was first
worked out by Eddington$^{[21]}$) being again  given in order of magnitude by the inverse
of the gravitational coupling constant $m_p{^2}$ i.e. by the {\it same} formula
(6.1) as was obtained a few years later by Chandrasekhar for the more sharply
definable upper mass limit for cold equilibrium: below this limit the dominant
pressure contribution is that of the non-relativistic particles, whose
substitution in the virial equilibrium condition (6.2) leads to a formula giving
the characteristic density corresponding to a given characteristic central
temprature $\Theta$ in the form
$$\rho\approx \left({\Theta^3\over m_p{^3}M^2}\right) \eqno(1.9)$$
whereas when the stellar mass exceeds the critical value (1.6) it can be seen
that it is the radiation gas pressure contribution that will be dominant so that
instead of (1.9) one will obtain a less strongly mass dependent result
expressible by 
$$\rho \approx \left( {\Theta^3\over M^{1/2}}\right) \eqno(1.10)$$

The explanation for the fact that the typical masses of ordinary observable
stars turn out to be comparable with the critical Eddington - Chadrasehkar
mass value given by (1.6) is to be found in terms of the criterion for
stability with respect to adiabatic variations, in which the pressure will 
vary as a function of density according to an approximately polytropic law of 
the form $P\propto \rho^\Gamma$ where the index value is $\Gamma=5/3$
in the non-relativistic limit to which (1.9) applies, but where $\Gamma=4/3$
in the radiation dominated limit to which (1.10) applies. It is immediately 
obvious from the form of the virial condition (1.2) that it is necessary for
stability that the effective polytropic index should exceed the critical value
$4/3$, i.e. precisely the same value that is characterises adiabatic 
perturbations of a radiation dominated gas. This means that stellar
configurations above the critical mass (1.2) can at best be stabilised only
marginally by their relatively small non-relativistic gas pressure 
contribution, and that for mass values a long way (more than two powers of ten)
above the critical value  $M\approx 1/m_m{^2}$ stable equilibrium will in
practice be extremely difficult to achieve. One would therefore expect that
(in accordance with what is actually observed) formation of stars by 
gravitational condensation (with central heating according to the law
$\Theta\propto \rho^{1/3}$ that is obtained from both (1.9) and (1.10))
from initially diffuse gas clouds would inevitably produce objects below
or not too far above the critical mass (1.6).     

As well as being limited in mass the conceivable range for ordinary stellar
equilibrium states is of course also limited in temperature, which must exceed
the Rydberg energy value, $\Theta\simeq e^4 m_e$, that is the threshold for the
ionisation of the gas that accounts for the opacity needed to delay the
radiation loss of the thermal energy: on substitution of this minimal Rydberg
temperature in (1.9) and comparison with the white dwarf equilibrium condition
(1.4) it can be seen (see figure 1) that the smallest possible mass for an ionised stellar
configuration with thermal pressure support is the same as the maximum possible
value, $M\approx e^3/m_p{^2}$ for a cold planetary configuration, i.e. about the
mass of Jupiter which is situated just at the lower end of the cold white dwarf
range.  At the opposite extreme the upper cut off to the conceivable range of
temperatures for ordinary stellar configurations is given by the electron
positron pair creation temperature $\Theta\approx m_e$  beyond which there is
longer  any possibility of stabilisation  by a non relativistic electron gas
contribution: on substitution of this pair creation temperature in (1.10) and
comparison with the white dwarf equilibrium condition (1.4) it can be seen (see
figure 1) that the highest possible characteristic density $\rho$ for an ionised
stellar configuration with thermal pressure support is obtained for a mass of
the order of the Chandrasekar limit value (1.6) and is the same as the maximum
possible value, $\rho\approx m_e{^3} m_p$, that is obtained at the upper end of
the white dwarf range.

Although energy loss by radiation from the outer (``chromospheric") surface
layers prevents them from lasting indefinitely, the ``excited" stellar
equilibrium states in the range delimited by the considerations of the
preceeding paragraphs can nevertheless survive over astrophysically long
timescales whose minimum value is determined by the minimal opacity contribution
that results from Thompson scattering of photons by electrons with effective
cross section given in order of magnitude by $\sigma\simeq (e^2/m_e)^2$, which
leads to a minimal evolution timescale  $\tau$, for stars in the radiative mass
range $M>>1/m_p{^2}$ to which (1.10) applies, that will be given by
$$\tau\approx{ \varepsilon e^4\over m_e{^2} m_p{^2}}\ , \eqno(1.11)$$
where $\varepsilon$ is the efficiency of conversion of rest mass into thermal
energy by nuclear reactions. The most efficient thermonuclear energy production
process is of course hydrogen burning which yeilds almost one per cent,
$\varepsilon\approx 10^{-2}$ at a ``main sequence" temperature $\Theta$ at which
stars spend most of the lifetime allowed by (1.11), which works out at about
$10^7$ years. For smaller stars with masses near or below the Eddington -
Chandrasekhar critical value (1.6), other mechanisms come into play which
increase the opacity and diminish the rate of energy loss by radiation, giving
timescales that for the smallest main sequence stars can greatly exceed even the
present age of the universe which is of the order of $10^{10}$ years. The value
of the relevant main sequence central temperature is derivable (by consideration
of the probability of coulomb barrier tunnelling by the ionic reactants) as
given in order of magnitude by the protonic analogue of the electronic Rydberg
energy, i.e. $\Theta\approx e^4 m_p$ which is logarithmic between the minimal
(ordinary electronic) Rydberg ionisation temperature $\Theta\simeq e^4 m_e$ and
of the maximal pair creation temperature $\Theta\approx m_e$.

The foregoing considerations lead to the prediction (in full agreement with
observation) not only that formation of stars in that mass range just above the
the Eddington - Chandrasekhar critical value should have been relatively common,
but also that most such moderately massive stars should already have passed the
ends of the thermonuclear lifetimes and so been already obliged to face the
issue of runaway gravitational collapse to densities in excess of the critical
Michell Laplace limit$^{[22][23]}$ value
$$\rho \approx {1\over M^2}   \eqno(1.12)$$
beyond which any description in Newtonian terms must be expected to break down,
the usual formula  for the scalar gravitational potential $\varphi\approx M/R$
with $R\approx (M/\rho)^{1/3}$ giving a result greater than unity, meaning that
the gravitational energy is greater than the rest mass energy and hence that the
escape velocity is greater than the speed of light, a situation that
corresponds, in the General Relativistic formulation described in the following
sections, to the light trapping mechanism that is the essence of the phenomenon
that is commonly referred to as the formation of a black hole. Since the speed
of light is normally supposed to represent an upper bound on the rate of
propagation of causal influences of any kind, the infalling matter within the
``horizon" (that is presumed to define the boundary of the region from which no
light escapes) will become causally decoupled from the outside region, which
thereby aquires the freedom to attain an equilibrium state of a new, essentially
non - material ``black hole" type, whose investigation will be the subject of
the discussion in the following sections. Assuming that the density would
retain its usual order of magnitude, i.e. that given by (1.1) Michell and
Laplace estimated  that light trapping would require a minimum mass of
the order of $10^7$ times that of the sun (a value so gigantic that it
was not taken seriously until, following up a suggestion by 
Lynden-Bell$^{[24]}$,  its potential relevance
to exotic quasar type phenomena in active galactic nuclei
 was pointed out by Hills$^{[25]}$ who noticed that it represents a
threshold value for tidal disruption of ordinary stars$^{[26]}$).

The paradox is that black holes in the relatively 
moderate mass range on which (by allowing for compressibility) our
attention has so naturally been focussed by the line of astrophysical
reasonning developped above i.e. a few times the Chandrasekhar
limit (1.6), would appear in practice to be very rare, despite the high
(predicted and observed) abundance of potential precursor stars. If these very
common massive main sequence stars do not become black holes, what happens
instead?

One much discussed idea which emphatically does {\it not} give the correct
explanation is that at the more than nuclear energy density that is attained at
the upper limit,  $\rho\approx m_p{^4}$, of the neutron star density range, the
equation of state deviates from the form (characterised by (1.7) and (1.5) ) on
which the above reasonning is based in such a way as to allow the existence of
ordinary material equilibrium states above the Chandrasekhar mass after all. The
theoretical objection to this idea is that it would require the pressure to
increase with density at a rate that would be incompatible with the causality
condition that presumeably requires the corresponding sound (compression wave)
speed with square given by $dP/d\rho$ to be less than unity. Since $P$ is small
compared with $\rho$ in the physically well understood low density regime,
respect for this causality requirement that we should have $P<\rho$ throughout
the entire range so that satisfaction of the virial equilibrium condition for a
given mass $M$ entails that the density $\rho$ cannot exceed the Michell limit
value given by (1.12). This consideration does not rule ou the possibility of
exotic ultra dense (e.g. quark nugget type) material equilibrium states with
$\rho>>m_p{^4}$ and correspondingly with $M<<1/m_p{^2}$, but it does rule out
their existence for higher mass values. A devil's advocate might still try to
argue that one could still get round Chandrasekhar's upper mass limit by
postulating some appropriately unorthodox relativistic gravitation theory for
which the virial condition (1.2) itself would be suitably modified, but such 
theoretical gymnastics would seem to be pointless in view of the the complete
absence of the slightest shred of observational evidence in favour of the
existence of any such weird states as would be produced that way. The conclusion
to be drawn from the intense astronomical activity of recent years is not just
that plausible black hole candidates in the mass range just above the
Chandrasekhar limit are comparitively rare, but also that there is no sign
whatsoever of any alternative non-black hole type of cold (as opposed to hot
stellar) equilibrium state at all in this mass range.

We thus get back to the basic question of what actually has happenned to the 
numerous stars in the moderately massive range $M>1/m_p{^2}$  that due to
of the comparitive shortness of the timescale (1.11) must have already
burned out by now.  The answer, which is implicit in the physics described in
the preceeding paragraphs, can be presented in terms of several successive 
steps. To start with, since they are never far from instability, the radiation
dominated stars inquestion will always tend to lose matter from their surface 
in the form of an outgoing winds which can carry away a very significant 
fraction of he original mass during the last stages of the thermonuclear 
lifetime. Secondly the dense burned out material that will accumulate in the 
core of the star will ultimately tend to evolve on its own almost 
independently of the comparitively diffuse (even if much more massive) outer 
envelope layers. As soon as there is a degenerate central core in excess of 
the Chandrasekhar limit it can be expected to collapse by itself without
waiting for the outer layers to be ready to follow. Surprised and shocked,
these outer layers will thus be vulnerable to being blown away in a supernova
type explosion by the energy released by the core collapse. The fact, refered to
above, that the neutron star mass limit is rather larger, perhaps about double,
that for the degenerate electron supported core, means that the core collapse
can be expected to be halted, with formation of the shock that acts back on the
outer layers, when the central density reaches that of neutron star matter.  The
conclusion (which is of course supported by a large amount of detailed numerical
calculations by many workers) is that while small main sequence stars can
obviously be expected to end up in white dwarf states, ones that are initially
much more massive can be expected to end up by forming only slightly more
massive neutron star remnants, the remainder of the mass being dispersed in the
form of a continous wind followed by an explosive burst, a picture whose broad
outline is fully consistent with what is actually observed.  Although the
details are complicated and still highly controversial, it is easy to see that
formation of a black hole is likely to be of more exceptional occurrence, due to
partial failure in a restricted parameter range of the supernova mass ejection
process, or to subsequent accretion from a binary partner onto the neutron star
remnant.

\medskip
\bigskip\parindent=0 cm
{\bf 2 The Example of Spherical Collapse.}
\medskip\parindent=1.2 cm

The collection of more or less well defined and physically plausible qualitative
notions - unstoppable collapse with formation of an event horizon hiding the
ensuing singularities - that constitute what may be referred to as the {\it
black hole paradigm} was originally derived from the relatively tractable
example provided by the spherically symmetric case, whose analysis will be the
subject of the present section. In the case of actual equilibrium states, a
considerable amount is now known about the non spherical generalisations that
will be the main subject of later sections, but as far as dynamical evolution is
concerned, although we can draw a few general qualitative conclusions (such as
the Hawking area theorem to be described in the next section) there is still
very little quantitative knowledge about what happens beyond the immediate
neighbourghood of spherical symmetry. This makes it necessary to rely rather
heavily on the spherical example despite of (and indeed even because of) the
fact that even today it is still far from clear to what extent the lessons
provided by the spherical model are generically valid.

The mathematical analysis in this and all the following sections will be based
on the standard Einstein theory of gravity as formulated in terms of a spacetime
manifold with local coordinates $x^{\mu}$ say ($\mu=0,1,2,3$) and Lorentz
signature metric field $g_{\mu\nu}$ (used for index lowering) that is governed
by dynamical equations of the form
$$G^{\mu\nu}\equiv  R^{\mu\nu}-{_1\over^2} R g^{\mu\nu}
=8\pi T^{\mu\nu}  \eqno(2.1)$$
where $ R_{\mu\nu}$ (with trace $ R= R^\mu{_\mu}$) is the Ricci
tensor of the spacetime metric $g_{\mu\nu}$ (see the appendix for definitions
and notation conventions) and $T^{\mu\nu}$ is an appropriately chosen
stress-momentum-energy density tensor whose form will depend on the kind of
matter under consideration but which, for consistency with (2.1), must of course
must always obey the ``covariant conservation" law
$$\nabla_\mu T^{\mu\nu}=0 \eqno(2.2)$$
where $\nabla_\mu$ is the standard operator of Riemannian covariant 
differentiation as defined with respect to $g_{\mu\nu} $.  

As explained in the appendix we shall use an underline whenever
necessary to distinguish quantities defined with respect to the geometry of an
imbedded surface under consideration from the analogous quantities as defined
with respect to the background geometry. As far as this present section is
concerned the relevant imbedded surfaces are to be understood as consisting of
the congruence of compact spacelike 2-surfaces  generated by the spherical
symmetry action, whose intrinsic Ricci curvature scalar will therefor, in
accordance with this convention, be denoted by $\underline R$ to distinguish it
from the background Ricci curvature scalar $R$. This allows the specification of
what we shall refer to as the Misner Sharp mass function, $M^\sharp$, by the
formula

$$(2\underline R)^{3/2} M^\sharp = 2\underline R-K_\mu K^\mu \eqno(2.3)$$
where $K_\mu$ is the extrinsic curvature vector of the spacelike two-surface, as
defined in the appendix.  This definition (whose right hand side is proportional
to the mean of Christodoulou's mutually conjugate ``mass aspect"
functions$^{[27][28]}$) has the advantage of being manifestly covariant and
giving a result that is well defined as a strictly local field for arbitrary
(not necessarily spherical) spacelike two surfaces, (in contrast with the
related but only semi-local Hawking mass$^{[28]}$, which involves surface
integration over the two-surface, but which, like  (2.3), was chosen so as to
agree with the original mass specification of Misner and Sharp$^{[29]}$ in the
spherical limit with which we are concerned here). 

The specially convenient feature of the scalar field defined by (2.3) is that 
in the spherically symmetric case its derivative is directly related to the 
Einstein tensor of the gravitational field equations by
(using square brackets to  denote antisymmetrisation) the  identity
$$  (2\underline R^3)^{1/2}\nabla_\mu M^\sharp=2\aG{^\rho}{_{ [\rho}} K_{\mu]}
\ , \ \ \ \ \ \aG_{\mu\nu}=\ag_\mu{^ \rho}
\ag_\nu{^\sigma}G_{\rho\sigma} \ ,\eqno(2.4)$$
using  
$\perp$ to indicate surface orthogonal projection 
(so that $\ag_{\mu\nu}\!=\!g_{\mu\nu}\!-\!\olg_{\mu\nu}$, where
$\olg_{\mu\nu}$ is the fundamental tensor of the
 spherical two surfaces - see appendix). This identity
was first derived (though not in such a manifestly covariant form) by
Misner and Sharp$^{[29]}$. It can be given a rather more explicit form by
introducing the usual circumferencial radius function $r$ of the spheres, in
terms of which their extrinsic curvature vector $K^\mu$ and Ricci scalar $R$
(whose surface integral is of course $8\pi$ by the Gauss Bonnet identity)
work out to be given simply by
$$K_\mu=-{2\over r}\nabla_\mu\ r \ , \ \ \ \ \ \ R={2\over r^2} \ .\eqno(2.5)$$
so that (2.4) reduces to the form
$$\nabla_\mu M^\sharp=r^2 G^\rho{_\nu}\ag{^\nu}{_{ [\mu} }
 \nabla_{\rho]} \ r ,\eqno(2.6)$$
whose derivation will now be described.

One of the convenient features of a spherical (as opposed to more general)
spacetime geometry is that it is possible to describe it in terms of
an orthonormal tetrad of covectors $\theta^{_{\Lambda}}{_\mu}$,
$\Lambda=0,1,2,3$, that is fully integrable in the sense that each one is
proportional to the gradient of a corresponding preferred coordinate. These may
be taken to be a provisionally unspecified space and time coordinate, $x^{_0}$
and $x^{_1}$ together with the usual spherical angle coordinates $\theta$ and
$\phi$, so that (using brackets to distinguish frame indices from coordinate
indices) one has
$$ \theta^{_{(0)}}_{\ \mu}dx^\mu=\si_{_0} dx^{_0} \ , \ \ \ \ \
\theta^{_{(1)}}_{\ \nu}dx^\mu=\si_{_1}dx^1  \ ,                  $$
$$\theta^{_{(2)}}_{\ \mu}dx^\mu=r\ d\theta   \ , \ \ \ \ \
\theta^{_{(1)}}_{\ \nu}=r\ {\rm sin}\theta\ d\phi  \ , \eqno(2.7)$$
of which the first pair of ``outer" frame vectors are orthogonal to the
two-spheres generated by the symmetry action while the last pair of ``inner"
frame vactors are tangential to them. The spherical symmetry is expressed by
the condition that the three unknown metric coefficients $r$, $\si_{_0}$,
$\si_{_1}$ are all functions of $x^{_0}$ and $x^{_1}$ only. In terms of these
quantities and of the fixed Minkowski frame metric $g_{_{\Lambda\Phi}}$ with
signature (-1,1,1,1) the metric form 
$$ds^2=g_{\Lambda\Phi}\theta^{_\Lambda}_{\ \mu}\theta^{_\Phi}_{\ \mu}\ dx^\mu\
dx^\nu
=\ag_{\mu\nu} dx^\mu dx^\nu
+\olg_{\mu\nu} dx^\mu dx^\nu \eqno(2.8)$$
where the ``outer" part is given by
$$\ag_{\mu\nu} dx^\mu dx^\nu =
-\si_{_0}{^2} dx^{_0\ 2} + \si_{_1}{^2} dx^{_1\ 2} \eqno(2.9)$$
while the metric within each two-sphere is given by the standard expression
$$\olg_{\mu\nu} dx^\mu dx^\nu = r^2( d\theta^{ 2} +
{\rm sin}^2\theta\ d\phi^{\ 2}) \ . \eqno(2.10)$$

Although the method most commonly given in textbooks proceeds by working out all
the (forty) Christoffel components, the quickest way of evaluating the curvature
tensor of a metric such as this is to use the Cartan technique$^{[30]}$
of  proceeding via the calculation of the connection forms 
$\varpi_\mu{^{_\Lambda}}{_{_\Phi}}$ which are got by solving the equations
$$\nabla_{[\mu}\theta^{_\Lambda}{_{\nu]}}=\theta^{_\Phi}{_{[\mu}}
\varpi_{\nu]}{^{_\Lambda}}{_{_\Phi}} \ , \ \ \ \ \ 
\varpi_{\mu_{(\Lambda\Phi)}}=0 \ , \eqno(2.11)$$
(using square and round brackets to indicate index antisymmetrisation and
symmetrisation respectively, and with the understanding that the fixed Minkowski
metric is used for lowering and raising of frame indices), the trick being that
due to the antisymmetrisation there is no need to know the Christoffel
connection components to carry out the covariant differentiation operations, the
result being obtainable simply by replacing the covariant differentiation
operators $\nabla_\mu$ by the corresponding partial differentiation operators
$\partial_\mu$.  The next step is to use this same (exterior
differentiation) trick again in evaluating the corresponding curvature form
$$R_{\mu\nu}{^{_\Lambda}}{_{_\Phi}}=2\nabla_{[\mu} 
\varpi_{\nu]}{^{_\Lambda}}{_{\Phi}} + 2\varpi_{_[\mu}{^{_{\Lambda\Theta}}}
\varpi_{\nu]_{\Theta\Phi}} \ . \eqno(2.12)$$

It will be convenient for what follows to use a systematic shorthand notation 
whereby a suffix $_{(0)}$ or $_{(1)}$ is used to indicate the effect of 
differentiation with respect to proper length in the space or time direction
respectively, i.e. to indicate the corresponding frame components of the
the covariant derivative when acting on a scalar, so that in particular,
for the circumferential radius function $r$ itself we have
$$\r_{_{(0)}}={1\over \si_{_0}}{\partial r\over\partial x^{_0}}\ , \ \ \ \ \
\r_{_{(1)}}={ 1 \over \si_{_1}}{\partial r\over\partial x^{_1}}\ . \eqno(2.13)$$
In terms of this notation scheme the six
independent connection forms are found from (2.15) to be expressible as
$$ \varpi_\mu{^{_{(0)}}}{_{_{(1)}}}={ \si_{_{0(1)}}\over \si_{_0}}
\theta^{_{(0)}}_{\ \mu}+{\si_{_{1(0)}}\over \si_{_1} }
\theta^{_{(1)}}_{\ \mu}\ , \ \ \ \
\varpi_\mu{^{_{(2)}}}{_{_{(3)}}}=- { {\rm cot}\theta\over r }
\theta^{_{(3)}}_{\ \mu} \ , $$
$$ \varpi_\mu{^{_{(0)}}}{_{_{(2)}}}={ r_{_{(0)}}\over r }
\theta^{_{(2)}}_{\ \mu} \ , \ \ \ \ \ \varpi_\mu{^{_{(0)}}}{_{_{(3)}}}=
{ \r_{_{(0)}}\over r } \theta^{_{(3)}}{_\mu}  \ , $$
$$ \varpi_\mu{^{_{(1)}}}{_{_{(2)}}}=- { \r_{_{(1)}}\over r }
\theta^{_{(2)}}_{\ \mu} \ , \ \ \ \ \ \varpi_\mu{^{_{(1)}}}{_{_{(3)}}}=
- { \r_{_{(1)}}\over r } \theta^{_{(3)}}{_\mu}  \ . \eqno(2.14)$$

The Cartan formula (2.16) can now be used for the direct evaluation of the
tetrad components $R_{_{\Theta\Psi}}{^{_\Lambda}}{_{_\Phi}}$ of the Rieman
tensor, the only ones that are independent (bearing in mind that the spherical
symmetry ensures that they are invariant under interchange of the indices
$_{(2)}$ and $_{(3)}$) being 
$$ R_{_{(0)(1)}}{^{_{(0)}}}{_{_{(1)}}} =
{\si_{_{1(0)(0)}}\over\si_{_1}}- {\si_{_{0(1)(1)}}\over\si_{_0}} \ ,
\ \ \ \ \ R_{_{(2)(3)}}{^{_{(2)}}}{_{_{(3)}}} ={1\over r^2 }\left( 1+
 \r_{_{(0)}}{^2} - \r_{_{(1)}}{}{^2}\right) \ ,  $$
$$ R_{_{(0)(2)}}{^{_{(0)}}}{_{_{(2)}}} ={\r_{_{(0)(0)}}\over r  }  
 -{ \r_{_{(1)}}\over r} {\si_{_{0(1)}}\over\si_{_0}}  \ ,
\ \ \ \ \ R_{_{(1)(2)}}{^{_{(1)}}}{_{_{(2)}}} =
{ \r_{_{(0)}}\over r}{ \si_{_{1(0)}}\over \si_{_1}} 
- {\r_{_{(1)(1)}} \over r} \ ,  $$
$$ R_{_{(1)(2)}}{^{_{(0)}}}{_{_{(2)}}} ={\r_{_{(0)(1)}}\over r } -  
{ \r_{_{(1)}}\over r} {\si_{_{1(0)}}\over \si_{_1}}  \equiv
{\r_{_{(1)(0)}}\over r  }-{\r_{_{(0)}}\over r}
{\si_{_{0(1)}}\over \si_{_0}} \ .\eqno(2.15)$$
The corresponding frame components of the Einstein tensor will be given
in terms of these by
$$G_{_{(0)(0)}}=2 R_{_{(1)(2)}}{^{_{(1)}}}{_{_{(2)}}} +
R_{_{(2)(3)}}{^{_{(2)}}}{_{_{(3)}}}  \ ,\ \ \ \
G_{_{(1)(1)}}=-2 R_{_{(0)(2)}}{^{_{(0)}}}{_{_{(2)}}} -      
R_{_{(2)(3)}}{^{_{(2)}}}{_{_{(3)}}}  \ ,$$
$$G_{_{(0)(1)}}=- 2 R_{_{(1)(2)}}{^{_{(0)}}}{_{_{(2)}}} \ , \ \ \ \
G_{_{(0)(2)}}= G_{_{(1)(2)}}= G_{_{(2)(3)}}= 0 \ ,$$
$$G_{_{(2)(2)}}=- R_{_{(0)(1)}}{^{_{(0)}}}{_{_{(1)}}} - 
R_{_{(0)(2)}}{^{_{(0)}}}{_{_{(2)}}} - R_{_{(1)(2)}}{^{_{(1)}}}{_{_{(2)}}} 
 \  . \eqno(2.16)$$

The resulting system can be considerably simplified by imposing that the
coordinates be {\it comoving} with respect to the flow congruence determined
by the eigenvector of the energy momentum tensor, which is equivalent
to the condition that the frame be such as to diagonalise the Einstein
tensor, i.e. 
$$ G_{_{(0)(1)}} = 0\ \ \ \ \ \Leftrightarrow \ \ \ \ \
\si_{_1} r_{_{(0)(1)}}= r_{_{(1)}}  \si_{_{1(0)}} \ , \eqno(2.17)$$
Subject to this requirement, which it is to be emphasized is not a physical 
restriction but just a gauge condition,  the only two  Einstein tensor
components still needed for the gravitational field equations can be seen to
be expressible directly in terms of the
 Misner Sharp mass function specified by (2.3) whose explicit form is
$$ M^\sharp={r\over 2}\left(1+r_{_{(0)}}{^2}-r_{_{(1)}}{^2}
\right) \ , \eqno(2.18)$$
as
$$G_{_{(0)(0)}}  =  {2M^{\sharp}_{_{(1)}}\over r^2 r_{_{(1)}}} \ , \ \ \ \ \ \
\ \ \ G_{_{(1)(1)}}=-{2M^{\sharp}_{_{(0)}}\over r^2 r_{_{(0)}}}\ . \eqno(2.19)$$
This can be seen to be just the frame component translation of the 
covariant version (2.6) of the Misner Sharp identity, whose derivation is
thus completed. It is to be remarked that $G_{_{(2)(2)}}$, the only remaining independent
Einstein tensor component not given by(2.6), is not needed, because the
equation in which it is involved will automatically hold as an identity whenever
the other Einstein equations and the consistency condition (2.2) are satisfied.

In considering the contributions to the right hand side of the Einstein 
equations (2.1) it is commonly convenient to work with a decomposition of 
the form
$$T^{\mu\nu}=T_{_M}{^{\mu\nu}}+T_{_F}{^{\mu\nu}} \eqno(2.20)$$
in which $T_{_M}{^{\mu\nu}}$ is a ``strictly material" contribution and
$T_{_F}{^{\mu\nu}}$ is an electromagnetic field contribution given by
$$ T_{_F}{^{\mu\nu}}={1\over 4\pi} \left(F^\mu{_\rho}F^{\nu\rho}
-{1\over 4}F_{\rho\sigma}F^{\rho\sigma}g^{\mu\nu}\right)\eqno(2.21)$$
in terms of an electromagnetic gauge curvature field
$$F_{\mu\nu} =2\nabla_{[\mu}A_{\nu]}   \eqno(2.22)$$
(where square brackets indicate antisymmetrisation of the included indices)
with (again necessarily) conserved source current,
$$J^\mu={1\over 4\pi} \nabla_\nu F^{\mu\nu}\ ,  \ \ \ \ \
\nabla_\mu J^\mu=0 \ . \eqno(2.23)$$
This formulation makes it possible to characterise the important ``electrovac"
case as that in which the source contributions $T_{_M}{^{\mu\nu}}$ and $J^\mu$
both vanish, the strict vacuum case being that in which the field $F_{\mu\nu}$
also vanishes. 

For many purposes, including those of the present section, it is sufficient
to use a treatment in which the source contributions are not
necessarily restricted to vanish but in which they are postulated to have the
particularly simple form describable as that of a non conducting perfect fluid,
meaning that there is a preferred timelike unit vector $u^\mu$ and associated
orthogonal projection tensor $\gamma^\mu{_\nu}$ as characterised by
$$\gamma_{\mu\nu}=g_{\mu\nu}+u_\mu u_\nu \ ,\ \ \ \ \ \ u_\mu u^\nu=-1
\eqno(2.24)$$
with respect to which the material and electromagnetic source fields
are spacially isotropic, meaning that they satisfy
$$T_{_M}{^{\mu\nu}}=\rho u^\mu u^\nu + P \gamma^{\mu\nu}\ , \ \ \ \ \
J^{[\mu}u^{\nu]}=0 \ , \eqno(2.25)$$
where the eigenvalues $\rho$ and $P$ are to be interpreted as the precise local
values of the mass density and pressure whose characteristic mean values were
the subject of discussion in the crude Newtonian order of magnitude treatment
of the previous section.  

It will be sufficient for the purpose of the present section to further restrict
our attention to the case of an adiabatic model, for which $P$ is determined 
as a (not necessarily uniform) function of $\rho$ allong each world line. As
far as spherical applications are conserned there will be no further loss of
generality in taking the model to be characterised by a pair of conserved number
currents
$$s^\mu=n u^\mu\ , \ \ \ \ \ n^\mu=n u^\mu\ , \ \ \ \ \ \nabla_\mu s^\mu=0 \ , \
\ \ \ \nabla_\mu n^\mu=0 \eqno(2.26)$$
with $n$ interpretable as the baryon number density and $s$ as an entropy
density, in terms of which the mass density $\rho$ is specified by a (uniform)
equation of state function whose derivatives, interpretable as the effective
temperature $\Theta$ say and the effective mass-energy per particle or chemical
potential $\mu$ say, determine the corresponding pressure function $P$ by the
familiar relation 
$$ P=s\Theta +n\mu-\rho\ ,\ \ \ \ \Theta={d\rho\over ds}\ ,\ \
\ \ \mu={d\rho\over dn} \ . \eqno (2.27)$$ 
In terms of the corresponding thermal and particle
four-momentum covectors, $$ \Theta_\mu=\Theta u_\mu\ , \ \ \ \
\mu_\rho =\mu u_\rho \ , \eqno (2.28)$$
and subject to the conservation laws (2.26)
the perfect fluid equations of motion obtained from (2.2) are 
expressible just as the momentum transport equation 
$$2u^\rho(s\nabla_{[\rho} \Theta_{\sigma]}+ n\nabla_{[\rho} \mu_{\sigma]} )
= F_{\sigma\rho}J^\rho \ .\eqno(2.29)$$ 
From a computational point of view, this latter
formulation$^{[31][16][32]}$ has the advantage  
(as compared with (2.2)) that, as in (2.11) and (2.12), the
antisymmetrised ``exterior" nature of the derivation involved makes it possible
to work it out directly, by direct substitution of the partial differentiation
operator $\partial_\mu\equiv$ $\partial/\partial x^\mu$ in place of the
Riemannian operation $\nabla_\mu$, thereby making it possible to avoid having to
go to the trouble of working out the Christoffel connection components.

As a prerequisite to applying the formulae (2.19) in 
this perfect fluid case, it is
necessary to impose the gauge restriction (2.17) to the effect that the
coordinates should be comoving, which means that the timelike frame vector $
\theta^{_{(0)}}{_\mu}$ is to be identified with the unit flow vector $u^\mu$ of
the fluid as introduced in (2.24), so that by (2.25) and (2.26) the particle
number and electric source currents will be given by
$$J^\mu = en^\mu\ , \ \ \ \ n_\mu= n\theta^{_{(0)}}_{\ \mu}\ , \eqno(2.30)$$
where the parameter $e$ represents the electric charge per particle.
It follows that the particle, entropy, and charge
conservation laws (2.26) and (2.23) will be expressible
simply by
$$(n r^2 \si_{_1})_{_{(0)}}=0\ , \ \ \ \left({s\over n}\right){_{_{(0)}}}
= 0 \ , \ \ \ \ \ e_{_{(0)}} =0 \eqno(2.31)$$
or equivalently by
$$N_{_{(0)}} =0\ , \ \ \ \ \ S_{_{(0)}} =0\ , \
 \ \ \ \ Q_{_{(0)}} =0 \ ,\eqno(2.32)$$
where $N$, $S$ and $Q$ are fields respectively representing
the total particle number, entropy, and electric charge within the the
corresponding sphere, which will be given as integrals over the interior of the
sphere at a fixed value of the coordinate time
$x^{_0}$ by 
$${N\over 4\pi}=\int r^2 n \si_{_1}dx^{_{_1}}\ , \ \ \ \ {S\over 4\pi}
=\int r^2 s \si_{_1}dx^{_1}\ ,\ \ \ \ {Q\over 4\pi}
=\int r^2 en \si_{_1}dx^{_1} .\eqno(2.33)$$
The corresponding frame components of the total energy momentum
tensor are then then obtainable from from (2.21) and (2.25) as
$$ T_{_{(0)(0)}} =\rho +{E^2 \over 8\pi} \ , \ \ \ \ \
T_{_{(1)(1)}} =P - {E^2 \over 8\pi}  \ , \eqno(2.34)$$
where the appropriate electric field magnitude,
 $E=F_{_{(1)(0)}}$,  
is obtainable by direct integration from the source equation
 (2.22) in the form
$$ E={Q\over r^2} \ . \eqno(2.35)$$
It can be seen from (2.21), by a similar integration, that the corresponding
magnetic field component $F_{_{(2)(3)}}$ is necessarily zero, i.e. there can be
no magnetic monopole moment, on the assumption that (initially at least)  there
is a well behaved spherical centre from which the integrals in (2.33) are
understood to be taken. It is to be remarked that the  other (crossed) field
components  $F_{_{(0)(2)}}$, $F_{_{(0)(3)}}$, $F_{_{(1)(2)}}$, $F_{_{(1)(3)}}$
all vanish trivially as a locall requirement for spherical symmetry.

Just as the classically familiar Coulombian form of the relation (2.35)
is due to an judicious choice of definitions of the variables involved, so
also the particularly astute Misner Sharp choice$^{[29]}$ for the definition
of the mass function $M^\sharp$ leads to a pseudo Newtonian form for
the integral relation expressing the spacial constraint resulting from the
first of the Einstein equations obtained from (2.19) which gives
$$M^\sharp=4\pi\int r^2 (\rho+{E^2\over 8\pi})dr \ ,\eqno(2.36)$$
in which, as in (2.33) it is to be understood that the integral is
taken over the interior of the relevant sphere at a fixed value of the comoving
time coordinate, (starting from a central origin that is assumed to be regular,
at least initially) which means that the radial variation will be expressible in
terms of that of the space coordinate $x^{_1}$ by the relation
$dr=\r_{_{(1)}}\Psi_{_1}dx^{_1}$.

Subject to the constraint (2.36) and the shift condition (2.17) which
can be rewritten in the symmetrically equivalent form
$$\si_{_0} r_{_{(1)(0)}} =r_{_{(0)}} \si_{_{0(1)}}\ , \eqno(2.37) $$
 the only other Einstein equation
that is needed is the dynamical equation
$$M^\sharp_{_{(0)}}= ({E^2\over 2}-4\pi P)r^2\r_{_{(0)}}\ .\eqno(2.38) $$
These two equations are to be solved in conjunction  with the constraint obtained from the
momentum transport equation (2.12 which takes the form
$$(\rho+P)\si_{_{0(1)}}=\si_{_0}\left( 
Een -P_{_{(1)}}\right)    \eqno(2.39)$$
in which the spacial pressure gradient is of course given by
$P_{_{(1)}}=n\mu_{_{(1)}}+s\Theta_{_{(1)}}$.

It is apparent at this stage that it will be convenient to introduce a 
modified, combined (electromagnetic as opposed to purely gravitational)
mass function $M$ say, given in terms 
of the original Misner Sharp mass function $M$ by
$$M=M^\sharp+{Q^2\over 2r}\eqno(2.40)$$
or equivalently by the more direct relation
$$1-{2M\over r}=r^2(K_\mu K^\mu-E^2) \ . \eqno(2.41)$$
It follows from the Misner Sharp identity  (2.4) in conjunction with the 
Einstein equations (2.1) that the gradient of this combined mass function will 
be given in terms of the purely ``material" contribution $T_{_M}{^{\mu\nu}}$ 
in the decomposition (2.20) by the manifestly covariant expression
$$\nabla_\mu M= {Q\over r} \nabla_\mu Q +8\pi r^2 T_{_M}{^\rho}{_\nu}
\ag{^\nu}{_{[\rho}}  \nabla_{\mu]} r
\ .\eqno(2.42)$$   For the case of
a nonconducting perfect fluid in the comoving frame, the constraint (2.36) and
the dynamical equation (2.38) can be rewritten in terms of this new combined 
mass variable as
 $$M=4\pi\int r^2 (\rho+neE{r\over r_{_{(1)}}})dr \ ,\eqno(2.43)$$
and
$$M_{_{(0)}}= -4\pi Pr^2\r_{_{(0)}}\ .\eqno(2.44) $$
of which the latter has the avantage of having the same simple form as that to 
which the original version (2.38) would reduce if no electromagnetic effects 
were present.  An analogous remark applies also to (2.44) in  any
external region where the charge density $ne$ vanishes.

Whichever formulation is used, the solution of the system  will in general
require numerical computation, as in the pionneering attempt at an
astrophysically realistic calculation by May and White$^{[23]}$ or the important
investigation of the possibility of naked singularity formation by Eardley and
Smarr$^{[34]}$. It is however possible to obtain analytic solutions in special
cases of which the most obvious are those in which the circumstances are such
that the right hand side of the constraint equation (2.39), and hence also that
of (2.37) is zero so that we obtain 
$$\si_{_0}=1\ , \ \ \ \ \ r_{_{(1)(0)}}=0\ ,  \eqno(2.45)$$ 
for a suitable normalisation of the time coordinate $x^{_0}$ which in this
particular case is adjustable to agree with the proper time allong the flow
lines. Such a possibility obviously occurs in layers of matter 
that are uncharged and
for which the pressure gradient $P_{_{(1)}}$ is zero, either because the
configuration is homogeneous as in the classic prototype collapse calculation of
Oppenheimer and Snyder$^{[35]}$ or because the matter has a pressure free (so
called ``dust") equation of state. A fluid of this latter uncharged dust type
(for which the flow will simply be geodesic) is characterisable by 
$$e=0\ ,\ \ \ \ \rho= mn   \ , \ \  \ \ m_{_{(0)}}=0 \,\eqno(2.46)$$
where $m$ is a constant mass per particle, so that in a layer of this type
$Q$ will be 
constant not just in time but also in space, while the combined mass function 
$M$ will at least be constant in time:
$$Q_{_{(1)}}=0 \ ,\ \ \ \ \ \ \ M_{_{(0)}} = 0 \ .\eqno(2.47)$$
This means that the radial evolution equation , which by the definition of 
$M$ will allways have the form
$$\
r_{_{(0)}}{^2}=r_{_{(1)}}{^2}-1+{2M\over r}-{Q^2\over r^2} \ ,
 \eqno(2.48)$$ 
will in this case be independently integrable for each flow line,
 since by (2.32), (2.45), and
(2.47) the quantities $Q$, $r_{_{(1)}}$, and $M$ appearing on the right
will all be constants allong each separate flow line. The simplest possibility
is the ``parabolic" case corresponding to zero radial velocity in the large 
radius limit which is got by taking
$$r_{_{(1)}}=1\ ,\ \ \ \ x^{_0} = c^{_0}-{(Mr+Q^2)\sqrt{2Mr-Q^2}\over 3M^2}
\ ,\ \ \ \ \ c^{_0}_{_{(0)}}= 0 \eqno (2.49)$$
where $c^{_0}$ like $M$ is an initially arbitrary constant along each flow line,
i.e. a function only of $x^{_1}$. This latter comoving space variable can now be
replaced (except in the special case for which both $c^{_0}$ and $M$ are
spacially uniform) by $r$ in the ``outer" part of the metric (2.8) which thereby
acquires the form
$$ \ag_{\mu\nu}dx^\mu dx^\nu =-
(1-{2M\over r}+{Q^2\over r^2})(dx^{_0})^2 +
2\sqrt{({2M\over r}-{Q^2\over r^2})}dx^{_0}dr +dr^2\eqno (2.50)$$
with the combined mass variable $M$ now determined implicitly through its 
functional dependence on $c^{_0}$ by the relation (2.49). 

The class of solutions specified by (2.49) and (2.50) is by no means simple
and it is only comparitively recently (with the work of Eardley and 
Smarr$^{[34]}$ on the uncharged $Q=0$ case)
that they have started to be examined seriously from the point of view of
questions such as naked singularity formation.  They do however include the 
genuinely simple electrovac case for which the mass coefficient $m$ in
(2.46) is set equal to zero, which implies the constancy in space as well as 
comoving time of the combined mass variable $M$ (but therefore not of the 
original Misner Sharp mass variable $M^\sharp$ except in the $Q=0$ case for 
which they coincide). In this special electrovac case, as characterised by
$$ m=0 \ , \ \ \ \ \ M_{_{(1)}}=0  \ , \eqno(2.51)$$
the flow just represents a geodesic test particle congruence, so there
in favor of the radius variable $r$ so that     there is no further loss
of generality in imposing the parabolicity condition (2.49). This means
that the form (2.50) with not just $Q$ but now also $M$ taken to be 
constant in space as well as time represents the most general spherical 
electrovac solution ( appart from the exceptional
Robinson-Bertotti case for which $c^{_0}$ is uniform,
 so that$^{[10]}$ one obtains a tubular 
universe with constant radius $r$ throughout).  This solution can be seen
to be automatically stationary since all dependence on $x^{_0}$ has dropped 
out, which in the spherical case means more particularly that it must be
static, i.e.  that it is unaffected not only by displacements but also by
reversals of a certain preferred time coordinate, $t$ say, that is determined
(modulo a constant of integration) by the differential relation
$$dt=dx^{_0}-{r\sqrt{2Mr-Q^2}\over r^2-2Mr+Q^2}  \ . \eqno(2.52)$$
Replacement of the comoving proper time coordinate $x^{_0}$ by this
preferred time coordinate $t$ leads to the manifestly static form 
$$ \ag_{\mu\nu}dx^\mu dx^\nu =-\left(1-{2M\over r}+{Q^2\over r^2}\right)dt^2 +
\left( 1-{2M\over r}+ {Q^2\over r^2}\right)^{-1} dr^2\eqno (2.53)$$
that was originally derived by Reissner and Nordstrom on the basis of the
postulate of staticity at the outset. The present approach, showing how 
staticity is obtained automatically as a consequence of spherical symmetry in 
the source free case, amounts to a demonstration of what is known as 
Birkhoff's theorem.

Our parabollically infalling version (2.50) has the significant
advantage over the algebraically simpler historic
form (2.53) that it remains well behaved on the ``Killing horizons"$^{[36]}$, 
i.e. stationary null hypersurfaces that occur, whenever $Q^2\leq M^2$,
at the roots
$$r=r_{\pm} \ , \ \ \ \ \ r_{_\pm}=-M\pm\sqrt{M^2-Q^2} \eqno(2.54)$$
whereas the manifestly static version (2.53) is singular there. However although
it is sufficient for describing the outside of a collapsing spherical charged or
neutral star model, even the more sophisticated version (2.50) has the
limitation of being geodesically incomplete even when extended over the full
coordinate range $0\leq r<\infty$, $\ -\infty< x^{_0}< \infty$. The
geometrically complete manifold was first described in the pure vacuum case,
$Q=O$ by Kruskal and Szekeres$^{[37]}$ and in the generic case $Q^2<M^2$ by
Graves and Brill$^{[38]}$, while for the special ``maximally charged" limit
case $Q^2=M^2$ the corresponding construction was first carried out rather later
by myself$^{[39]}$. 

It was for the purpose of describing such extensions  that I first introduced
$^{[39][40]}$ the representational technique of {\it conformal projection} (the
space time analogue of the beloved Mercator projection of terrestrial
navigators) that has since been generally adopted as a standard tool for
understanding the topological and causal structure of any timelike two
dimensional manifold or submanifold, the idea being to first convert the metric
into null coordinate form
$$ \ag_{\mu\nu}dx^\mu dx^\nu =-\Psi du^{_+} du^{_-} \eqno(2.55)$$
which is always locally possible for some conformal factor $\Psi$ determined
as a function of the null coordinates $u^{_+}$ and $u^{_-}$, and then to take 
advantage of the fact that this null form is preserved, only the functional 
dependence of the conformal factor being altered,  by a transformation
$u_+\mapsto\tilde u^{_+}$, $u_-\mapsto\tilde u^{_-}$, $\Psi\mapsto\tilde \Psi$
whereby each of the null cordinates is replaced by an arbitrary function only of
itself, which one is free to choose in such a way as to cover what from a metric
or affine point of view might be an infinite region by a finite coordinate range
which can thus be plotted directly as a diagram.

\begin{figure}
\centering
\epsfig{figure=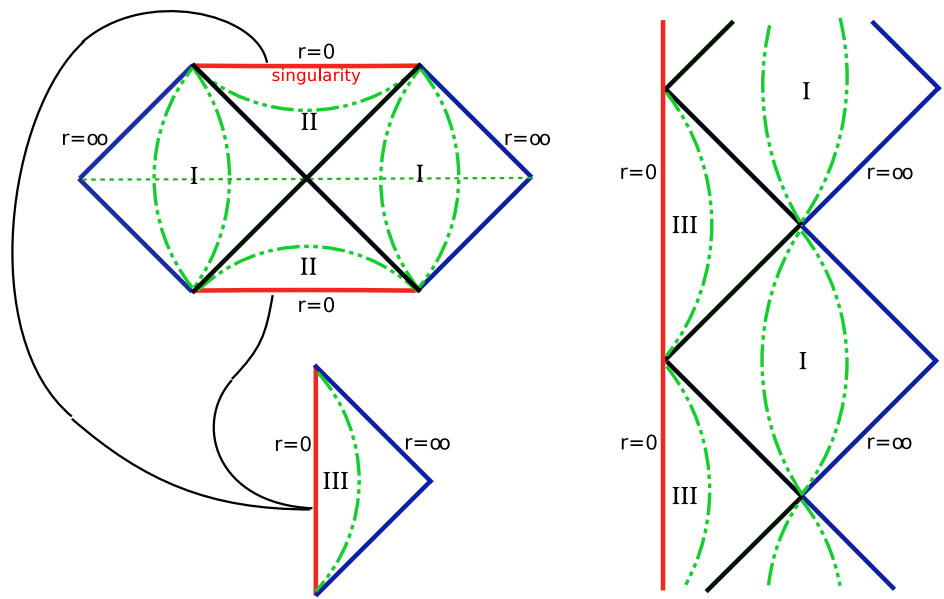, width=12.4 cm}
\caption{\color{navy}
\label{Schwarz}
{\it Facsimiles of the original C.P. diagrams$^{[39][40]}$ for
the limiting cases $Q^2=0$, i.e. Kruskal-Schwarzschild (including 
disconnected branch representing analyutic extension to region $r<0$),
and $Q^2=M^2$, i.e.  ``maximally charged" Reissner-Nordstrom.  }
}
\end{figure}

 The use of this method of
representation of event horizons in two dimensional manifolds was suggested to
me by the example from a related but rather different context (namely  the study
of the distant assymptotically flat outer regions) of Penrose's conformal
boundary procedure procedure$^{[41]}$ (as used for setting up what is known to the
initiated by the term prounounced as ``Scri"). However the concepts of a
conformal boundary and a conformal projection should not be confused. The
Penrose conformal boundary (Scri)  concept is not restricted to two dimensions,
but on the other hand it is dependent on rather severe assymptotic regularity
conditions that may fail in many relevant applications, whereas the conformal
projection (C.P.) technique is limited to two dimensional sections but not
otherwise restricted, so that in particular it is very useful for the analysis
of boundaries that may be singular.

In the present case the most obvious choice to start with is to take the
the incoming and outgoing stationarity preserving null coordinates which
 are defined (modulo an arbitrary choice of origin ) by
$$du^{_\pm}=dt\pm{r^2\ dr\over r^2-2Mr+Q^2}\eqno(2.56)$$
whose separate substitution gives the forms
$$\ag_{\mu\nu}dx^\mu dx^\nu =-{r^2-Mr+Q^2\over r^2}(du^{_\pm})^2
\pm du^{_\pm}dr \ .\eqno(2.57) $$
These null inflowing and outflowing coordinate forms (like our original
parabollically inflowing form (2.53)) are locally well behaved on the
Killing horizons at $r=r_\pm$ but nevertheless still incomplete. When both
are substituted together, with the radius variable now considered no longer
as a coordinate in its own right but just as a function of the null
 coordinates, the metric aquires the required doubly null form (2.55)
with the conformal factor given by
$$\Psi={r^2-2Mr+Q^2\over r^2} \eqno(2.58)\ . $$

In the generic case $Q^2<M^2$, the integration of (2.56) gives the
explicit expressions
$$u^{_\pm} =t\pm r \pm {1\over 2\kappa_+}{\rm ln}\vert r-r_+\vert\pm
{1\over 2\kappa_-} {\rm ln}\vert r-r_-\vert \ , \eqno(2.59)$$
from which, in terms of the {\it decay constants} of the Killing horizons which
are given by
$$\kappa_\pm=\pm{(M^2-Q^2)^{1/2}\over r_\pm^{\ 2}} \ ,\eqno(2.60)$$
one obtains the functional dependence of the variable $r$
in (2.59) in the implicit form
$$ 2r + {1\over\kappa_+}{\rm ln}\vert r-r_+\vert - {1\over\kappa_-} 
{\rm ln}\vert r-r_-\vert = u^{_+}-u^{_-}\ , \eqno(2.61)$$
whose unambiguous solution requires the specification that $r$ should lie in
some particular one of the three possible ranges characterised by the condition
that neither, just one, or both of the quantities $r-r_\pm$ be positive. Except
for the first of these three possibilities, which includes the value $r=0$ that
correponds to an irremovable geometric singularity, the resulting conformal
factor $\Psi$ will be regular over the full coordinate range
$\infty < u^{_\pm}< \infty$, but from the point of view of completeness the
metric version given by (2.58) and (2.61)  is no improvement on the traditional
manifestly stationary version (2.53): all that has been achieved is to push the
Killing horizons out of the coordinate chart, but not to regularise them.
However a genuine regularisation is now easily obtainable by a conformal
coordinate transformation to a a new null coordinate form
$$ \ag_{\mu\nu}dx^\mu dx^\nu =\tilde\Psi
 d \tilde u^{_+} d\tilde u^{_-} \ . \eqno(2.62)$$
Depending on whether it is an ``outer" Killing horizon at $r=r_+$
or an ``inner" one at $r=r_-$ that one wishes to cover,
it suffices to take
$$\tilde\Psi=  {\pm e^{-2\kappa_\pm}(r-r_\pm)\over
\kappa_\pm^{\ 2}r^2 \vert r-r_\mp\vert ^{\kappa_\pm/\kappa_\mp}} \ ,\eqno(2.63)$$
where $r$ is now given implicitly  by
$$\pm(r_\pm-r)e^{2\kappa_\pm r}\vert r-r_\mp\vert^{\kappa_\pm/\kappa_\mp}
=\tilde u^{_+}\tilde u^{_-}  . \eqno(2.64)$$
This relation shows in particular how, at the outer horizon the decay parameter
$\kappa_+$ is interpretable as measuring the exponential relation between
the affine time parameter $\tilde u^{_-}$ and the group parameter $u^{_-}$.

The unambiguous solution  of (2.64) requires the specification that $r$ should
lie in one or other of just two possible ranges characterised respectively by
$r<r_\mp$ and $r>r_\pm$, and except for the irremovable geometric singularity at
$r=0$ in the first of these two ranges, the new conformal factor $\tilde\Psi$ will
be regular over the full coordinate range $\infty<\tilde u^{_\pm}< \infty$, of
the new null coordinates, including the locus $r=r_\pm$ which can be seen to
consist of two intersecting Killing horizons characterised in the new
coordinates by $\tilde u^{_+}=0$ and $\tilde u^{_-}=0$ respectively.    The
stationarity group transported coordinates  $u^{_+}$ and $u^{_-}$ of the
original system in the more restricted patches on either side of these now
regularized horizons are given in terms of the new ones, which can be seen to
be characterised by the property of measuring {\it affine} distance allong the
regularised horizon at $r=r_\pm$, by relations of the simple exponential form
$$\kappa_\pm u^{_+}= {\rm ln}\vert \tilde u^{_+}\vert \ , \ \ \ \ \
\kappa_\pm u^{_-}=-{\rm ln}\vert\tilde u^{_-}\vert \ . \eqno(2.64)$$

The use of transformations of the simple form (2.64) allows us, according to
choice, to cover either the locus $r=r_+$ or the locus $r=r_-$ with a regular
coordinate chart, but it does not allow us to cover both at once. Nevertheless
since the alternative kinds of chart overlapp (in the intermediate range $r_-<
r<r_+$ where both are perfectly regular) they can be used as successive patches
to build up a maximally estended manifold in the manner first described (in
terms of a somewhat different system) by Graves and Brill$^{[38]}$. It is for the
purpose of visualising the final result of such successive extensions that the
C.P. (conformal projection) technique$^{[39][40]}$ is particularly useful. If one is
willing to sacrifice the desideratum of having a simple analytic expression such
as (2.64), there is no obstacle in principle to the introduction of further
modified null coordinates $\tilde{\tilde u}{^{_\mp}} $say whose range covers the
entire maximally extended manifold.

\begin{figure}
\centering
\epsfig{figure=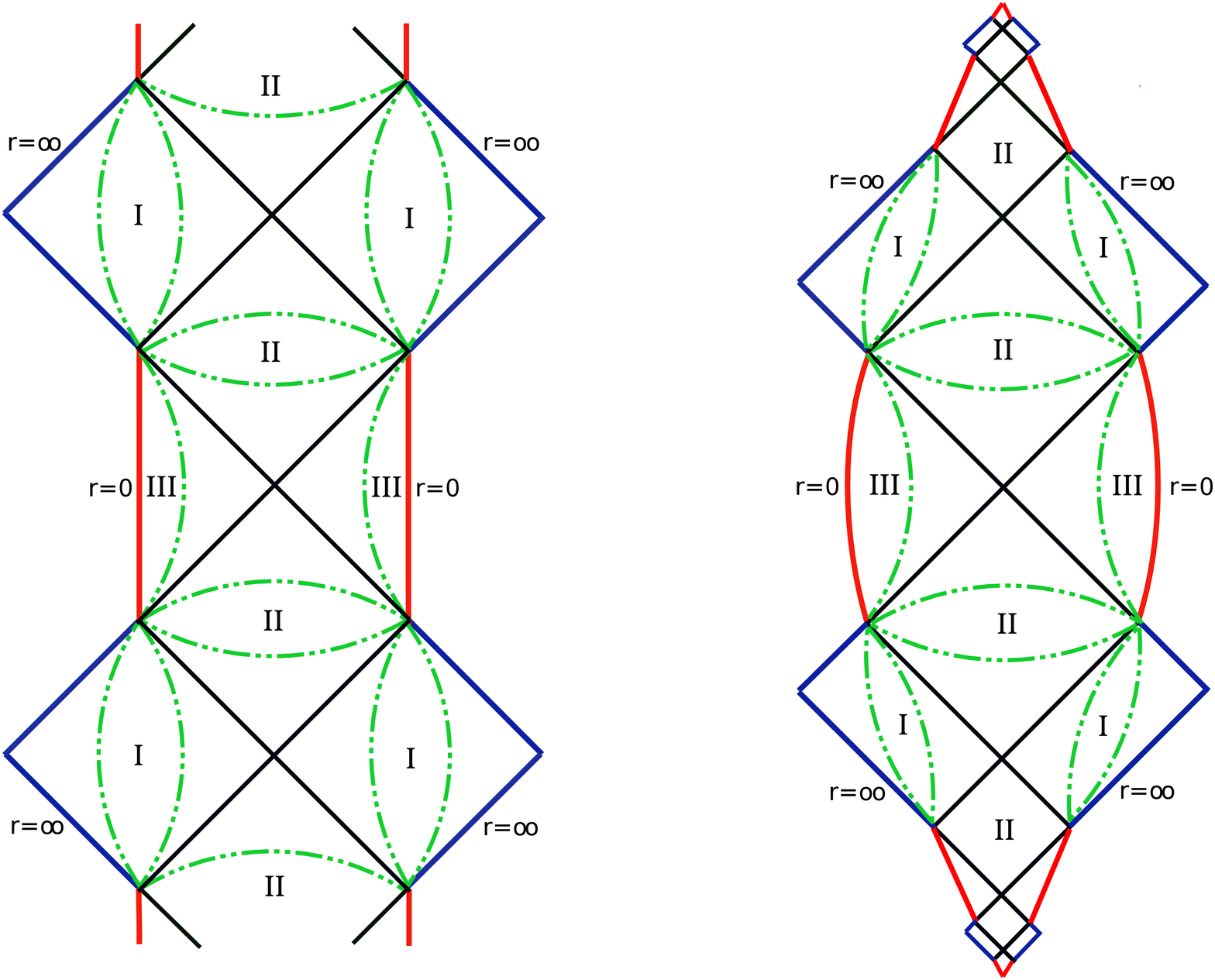, width=12.4 cm}
\caption{\color{navy}
\label{Comp}
{\it Facsimile of the original C.P. diagram$^{[40]}$ for
the Graves - Brill extension of Reissner Nordstrom  for $0 <Q^2 <M^2$
together with previously unpublished ``new look" version in which
complete compactification is achieved by letting the scale for successive
universes  tend to zero at the extremities of the chain.}
}
\end{figure}

An invaluable practical feature of a C.P. diagram of the kind obtained by
plotting such coordinates directly (traditionally with a diagonal
orientation) on a flat screen or page (as in figures 2 and 3 ) is that, as far as
the essential causal and topological features are concerned, it does not
matter whether or not one knows the precise functional form of the
functional relation between the original (restricted) and new (extended)
null coordinate systems: provided the linear (diagonal) representation of
the null congruences is preserved, any smooth (not necessarily analytic)
deformations are admissible. This means that (provided it is not restricted
by the inclusion of too much detail) any C.P. diagram that has been
constructed as a rough free-hand sketch has the beautiful feature of being
interpretable post facto as an accurate representation in terms of null
coordinates whose precise specification (if one were interested) could in
principle by found out later by carrying out empirical measurements on the
sketch.

\vfill\eject

\parindent = 0 cm \
{\bf 3. Qualitative theory of non-spherical Black Hole formation.}
\medskip
\parindent=2 cm

Whereas a considerable amount is known about non spherical black hole
equilibrium states (to which the subsequent sections will be devoted) as also
about non stationary states of spherical collapse (the subject of the previous
section) the subject of generic nonspherical gravitational collapse and black
hole formation still consists mainly of a few vague, qualitative, and for the
most part far from rigourously established notions, that are largely inspired by
the spherical example. The question of the extent to which various features of
spherical collapse scenarios may be taken over to more general situations has
long been and still remains a subject of animated debate.  The unreliability of
the spherical example as a guide to more general cases is shown by the case of
Birkhoff's theorem, to the effect that (as was demonstrated in the previous
section) the source free (strict or electromagnetic) vacuum outside a collapsing
spherical object must necessarily be static (i.e.  not only time independent but
even time reversal invariant) whereas in the non spherical case it need not even
be stationary (i.e. time independent) in view of the possibility of
gravitational and electromagnetic radiation whose absence, exceptionally, in the
spherical case is due to the absence of any scalar  part of either the
electromagnetic field which is purely vectorial or the gravitational field which
is purely (i.e. tracelessly) tensorial, at least in Einstein's theory to which
our discussion here  is restricted.

\begin{figure}
\centering
\epsfig{figure=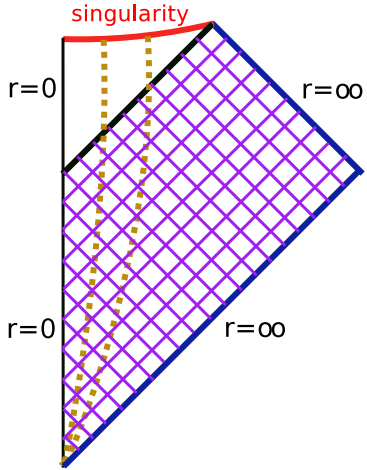, width=4.8 cm}
\caption{\color{navy}
\label{ColDOC}
{\it Illustration of the standard black hole paradigm by  a C.P. diagram representing
a two dimensional radial section through a collapsing body with cross hatched shading 
of the D.O.C. (domain of outer communications).}
}
\end{figure}

Among the features that are generally thought to survive the the breaking of
spherical symmetry of the collapse, some of the most important may be listed as
follows:

 \parindent =0 cm
(1) The ultimate formation of a {\it singularity} of some kind ( not necesarily
just a simple density singularity, but something sufficient to prevent affine
completeness) was shown by the work of Penrose and Hawking to occur very
generally$^{[42][43][7]}$, but its generic nature is still not well established.

(2) The phenomenon for which Penrose$^{[44]}$ coined the term {\it cosmic
censorship}, whereby the singularities are hidden from the outside
asymptotically flat universe  behind a regular {\it event horizon} bounding the region for
which Wheeler coined the term {\it Black Hole} would appear to be stable against
moderate perturbations from spherical symetry and from the uniformity of the
homogeneous Oppenheimer Snyder collapse scenario that provides its simplest
example. Nevertheless much recent work$^{[34][27][28][45][46]}$ has made it
clear that sufficiently (one might be tempted to say unnaturally) large
deviations from uniformity can bring about the occurrence of non-trivial {\it
naked} singularities, i.e. ones from which light can escape to large asymptotic
distances, so although the regular black hole scenario, as governed by the
cosmic censorship postulate, may plausibly provide a generic description of
astrophysically realistic collapses, its mathematical generality would seem to
be more severely circumscribed than was once thought.

(3) Although the vacuum region outside a generic  collapsing body will not
become immediately static (as it must, by Birkhoff's theorem in the spherical
case) it is nevertheless to be expected that the energy of non stationary
oscillatons will ultimately be radiated away so that in the end the vaccuum
region outside an (isolated) collapsing  body will settle down asymptotically
towards an ultimate {\it equilibrium} that is stationary at least in the weak
sense of being invariant under the action of a Killing vector field that is
timelike at large distances even if not everywhere outside the horizon.

\parindent=2 cm
Experience with the Schwarzshild and Reissner Nordstrom examples (as described
in the previous section) shows however that whereas the physical collapse
situation may be regular in the past, starting with an ordinary well behaved
assymptotically flat Cauchy initial value hypersurface, the asymptotically
approached equilibrium metric may have a ``white hole" region including
singularities in the past, so the strongest regularity condition it can be
expected to satisfy is assymptotic predictability, meaning that there exists a
partial Cauchy surface (a not necessarily complete globally spacelike, i.e.
achronal, hypersurface) extending in from outer infinity to the black hole
horizon and at governing (i.e. intercepting all sufficiently extended past
directed timelike lines from) not necessarily all of its future (as would be
required for a strict Cauchy surface) but at least a part consisting of a
regular asymptotically flat domain of outer communications with inner bound on a
regular black hole horizon.

\begin{figure}
\centering
\epsfig{figure=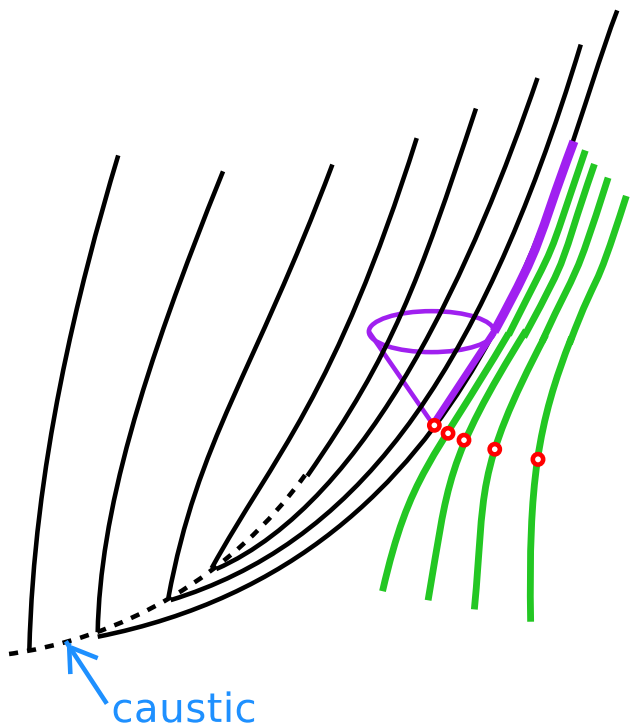, width=8.4 cm}
\caption{\color{navy}
\label{horicusp}
{\it Illustration of possibility of cusp on past event horizon. It is shown
how a null generator is attainable as limit of timelike lines escaping from 
sequence of points approaching horizon from outside.}
}
\end{figure}

The future boundary of the region governed by the partial Cauchy surface (which
in the standard Reissner Nordstrom example occurs at $r=r_-$) is called its
Cauchy horizon, and is an example of what is commonly described as a local
``future event horizon", whereas a black hole horizon (which in the Reissner
Nordstrom example occurs at $r=r_+$), i.e. the boundary of the region from which
a future directed timelike line can be extended to the outer asymptotically flat
region, is analogously describable as being a  local ``past event horizon". Both
kinds obviously belong to the category of ``achronal boundaries", meaning
boundaries of which no two points are connectable by a strictly timelike curves.
Local future and past event horizons are characterised more particulary by the
property (which was first systematically exploited by Penrose$^{[42][43]}$ and
Hawking$^{[47][7]}$) of being generated locally by by null geodesics with no
respectively past or future end points (which means that they are ordinary null
hypersurfaces wherever they are smooth, but that there null generators may reach
caustics when extrapolated to respectively the future of the past). In the case
of a black hole horizon this property (see figure) is not just local but global
i.e. its null generators can never reach a future end pont no matter how far
they are extrapolated.

To draw quantitative conclusions from these considerations it is necessary
to recapitulate some of the standard kinematic properties of
generating congruences.
 To start with we recall  that for
any vector field $\el_\mu$ has an (unnormalised) acceleration  $  \el^\mu$
that is related to its Lie derivative with respect to itself by
 $$\vec\el{\cal L}\el_\mu=\dot l_\mu+{_1\over^2}\nabla_\mu(\el^\nu\el_\nu)  \
, \ \ \ \ \ \
 \dot \el{^\mu}=\el^\nu\nabla_\nu \el^\mu \ . \eqno(3.1) $$
In order for the field to be normal to a hypersurface,
 it must satisfy the Frobenius
integrability condition
$$\el_{[\mu}\nabla_\nu \el_{\rho]}=0\ , \eqno(3.2)$$
which implies
$$\el_{[\mu}\vec\el{\cal L}\el_{\rho]}=\el^\nu \el_\nu \nabla_{[\mu}\el_{\rho]}
\ \ . \eqno(3.3)$$

It is apparent from (3.3)
 that in the particular case of a hypersurface that is locally
null, i.e. whose normal satisfies
$$\el^\nu \el_\nu=0 \ , $$
this normal must automatically satisfy the {\it geodesic equation}
$$\el_{[\mu}\dot\el_{\rho]}=0\ .\eqno(3.4)$$
Let us introduce a second null vector $\tilde\el^\mu$ say, transverse to the
null hypersurface and normalised so that
$$\tilde\el^\nu\tilde\el_\nu=0\ , \ \ \ \ \ \tilde\el^\nu\el_\nu=-1 \ . \eqno(3.5)$$
Such a vector can be uniquely specified by the condition that it be
normal to some given spacelike 2-surface ${\cal S}$ say in the horizon at the
point under consideration, in which case the corresponding rank 2 projection
tensor
$$\olg{^\mu}{_\nu}=g^\mu{_\nu}+\el^\mu\tilde\el_\nu+\tilde\el^\mu\el_\nu\
, \ \ \ \ \ \olg{^\nu}{_\nu}=2 \eqno(3.6)$$
will be interpretable (see appendix) as the (first) {\it fundamental tensor} of
the spacelike 2-surface.

With respect to any previously specified normalisation, corresponding to a time
parametrisaton such that $\el^\mu=dx^\mu/dt$, we can define a corresponding
affine time parametrisation $\tau$ say whose relation to the original time
parameter $\tau$ specifies a corresponding {\it decay coefficient} $\kappa$ in
terms of which the non affine geodesic equation (3.4) takes the form
$$\dot\el_\mu=\kappa \el_\mu   \eqno(3.7)$$
where explicitly
$$\kappa=({\rm ln}\ \dot\tau)^\cdot=\ddot\tau/\dot\tau=-\tilde\el
^\nu\dot\el_\nu\ . \eqno(3.8)$$

The usual way of defining the divergence $\theta$ and the (automatically real)
magnitude $\sigma$ of the (automatically symmetric) shear rate tensor
$\sigma_{\mu\nu}$ of the null generators is via the projection
$$\theta_{\mu\nu}=\olg{^\rho}{_\mu}\olg{^\sigma}{_\nu}\nabla_\rho\el_\sigma\ ,
\ \ \ \ \theta=\theta^\nu{_\nu}\ , \ \ \  \ \sigma_{\mu\nu}=\theta_{\mu\nu}
-{_1\over^2}\theta\olg_{\mu\nu}  \  ,
\ \ \ \ \sigma^2=2\sigma_{\mu\nu}\sigma^{\mu\nu}\ . \eqno(3.9)$$
Taking the contraction of the defining identity of the Riemann tensor,
which for any field $\el_\mu$ whatsoever gives the consequent identity
$$\el^\mu\nabla_\mu(\nabla_\nu\el^\nu)=\nabla_\mu(\el^\nu\nabla_\nu\el^\mu)
-(\nabla^\mu\el\nu)\nabla_\nu\el_\mu- R_{\mu\nu}\el^\mu\el^\nu ,\eqno(3.10)$$
one obtains, for the case of a congruence of null geodesic generators
as characterised above, the famous Penrose$^{[42]}$ null version
$$\dot\theta-\kappa\theta=-{_1\over^2}(\theta^2+\sigma^2)-R_{\mu\nu}
\el^\mu\el^\nu \ , \eqno(3.11)$$
of the equation whose analogue for a timelike congruence was first brought to
attention by Raychaudhuri$^{[48]}$, where a dot denotes differentiation with
respect to an arbitrary time parametrisation, whose adjustment to be affine can
be used to get rid of the $\kappa$ term on the left hand side. The special
importance of this equation comes from the fact that one is then left with a
right hand side that is strictly non positive provided the Ricci tensor is
determined by the Einstein equations (2.1) either for a vacuum or at least with
an energy tensor $T^{\mu\nu}$ that, as is the case for all the usual macroscopic
matter models, is such as to satisfy the appropriate energy inequality
$T^{\mu\nu}\el_\mu\el_\nu =0$.

The Penrose inequality
$$(\theta^{-1})^\cdot\geq {_1\over 2}  \eqno(3.12)$$
that is obtained for the rate of affine variation under such conditions can be
immediately used to see that if $\theta$ is ever negative then there will
inevitably be a caustic where it diverges to infinity within an affine distance
bounded above by $2/\vert\theta\vert$ in the future. Penrose's original
exploitation of this result was for the purpose of demonstrating the
inevitability of some kind of singularity formation to thr future of any {\it
closed trapped surface} on a well behaved initial hypersurface by showing the
affine bounedness (which would be impossible in the absence of a singularity) of
the future event horizon bounding the future of the closed trapped surface,
where this term is understood to mean a compact spacelike topologically
spherical 2-surface ${\cal S}$ for which the divergence $\theta$ of the null
normals is everywhere negative.

In terms of the background tensor curvature formalism$^{[15]}$ described in the
appendix, it can be seen (from (A9)) that the specifications (3.9) for the
divergence and shear of the outgoing null congruence $\el^\mu$  can be rewritten
in terms of the {\it second fundamental tensor} $K_{\mu\nu}{^\rho}$ (which is
equivalent to what is  referred to by Hartley and Tucker$^{[49]}$ as the shape
tensor) of the spacelike 2-surface ${\cal S }$ and of the corresponding {\it
curvature vector} $K^\mu$ and trace free (and conformally invariant)
conformation tensor  $C_{\mu\nu}{^\rho}$ as
$$\theta_{\mu\nu}=-K_{\mu\nu}{^\rho}\el_\rho\ ,\ \ \ \ \theta=-K^\nu\el_\nu\ ,
\ \ \ \sigma_{\mu\nu}= -C_{\mu\nu}{^\rho}\el_\rho \ .\eqno(3.13)$$
Similarly for the ingoing null congruence as specified by $\tilde\el^\mu$
(for which we are still assuming the normalisation condition (3.5))
the corresponding divergence and shear will be given by
$$\tilde\theta_{\mu\nu}=-K_{\mu\nu}{^\rho}\tilde\el_\rho\ ,\
 \ \ \ \ \tilde\theta=-K^\nu\tilde\el_\nu\ ,\ \ \ \ \
\tilde\sigma_{\mu\nu}= -C_{\mu\nu}{^\rho}\tilde\el_\rho \ .\eqno(3.14)$$

The usual situation for an approximately spherical 2-surface at approximately
constant time in an approximately flat background is to have ingoing
null normals that converge, $\tilde\theta <0$  but outgoing ones that diverge,
$\theta>0$ so that the product, which can be seen from (3.6) to be expressible
in manifestly the normalisation independent form
$$\tilde\theta\theta=-{_1\over^2}K_\nu K^\nu\ ,   \eqno(3.15)$$
will be negative,
whereas if $\theta$ changes sign and becomes negative, with $\tilde\theta$
still also negative, then the product will change sign also, i.e. the curvature 
vector $K^\mu$ will change from being spacelike to being timelike. A special
interest applies to the marginally trapped case characterised by $\theta=0$
everywhere, for which Hawking$^{[47][50][7]}$ has introduced the term apparent
horizon. Evidently such a marginally trapped surface may be described as one for
which the curvature vector $K^\mu$ is null.

         Following the Penrose application of (3.12) to the future
of a closed trapped surface, Hawking$^{[47][50][17]}$ pointed out that a very
powerful result can be obtained by applying it to the black hole horizon
itself, using the condition that the future generators of a black hole
horizon can never terminate, which implies that the generators of such
a horizon can never have negative divergence $\theta$. Noting that $\theta$
is interpretable as specifying the rate of variation of the measure
of a 2-surface element dragged along by the generators according to the
formula
$$(d{\cal S})^\cdot=\theta d{\cal S} \ ,\eqno(3.16) $$ 
and applying this to the integrated area
$${\cal A} =\oint d{\cal S}  \eqno(3.17)$$
of a 2 dimensional spacelike section through the black hole horizon, Hawking
obtained the (now famous) law to the effect that the horizon must evolve with
time according to the inequalities
$$\dot{\cal A}\geq \oint \theta d{\cal S}\geq 0 \ , \eqno(3.18)$$
the extra inequality on the right being to allow for the fact that in addition
to the area increase resulting from smooth expansion there is also the
possibility of an additional increase due to the branching off of new generators
from a caustic (see diagram). In particular if two black holes with areas ${\cal
A}_{\ 1}$ and ${\cal A}_{\ 2}$ merge to form a combined black hole with area ${\cal
A}_{\ 3}$ then we must have the strict inequality
$${\cal A}_{\ 3}>{\cal A}_{\ 1}+{\cal A}_{\ 2}\ . \eqno(3.19)$$

Before closing this section and going on to restrict our attention to states of
stationary equilibrium, it is to be remarked that though the locality of a
marginally trapped surface or ``aparent horizon", as characterised locally by
$K_\mu K^\mu=0$, is merely an inner bound on the location of (teleologically
defined)  true horizon, it may nevertheless give a very good approximation to
the localisation of the true horizon in the almost stationary limit when the
situation is not too strongly different from its ultimate equilibrium state,
in which case the approximate stationarity will determine a corresponding
approximately 
well defined and in general non affine time parametrisation on the horizon,
so that there will be a correspondingly well defined {\it decay} constant
$\kappa=\ddot\tau/\dot\tau$ where $\tau$ is the corresponding proper time.
Under such conditions the ratio between the values $d{\cal S}_0$ and
$d{\cal S}_1$ of the measure of a generator transported surface
element $d{\cal S} $ between times $t_0$ and $t_1$ can be seen$^{[11][12]}$
to be given approximately by an expression of the non-teleological form
$${\rm ln} \left({d{\cal S}_1\over d{\cal S}_0}\right)=\int_{t_0}^{t_1}
\theta dt\approx {8\pi\over\kappa}D dt \eqno(3.20)$$
which gives the Hartle Hawking formula$^{[51]}$
$$\dot{\cal A}\approx\oint{D\over \kappa}d{\cal S}\ , \eqno(3.21)$$
where the rate of effective dissipation is given by
$$D={\sigma^2+\theta^2\over 16\pi}+ T^{\mu\nu}\el_\mu\el_\nu \ .
\eqno(3.22)$$

The discovery$^{[47][51]}$  of the laws (3.18) and (3.22) suggested an obvious
thermodynamic analogy, with ${\cal A}$ proportional to the entropy and $\kappa$
to the temperature. The deeper significance of this analogy was first guessed by
Beckenstein$^{[53]}$ and later established by the discovery of Hawking
radiation$^{[20]}$, which is associated with a temperature given exactly by
$\kappa/2\pi$ in Plank units, corresponding to an entropy given exactly by
${\cal A}/4$.  The crux of the analogy is constituted by the ``zeroth law"
that will be established in the next section.

\vfill\eject

\parindent = 0 cm
{\bf 4. Rotating Equilibrium States: zamos and local properties of
a Killing horizon.}
\medskip\parindent=1.2 cm

The concept of {\it stationarity} that is relevant to the theory of black hole
equilibrium states means the condition the the spacetime is invariant under the
action generated by an assymptotically timelike Killing vector
$$k^\mu\ \ \leftrightarrow \ \ {\partial\over \partial t} \ , \ \ \ \ \
\nabla_{(\mu}k_{\nu)}=0  \eqno(4.1)$$
It is to be noticed that this definition is slightly weaker than the one
commonly used in  other contexts where it is stipulated tht the stationarity
Killing vector be timelike not just at large asymptotic distances but
throughout, which would exclude the existence of  the ``ergorgions"
which are of importance not just in black hole thory but even in the theory
of ultrarapidly rotating stars.

All that follows will be based on the postulate that I decided to adopt when I
first looked into this area of work in the 1960's, namely that the stationary
spacetime under consideration is also characterised by {\it axisymetry}, meaning
that it is also invariant under the action generated by a spacelike  Killing
vector
$$m^\mu\ \ \leftrightarrow \ \ {\partial\over \partial \phi} \ , \ \ \ \ \
\nabla_{(\mu}m_{\nu)}=0  \eqno(4.2)$$
whose action is periodic, with closed circular (or, on the axis itself, fixed
point) trajectories allong which the group parameter $\phi$ is therfore also
periodic, with period $2\pi$ for the standard normalisation. Relaxation of the
requirement that $m^\mu$ be spacelike is mathematiclly conceivable but
physically inappropriate since, in view of the periodicity, it obviously violate
the causality requirement that there exist no closed timelike or null lines.
In an asymptotically flat background it is inevitable$^{[54]}$ that this
second symmetry commutes with the first, i.e.
$$k^\nu\nabla_\nu m^\mu-m^\nu\nabla_\nu k^\mu=0 . \eqno(4.3)$$

Just as it is  plausible that a isolated  system with or without a central black
hole should tend towards a stationary equilibrium state so also it is plausible,
particularly in a context where gravitational radiation needs to be taken into
account, that under natural conditions the stationary state would also have to
be axisymmetric. It is of course possible mathematically to construct artificial
counterexamples (such as the Dedekind ellipsoids to which
Chandrasekhar$^{[55]}$ has drawn attention),
but under natural astrophysical conditions it is very hard to imagine
stationary black hole scenarios for which the axisymmetry assumption would be in
danger of failing. (For the case of of an isolated black hole with a vaccuum or
simple gaseous exterior  considerable effort has been invested, most notably by
Hawking$^{[47][50][7]}$, in attempts to prove that axisymmetry of equilibrium states is not just
physically plausible but mathematically necessary. However the crucial result,
describable as Hawking's ``Strong Rigidity Theorem", to this
effect is itself based on a postulate of analyticity that is also physically
plausible but whose mathematically justification requires assuming the
axisymmetry one wanted to prove in the first place, so that as a mathematical
challenge the problem remains wide open.)

In any study of stationary axisymmetric systems  an important role is played
by the Killing vector invariants
$$ V=-k^\mu k_\mu\   , \ \ \ \ \ \ W = k^\mu m_\mu \ ,\ \ \ \ \
  X=m^\mu m_\mu \eqno(4.4)$$
and  by the determinant
$$\varpi^2=VX+W^2={1\over 2}\varpi_{\mu\nu}\varpi^{\nu\mu}\ ,\ \ \ \
\ \ \varpi_{\mu\nu} = 2 k_{[\mu}m_{\nu]} \ . \eqno(4.5)$$
and the ratio
$$\omega=-{W\over X}  \ , \ \ \ \  \ X>0 \eqno(4.6)$$
which is well defined wherever $X$ is strictly positive, which
by the causality condition that $m^\mu$ is spacelike, will hold
everwhere except on the symmetry axis itself where $m^\mu$ reduces to
a zero vector so that $X$ and $W$ both vanish, making $\omega$ undefined.

Other contractions of interest are the energy, $E$ say,
 and angular momentum $L$ say, of a particle with momentum covector
$p_\mu$, as given by
$$ E=-k^\nu p_\nu\ ,\ \ \ \ \ \ L = m^\nu p_\nu  \ , \eqno(4.7)$$
which are of course conserved for free orbits:
$$u^\nu\nabla_\nu p_\mu=0\ ,\ \ \ \  p^\mu= m u_\mu\ ,\
\ \ u^\mu u_\mu=-1 \ \ \
\Rightarrow\ \ \ u^\nu\nabla_\nu E=0 \ , \ \ \ u^\nu\nabla_\nu L=0 \
.\eqno(4.8)$$

The quantities $\varpi$ and $\omega$ defined by (4.5) and (4.6) are of
particular interest in the context of {\it circular flow }, meaning flow allong
trajectories on circles generated by the Killing vectors, i.e. with unit flow
vector $u^\mu$ of the form
$$u^\mu=\alpha(k^\mu+\Omega m^\mu)  \eqno(4.9)   $$
where the coefficient $\Omega=d\phi/dt$ is the {\it angular velocity} of the
trajectory.  The important ``Kepplerian" special class of circular trajectories
consists of those that are free in the sense of (4.8), a possibility which
typically will exist only in a restricted equatorial plane. A more generally
defined class that is of more immediate (though mathematical rather than
physical) interest for our present purposes consists of what Bardeen$^{[56]}$
has called ``zamo" trajectories (short for zero angular momentum orbiters) which
are characterised by the (obviously non Kepplerian) condition of having $L=0$,
which can be seen to be equivalent to the condition that their angular velocity
be given directly by $\Omega=\omega$. The obvious interest of $\varpi$ as
defined by (4.5) in this context is that its reality, i.e. the positivity
condition
$$\varpi^2>0 \eqno(4.10)$$
is evidently the necessary and sufficient condition for the existence of a
strictly timelike zamo at the position in question. The importance of this
purely local condition is that subject to very weak hypotheses it can also be
shown$^{[36][10][57][12]}$, as described later on below, to characterize the
domain of outer communications, whose  (globally defined) boundary at the
surfaces of the black hole region, will be characterisable (locally, this is
what is so convenient) as a ``zamosphere", where $\varpi=0$, on which the zamo's
become null.

A more frequently discussed$^{[98]}$ but in the final instance less 
important analogue of
the zamosphere is the ``ergosphere", where $V=0$, i.e. on which the stationarity
generator $k^\mu$ becomes null. The interest of this is that it bounds the
``ergoregion" characterised by $V<0$ within which a free particle energy $E$, as
defined by (4.7) can become negative, whereas outside the ergo region, i.e.
wherever $k^\mu$ is timelike, the free particle energy is bounded below by the
condition $E\geq m\sqrt V$. The existence of an ergoregion (unless confined
within the horizon as in non rotating case) makes possible the extraction of
energy from the background by the mechanisms such as the Penrose
process$^{[44]}$ 
whereby a particle coming in with energy $E_1$ splits into a
part with negative energy, $E_2 <0$, and an outgoing part with energy $E_3$
which, by conservation of the sum, $E_2+E_3=E_1$ must exceed the initial energy,
$E_3>E_1$.

For charged orbits, as given by
$$u^\nu\nabla_\nu p_\mu=F_{\mu\nu}u^\nu\ , \eqno(4.11)$$
in a stationary field,
$$\vec k{\cal L}A_\mu=0 \ \ \ \Leftrightarrow\ \ \ F_{\mu\nu}k^\nu+\nabla_\mu
\Phi= 0\ ,\ \ \Phi=-k^\nu A_\nu \  \eqno(4.12)$$
one gets conservation not of the ordinary energy $E$ but of a generalised gauge
dependent generalisation ${\cal E}$ say, constructed from the gauge dependent
generalised momentum covector 
$$\pi_\mu=p_\mu+eA_\mu \ , \eqno(4.13)$$
i.e. one gets 
$$u^\nu\nabla_\nu{\cal E}=0 \ , \ \ \ \ {\cal E}=-h^\nu\pi_\nu= E+e\Phi \ ,
\eqno(4.14)$$
This generalised evergy can be negative even outside of the ergosphere, where
$V>0$ so that the lower bound will be given by ${\cal E}\geq m\sqrt V+e\Phi$. It
is apparent that there will be an extended electric ergoregion$^{[58]
[12][59]}$
characterised for a given charge to mass ratio by the possiblity of negative
energy for at least one sign of the charge $\pm e$, for which the condition is
just 
$$ V<({e\over m})^2 \Phi^2  \ . \eqno(4.15)$$
The significance of this relation is of course dependent on how the energy is
calibrated, the usual assymptotic specification being not the only one of
interest: another possibility of particular interest$^{[58][12][60]}$ 
for the specialised theory of non rotating black
holes$^{[7][61][62][63][64][65][66]}$ is to calibrate with respect to the
horizon which is possible in that case because of its uniform potential
condition which will be demonstrated below.

The properties of a stationary axisymmetric system simplify enormously under
conditions of what I call {\it circularity} which in practice are almost sure to
be satisfied in the applications that are relevant to black holes. In the case
of an electromagnetic source current $j^\rho$ this means just that it should 
be a linear combination of the Killing vectors which is equivalent to
the requirement 
$$j^{[\mu}\varpi^{\rho\sigma]} =0 \ , \eqno(4.16) $$
while for a gravitational source it means inthe case of a perfect fluid just
that the corresponding flow vector $u^\mu$ should satisfy the ananlogue of (16)
which is equivalent to the condition (4.9) for the corresponding flow
trajectories to be simple circular orbits.  For a more general material source
(for which a preferred reference vector $u^\mu$ might not be defined)
circularity is to be understood as meaning that the relevant material energy and
angular momentum flux vectors $k_\nu T_{_{\rm M}}{^\nu}{_\mu}$ and $m_\nu
T_{_{\rm M}}{^\nu}{_\mu}$ should have the same property, this general
circularity condition being expressible as
$$k_\nu T_{_{\rm M}}{^\nu}{_{[\mu}}\varpi_{\rho\sigma]}=0\ , \ \ \ \ \ \
m_\nu T_{_{\rm M}}{^\nu}{_{[\mu}}\varpi_{\rho\sigma]}=0\ . \eqno(4.17)$$

This condition can in principle fail in a star that is partly solid$^{[99]}$
(as in the case of a neutron star crust ) or even in a stricty perfect inviscid
fluid star where there is convection, but for the more plausibly relevant case
of  viscous fluid the possibility of other than circular motion can be ruled out
because it would inevitably produce thermal dissiption and thus violate the
requirement of strict stationarity.

The crucial simplification  that one gets in such circumstances is provided by
the {\it circularity theorem}$^{[36][10][12]}$ which (generalising a result first
demonstrated in the case of a pure vaccum by Papapetrou$^{[68]}$
and for an uncharged perfect fluid by Kundt and Trumper$^{[69]}$)
tells us that the
system will be {\it orthogonally transitive}, meaning that the circular
trajectories generated by the two Killing vectors will be orthogonal to a
congruence of two dimensinional surfaces which (must obviously be spacelike
where, and only where, the zamos are timelike) throughout any continuous region
connected to the rotation axis within which the source circularity conditions
(4.16) and (4.17) are satisfied. Thus if the source circularity conditions are
satisfied everywhere one gets orthogonal transitivity everywhere. The well known
Frobenius condition for such orthogonal transitivity is expressible as the
requirement that the twist vectors
$$\omega^\mu={_1\over^2}\varepsilon^{\mu\nu\rho\sigma}k_\nu\nabla_\rho k_\sigma\
,  \ \ \ \  \ \psi^\mu={_1\over^2}\varepsilon^{\mu\nu\rho\sigma}
m_\nu\nabla_\rho m_\sigma\ ,  \eqno(4.18)$$ be orthogonal to the Killng vectors,
i.e. that we should have
$$\omega^\mu m_\mu = 0 \ , \ \  \ \ \ \psi^\mu k_\mu=0 \ . \eqno(4.19)$$
Since these contractions must both vanish identically on the axis where $m^\mu$
is zero, it is sufficient to obtain the required result that these contractions
should have the property of {\it uniformity}, i.e. that the should be constants,
over the region in question.

The proof the circularity theorem to this effect uses the fact that the Killing
equations (4.1) and (4.2) imply corresponding higher order conditions
$$\nabla_\nu\nabla^\nu k^\mu=-R^\mu{_\nu}k^\nu  \ , \ \ \ \ \ \ \
\nabla_\nu\nabla^\nu m^\mu=-R^\mu{_\nu}m^\nu \ , \eqno(4.20)$$
from which, with the aid of (4.3) one gets
$$\nabla_\mu(\omega^\nu m_\nu)= {_1\over^2}\varepsilon_{\mu\nu\rho\sigma}
k^\nu m^\rho R^\sigma{_\tau}k^\tau\ ,  \ \ \
 \nabla_\mu(\psi^\nu k_\nu)= {_1\over^2}\varepsilon_{\mu\nu\rho\sigma}
m^\nu k^\rho R^\sigma{_\tau}k^\tau\ .  \eqno(4.21)$$
In the absence of electromagnetic source contributions the material
circularity conditions (4.17) alone are sufficient to ensure that the right
hand sides of the foregoing pair of equations vanishes, which is evidently
sufficient to establish the required uniformity. To show that the result 
remains valid in the presence of electromagnetic effects requires a little
more work, starting from the group invariance conditions
$$k^\nu\nabla_\nu F_{\rho\sigma}=2F_{\nu[\rho}\nabla_{\sigma]}k^\nu\ , \ \ \ \
m^\nu\nabla_\nu F_{\rho\sigma}=2F_{\nu[\rho}\nabla_{\sigma]}m^\nu\ 
\ . \eqno(4.22)$$
These conditions, together with the Maxwellian field equations (?.?)
imply that the field vectors
$$E_\mu= F_{\mu\nu}k^\mu \ , \ \ \ \ B^\mu={_1\over^2}\varepsilon^{\mu\nu
\rho\sigma}k_\nu F_{\rho\sigma}\ , \eqno(4.23)$$
 will satisfy
$$\nabla_\mu(E_\nu m^\nu )=0 \ , \ \ \ \ \nabla_\mu(B^\nu m_\nu)=
4\pi\varepsilon_{\mu\nu\rho\sigma}k^\nu m^\rho j^\sigma \ , \eqno(4.24)$$
in which not only the first but also the second of the right hand sides
will obviously vanish wherever the current singularity condition (4.16)
is satisfied, with the implication that $E_\nu m^\nu$ and $B^\mu m_\mu$
will also both be uniform and therefore vanish
$$E_\nu m^\nu=0 \ , \ \ \ \ B^\mu m_\mu =0    \eqno(4.25)$$
since they both obviously must vanish on the axis. The conditions (4.25) can
appropriately be described as field circularity conditions, since they are
sufficient for the corresponding electromagnetic contribution to  the
gravitational source to satisfy the analogue of the material circularity
condition (4.17) so that its effect does not invalidate the conclusion
that $\omega^\nu m_\nu$ and $\psi_\nu k^\nu$ will vanish also.

The orthogonal transitivity property that is established in this way means that
it will be possible, except where $\varpi=0$, to choose the spacetime
coordinates in uch a way as to expess the metric in the standard Papeptrou form:
$$ds^2=g_{ij}dx^i dx^j + X\{ (d\phi-\omega  dt)^2-\varpi^2 dt^2\} ,
\eqno(4.26)$$
where the coefficients $X$, $\omega$ $\varpi$ (as defined by (4.5) and (4.6) are
independent of the ``ignorable" coordinates $\varphi$ and $t$, but functions of
the two other coordinates $x^i$, $i=1,2$ whose locci of constancy are the
orthogonal two surfaces predicted by the theorem.  It can be seen that under
these conditions the zamo trajectories themselves are orthogonal to the
hypersurfaces on which $t$ is constant, this hypersurface orthogonality
condition (which, like (4.26) itself, would fail if the circularity conditions
(4.16) and (4.17) were not satisfied) is what is meant by the statment
that the zamos congruence is irrotational.

To understand what happens on the {\it zamosurface} where the zamo
worldlines become null so that $\varpi$ vanishes and the metric form (4.22)
becomes singular,  we use the fact that the Frobenius conditions (4.19)
on which it depends imply
$$2\varpi_{\mu[\nu}\nabla_{\rho]}\varpi_{\sigma\tau}=\varpi_{\sigma\tau}
\nabla_{[\rho}\varpi_{\sigma\tau]} \ . \eqno(4.27)$$
This gives an equation for the gradient of $\varpi^2$ orthogonal to the
Killing vector surfaces of transitivity that is somewhat analogous to the
one obtained for the gradient of the zamo angular velocity $\omega$
directly from its definition, i.e.
$$\varpi_{[\sigma\tau}\nabla_{\mu]}\varpi^2=\varpi^2\nabla_{[\mu}
\varpi_{\sigma\tau]}\ , \ \ \ \  X^2\varpi_{[\sigma\tau}\nabla_{\mu]}\omega
=2\varpi^2 m_{[\mu}\nabla_\sigma  m_{\tau]}\ , \eqno(4.28)$$
the noteworthy thing about both these equations being that their right hand 
sides vanish on the zamosurface where $\varpi^2=0$. Since both
$\varpi$ and $\omega$ are invariant under the group action their gradients
everywhere must be orthogonal to  the surfaces of transitivity generated by
 the Killing vectors, whereas according to (4.28)
 they must actually be aligned with these surfaces of transitivity
on the zamosphere, conditions which can only be reconciled if they
are both aligned with the unique combination of Killing vectors
that is null on the zamosphere, i.e. with the zamo direction itself,
and hence that they are aligned with each other. Provided the gradient
$\nabla_\mu\varpi^2$ is non zero, and so defines the direction
normal it the zamosphere it obviously follows that (as can be shown
with a little more care to be true in any case$^{[36]}$)  the
the zamosurface is a {\it null} hypersurface and that the
zamo angular velocity $\omega$ has a {\it uniform} value $\Omega^{_{\rm H}}$
say on it. This is the result that I refer to as the weak
{\it rigidity theorem}$^{[70]}$ (the qualification weak being because
 it is bases
on a line of argument assuming axisymetry at the outset, in contrast
with the stronger rigidity theorem of Hawking$^{[47][7]}$ based only
on an assumption of analyticity).

The uniform angular momentum value whose existence is thus established
can be extrapolted off the zamosurface to a uniform value throughout
space,
$$\nabla_\mu\Omega^{_{\rm H}}=0 \ , \eqno(4.29)$$
in terms of which can construct a unique Killing vector combination
$$\el^\mu=k^\mu+\Omega^{_{\rm H}} m^\mu \ , \ \ \ \ \nabla_{(\mu}\el_{\nu)}=0
\eqno(4.30)$$
which is characterised by the property of becoming aligned with the zamo
direction where this direction becomes null, i.e. on the zamosurface which we
now know to be itself null. This shows that (unlike an ergosurface which is
typically timelike, and subject of course to the postuate that the circularity
conditions (4.16) and (4.17) are satisfied) the zamosurface is automatically
what I have called a {\it Killing horizon}$^{[36]}$, i.e. a null hypersuface
whose null generator conicides with the generator of an isometry.

Before going on to consider the global question of the identification of the
locally defined zamosurface Killing horizon with the globally defined black hole
event horizon, there are some further local properties of Killing horizons that
can logically be derived at this stage.  To start with it is apparent
from the Penrose Raychaudhuri equation  (3.11)
for the null generator that since all the other terms vanish we must
also have
$$R_{\mu\nu}\el^\mu \el^\nu =0\eqno(4.31)$$
which subject to the material energy positivity postulate means that we must
separately have
$$T_{_{\rm M}}^{\mu\nu}\el_\mu\el_\nu=0   \ ,\ \ \ \
T_{_{\rm F}}^{\mu\nu}\el_\mu\el_\nu=0  \eqno(4.32)$$
where the electromagnetic part is given by
$$T_{_{\rm F}}^{\mu\nu}\el_\mu\el_\nu ={1\over 8\pi}
(E^{\dag}{_\mu}E^{\dag\mu}
+B^{\dag}{_\mu}B^{\dag\mu} )\ , \ \ \ \ \ E^{\dag}{_\mu} =F_{\mu\nu}\el^\nu\ ,\
\ \ \ \ B^{\dag\mu}={_1\over^2}\varepsilon^{\mu\nu\rho\sigma}F_{\nu\rho}
\el_\sigma \ , \eqno(4.33)$$
where a dagger symbol is used to distinguish quantities defined with
respect to the {\it corotating} Killing vector field (4.30) from their
analogues as defined with respect to the ordinary (asymptotically
timelike) stationarity Killing vector (4.1) ( a distinction that
is not necessary in the static case for which they both coincide).
If the material contribution is simply of perfect fluid type (2.25) subject
to the inequalities $\rho\geq 0\ $, $P\geq 0$ the first of the conditions
(4.28) can be seen to give
$$\rho+P=0 \ \ \ \ \Rightarrow \ \ \ \ \rho=0 \ , \ P=0 \ , \eqno(4.34)$$
i.e. there must be a vacuum at the horizon, while since
both $E^{_\dag}{_\mu}$ and $B^{_\dag}{_\mu}$ are both by construction orthogonal
to $\el^\mu$ they cannot be timelike on the Killing horizon, so the 
reconciliation of (4.32) with (4.33) requires that they both be null
there and hence proportional to the generator itself, i.e.
$$E^{\dag}{_{[\mu}}\el_{\nu]}=0 \ ,\ \ \ \ \ \ B^{\dag}{_{[\mu}}\el_{\nu]}=0
\ , \eqno(4.35)$$
of which the first tells us$^{[10]}$ that the horizon is like a conductor whose
equilibrium requires uniformity of the corresponding potential,
an analogy that, since it was first noticed, has been developped in
considerable detail$^{[6][100][101]}$.
Explicitly we have
$$ E^{\dag}_\mu=\nabla_\mu\Phi^{\dag}\ ,\ \ \ \ \
\Phi^{\dag}=A_\mu\el^\mu \eqno(4.36)$$
everywhere, with the potential $\Phi^{\dag}$ necessarily uniform over
the horizon. Formally, for any tangent vector $\xi^\mu$ to the Killing
horizon we have
$$\xi^\mu\el_\mu=0 \ \ \ \Rightarrow \xi^\mu\nabla_\mu \Phi^{\dag}=0 \
.\eqno(4.37)$$

This uniformity property of the angular velocity $\Omega^{_{\rm H}}$
and of the potential $\Phi^{\dag}$ on a Killing horizons are
prototypes$^{[36][70][10]}$ for a less intuitively obvious 
uniformity$^{[10][50][56][52]}$, namely that
of the decay parameter $\kappa$ whose definition
by the general formula (3.7) is unambiguous now that the normalisation
of $\el^\mu$ is fixed by (4.30). The fact that $\el^\mu$ must satisfy the
Frobenius orthogonality condition (3.2) on the horizon can be seen
to mean that there must exist some vector $q_\mu$ on the horizon such that
$$\nabla_\nu\el_\mu=2q_{[\nu}\el_{\mu]} \ ,\ \ \ \ \tilde\el^\nu q_\nu=0
\eqno (4.38)$$
where $\tilde\el^\mu$ is an ingoing nullvector as introduced in (3.5).
It is easy to see from the expression (4.38) that for any vectors
$\xi^\mu$, $\eta^\mu$ lying in the horizon we shall have
$$\xi^\mu\el_\mu=0 \ , \ \ \ \eta^\mu\el_\mu=0  \ \Rightarrow \ \
\xi^\mu\eta^\nu\nabla_\nu\el_\mu=0 \ \Rightarrow
\el^\mu\eta^\nu\nabla_\nu\xi_\mu=0  \ , \eqno(4.39)$$
which shows that the Killing horizon is extrinsically flat (geodesically
generated) since it shows that $\eta^\nu\nabla_\nu\xi^\mu$ will
automatically be tangenial to the horizon.   A further derivation
now leads (using $\el_\nu\nabla_\rho\xi^\nu=\xi^\nu q_\nu \el_\rho$) to
$$\xi^\mu\eta^\nu\nabla_\rho\nabla_\mu\el_\nu =-(\nabla_\mu\xi_nu)
(\xi^\nu\nabla_\rho\eta^\nu+\eta^\nu\nabla_<rho\xi^\mu)=0 \eqno(4.40)$$
Since the Killing vector property (4.30) by itself implies
$$\nabla_\rho\nabla_\mu\el_\nu=R_{\mu\nu\rho}{^\tau}\el_\tau\ , \eqno(4.41)$$
we end up with
$$R_{\mu\nu\rho\tau}\el^\mu\xi^\rho\eta^\tau=0\ ,\ \ 
\Rightarrow\ \ \olg{^{\nu\tau}}R_{\mu\nu\rho\tau}\el^\mu\xi^\rho=0\  .
\eqno(4.42)$$
where $\olg{^\mu}{_\nu}$ is the projection tensor given by ((3.6),
whose substitution then gives
$$R_{\mu\rho}\el^\mu\xi^\rho=R_{\mu\nu\rho\tau}\el^\mu\tilde\el^\nu
\el^\rho\xi^\tau\ . \eqno(4.43)$$
Since the defintion (3.7) is clearly equivalent to
$$\kappa=-\tilde\el^\mu\l^\nu\nabla_\nu\el_\mu=\el^\nu q_\nu \eqno(4.44)$$
direct differentiation gives
$$\xi^\nu\nabla_\nu \kappa=-\xi^\rho\left(\tilde\el^\mu\el^\nu\nabla_\rho
\nabla_\nu\el_\mu+ \tilde\el^\mu(\nabla_\rho\el^\nu)\nabla_\nu\el_\mu
+\kappa\el_\mu\nabla_\rho\tilde\el^\mu\right)  \eqno(4.45)$$
in which the last two terms cancel since they are respectively equal
and opposite to $\kappa\xi^\nu q_\nu$, so using (4.43) one finally obtains
the simple result
$$\xi^\nu\nabla_\nu\kappa=-R_{\mu\nu}\el^\nu\xi^\nu\ . \eqno(4.46)$$
The reasonning up to this point has been purely kinematic. If we now invoke
the Einstein equations we immediately obtain the required uniformity
condition that for an arbitrary tangent vector $xi^\mu$ to the horizon
$$\xi^\nu\nabla_\nu\kappa=0 \eqno(4.47)$$
in the pure vacuum case.
This ``zeroth law of black hole mechanics" result$^{[52]}$
can easily be seen to remain valid
for a source free Einstin Maxwell vacuum$^{[10][11]}$ since in that case, although
$R_{\mu\nu}$ will not be zero $R_{\mu\nu}\el^\mu$ will be be proportional
to the null tangent covector $\el_\mu$ which is all that is required to get
(4.47) from (4.46).

\bigskip

\parindent = 0 cm
{\bf 5. Rotating Equilibrium States: the global problem.}
\medskip\parindent=1.2 cm

The kind of stationary equilibrium state towards which, subject to the cosmic
censorship hupothesis, it seems reasonable to suppose  an isolated
gravitationally collapsing system would evolve, and that we shall understand to
be meant by the qualification ``well behaved black hole equilibrium state" is a
stationary spacetime whose domain of outer communications (D.O.C.) is bounded to
the future by a well behaved black hole event horizon.

In the concrete example of the Schwarzschild and Riessner Nordstrom solutions we
have seen that  the D.O.C. is also bounded to the past by a well behaved ``white
hole horizon", but the latter corresponds to nothing that exists in a dynamical
system that collapses from well behaved initial condition (being merely an
artefact of analytic extrapolation to the past) whereas the black hole horizon
in the stationary state really does correspond to the limit of the black hole
horizon of the dynamically collapsing state. This is why only the latter can
appropriately be postulated to exist as a defining characteristic of a well
defined black hole equilibrium state, the existence of any other horizon in the
past being something to be proven (if it can be) subsequently, but not to be
postulated in advance. If it is already known in some particular case not only
that both past and future event horizons exist but also that they have a well
behaved Kruskal type crossover on a spacelike two surface then one can prove a
result such as the ``zeroth law" obtained at the end of Section 4 by
a much shorter argument than was given there (see the accompanying lectures of
Wald) but it is important for our present purpose to have shown that the result
(i.e. the uniformity of $\kappa$) can be established independently of any such
assumption since it is needed as an an intermediate step in the line of
reasonning that ultimately shows (at least in the generic vaccuum case, the more
general question remaining open) that the assumed crossover really does occur.

In order to have the right to utilise the local properties established for a
zamosurface Killing horizon in the previous section we must show that the
globally defined black hole event horizon forming the future boundary of the
D.O.C. really is of this type. As a step towards this it, is convenient to use a
lemma$^{[12][10][57]}$ giving a pseudo local characterisation of the D.O.C. whose
original global definition is expressible for a stationary state as the
specification that it consists of the intersection of the past and the future of
the outer region where the trajectories of the stationary generator,
$\partial/\partial t=k^\mu\partial/\partial x^\mu$ are actually timelike. The
pseudo local characterisation is based on consideration of where these
generators are globally bradyonic, using this term to qualify any curve in
spacetime with the property that  any point $x$ within it determines
corresponding points $x_+$ and $x_-$ such that the part of the curve preceeding
$x_-$ lies entirely in the past of $x$ and the part of the curve subsequent to
$x_+$ lies entirely in the future of $x$. This condition is satisfied trivially
by an ordinary timelike curve (for which $x_-$ and $x_+$ can be identified with
the original point $x$ itself) but it also includes a curve which, though
locally spacelike, turns back towards itself in such a way that its evolution is
effectively timelike in the long run. 

It is not difficult to see$^{[12][10][57]}$ that although there may be an
ergoregion in which the Killing vector $k^\mu$ ceases to be timelike, its
trajectories must always remain globally bradyonic throughout the D.O.C. It is
even more obvious that in any {\it connected} region such that the trajectories
there are all globally bradyonic, any one of them can be conected to any other
by a timelike line with either orientation. This leads to a {\it
lemma}$^{[12][10][57]}$ characterising the D.O.C. as the {\it maximal connected
extension} of the outer region where the stationary trajectories generated by
$k^\mu$ are timelike {\it such that the stationary generators remain at least
globally bradyonic}.  As an immediate corollary it follows that $k^\mu$ can
never be timelike on the boundary (including the black hole horizon) of the
D.O.C., while on the other hand  the stationarity generator $k^\mu$ can never
vanish nor have any closed trajectory within the D.O.C. The latter conclusin
means that $k^\mu$ must always be linearly independent of $m^\mu$ in the D.O.C.
and hence that within the D.O.C. the Killing bivector $\varpi^{\rho\sigma}$ can
never be degenerate except on the axis $m^\rho=0$, which means that $\varpi^2$
can vanish only on the axis or where the bivector, and thus also the
corresponding zamo direction, becomes null.

Let us now designate by ${\cal Z}$ the maximal connected extension of the outer
region where the stationary trajectories generated by $k^\mu$ are timelike such
that the local condition {\it that the zamo trajectories be timelike} within it
is satisfied. Our aim is to show, subject to very weak assumptions that this
locally defined domain will be identifiable with the globally defined D.O.C.
What is obvious is that, since any connecting curve within ${\cal Z}$ will be
continuously deformable into a timelike curve (by the group action along the
zamo direction at each point)  ${\cal Z}$ must certainly lie entirely within the
D.O.C. The definition of ${\cal Z}$ means, according to (4.10), that ${\cal Z}$
is characterised by $\varpi^2>0$ except on the rotation axis where $\varpi^2=0$
and that we must have $\varpi^2=0$ everywhere on the boundary $\dot{\cal Z}$
of ${\cal Z}$. By the conclusion of the previous paragraph, this implies
that except on the axis $\dot{\cal Z}$ lies on the locus where the zamo
direction is null, and hence by the results of the previous section that
provided the circularity postulate is satisfied; the connected components of
$\dot{\cal Z}$ must be Killing horizon and thus {\it null hypersurface}
segments, each, by continuity, with  uniform time orientation.

To complete the demonstration that ${\cal Z}$ can be identified with the D.O.C.,
we must now invoke the further  postulate that the latter be {\it simply
connected}, which means that a (hypothetical) connected component ${\cal D}$ say
of the complement of ${\cal Z}$ in the D.O.C. must have a boundary $\dot{\cal
D}$ within the D.O.C. that is itself connected. It follows that $\dot{\cal D}$
would have to be a null hypersurface segment with uniform time orientation which
means that ${\cal D}$ could be reached from ${\cal Z}$ only by future directed
timelike lines, or only by past directed ones, but not by both kinds as would be
required for ${\cal D}$ to lie within the D.O.C. so in order to avoid a
contradiction it must be concluded that ${\cal D}$ is empty,  and thus that
${\cal Z}$ covers the whole of the D.O.C. as required. 

The foregoing demonstration to the effect that the Killing bivector
$\varpi^{\rho\sigma}$ is timelike throughout the D.O.C. means that the two
surfaces orthogonal to the Killing vectors that were shown to exist by the
circularity theorem of the previous section will correspondingly be strictly
spacelike there. Moreover the reinforcemnt of the circularity postulate
introduced in Section 4 section by the simple connectivity postulate introduced
in the previous paragraph implies that these orthogonal two surfaces will be
constructible not just locally but globally. Since any such surface differs from
flatness only by a locally variable conformal factor, $\Sigma$ say, it follows
that the space coordinates in the general Papaptrou form (4.26)  may be chosen
more specifically to be cylindrical type coordinates $\rho$, $z$ say, in such a
way that the  metric will be expressible in the form
$$ds^2=\Sigma(d\rho^2+dz^2)+X\{(d\phi-\omega dt)^2 -\varpi^2 dt^2\}
\eqno(5.1)$$
which will be {\it globally} valid over the entire D.O.C. except for the
familiar degeneracy on the axis where $X$  and $\varpi^2$ vanish, their vakues
elsewhere being strictly positive, as is the value of $\Sigma$ everywhere. An
analogous form is obtainable for the vector potential using the invariance
conditions (4.22) which imply the existence locally, and hence by the simple
connectivity postulate globally, of scalars $\Phi$ and $B$ such that
$$F_{\rho\sigma}k^\sigma=\nabla_\rho \Phi \ , \ \ \ \ \
F_{\rho\sigma}m^\sigma=\nabla_\rho B\ , \eqno(5.2)$$
in terms of which, using the consequence (4.25) of the circularity postulate,
it can be verified that the gauge may be chosen in such a way that
$$A_\rho dx^\rho=\Phi dt+B d\phi\ . \eqno(5.3)$$
It is of course to be understood here that the new coefficients $\Sigma$
$\Phi$, $B$, like $X$, $\omega$, and $\varpi$ as introduced previously,
areall functions only of $\rho$ and $z$ only, i.e. the stationarity and
axisymety is mad manifest by their independence of the ignorable
coordinates $\phi$ and $t$.

Up to his stage the analysis has been sufficiently general to cover a wide range
of conceivable black hole configurations with external matter rings for which
explicit analytic solutions are not available, but from this point on we shall
restrict our attention to the globally source free case for which it has long
been known$^{[71][72][40][73][74]}$ 
that the Kerr Newman class of solutions provide explicit
examples. The purpose of the systematic step by step approach whose development
I have been describing  is to solve the problem of whether there can exist any
others. It will be shown below that subject to the preceeding assumptions,
including notably that of a simply connected topology for the D.O.C., it can be
shown, by an argument that has been able to be made completely watertight only
comparitively recently$^{[92][93][94]}$, that these known solutions are indeed the only
ones for the strictly source free Einstein Maxwell equations. However the
problem remains wide open$^{[75]}$ for the slightly more general case of
solutions of the source free Einstein Maxwell equations with cosmological
$\Lambda$ term: such solutions cannot of course be asymptotically flat, but I
have discovered$^{[76][77][10]}$ a wide class of asymptotically De Sitter
solutions (one of the first cases for which the C.P. technique described in
Section 2 proved quite indispensible for providing an understandable global
description$^{[10][78]}$. The problem that remains unsolved is the extent to
which these known asymptotically De Sitter black rotating black  be whole
solutions are unique.

The reason why the results that follow have not yet been generalised to allow
for a cosmological $\Lambda$ term in the generalised source free
Einstein Maxwell system
$$R^\mu{_\nu}-{_1\over^2}R g^\mu{_\nu}=8\pi T_{_{\rm F}}{^\mu}{_nu}+\Lambda
g^\mu{_\nu}\eqno(5.4)$$
is that the next step uses the trace, not over the full four dimensional
system (5.4) but over its restriction to the two dimensional surface
of transitivity generated by the Kiling vectors for which due to the
cicularity condition te electromagnetic contribution cancels out
so that substitution of the form (5.1) simply gives
$$ {-1\over\sigma\varpi}\nabla^2\varpi=2\Lambda \ , \eqno(5.5)$$
where $\nabla^2$ is the Laplacian that is defined with
respect to of the  {\it flat} two dimensional
metric $d\rho^2+dz^2$.

In the absence of the cosmological term, (5.5) tells us the $\varpi$
is a harmonic function on the conformally flat spacelike two surfaces
which means that using the freedom to make conformal adjustments
one can choose the coordinate system so as to identify it with $\rho$:
$$\Lambda=0 \ \ \Rightarrow \ \ \varpi=\rho\ . \eqno(5.6)$$
In such a coordinate system (5.1) reduces to the specialised
Papapetrou form
$$ds^2=\Sigma(d\varpi^2+dz^2)+X\{(d\phi-\omega dt)^2 -\varpi^2 dt^2\} \ .
\eqno(5.7)$$
On the other hand the presence of a $\Lambda$ term suffices to block the
apparently innocent but actually crucial step from (5.1) to (5.7), without which
none of the work that follows can be carried through, the discovery
of an alternative route being thus left as a challenge for future work.

As has been well known since early work$^{[79][80]}$  on stationary axisymmetric
systems in other contexts, after the variable $\varpi$ has been thus taken out
of the list of unknown variables by its promotion to the status of a ``known"
coordinate variable, the system of source free field equations reduces to a
decoupled system just for the two metric variables $X$ and $\omega$ together
with the two electromagnetic potentials $\Phi$ and $B$, together with a separate
equation that can be solved afterwards to obtain the remaining variable, i.e.
the conformal factor $\Sigma$ by a direct quadrature, with the constant of
integraton fixed by the condition $\Sigma\rightarrow 1$ at large asymptotic
distance. This means that the four dimensional black hole equilibrium problem
from which we started reduces now to a two dimensional boundary problem for the
fields $X$, $\omega$, $\Phi$, $B$ as functions of the independent variables
$\varpi$ and $z$.

The conclusion$^{[10]}$ that (5.7) is not just valid  locally but covers
the entire D.O.C. with $\varpi$  (ranging from $0$ to $\infty$)
and $z$ (ranging from $-\infty$ to $\infty$) as globally well behaved
cylindrical type coordinates, depends on the simple connectivity
postulate and the use of a specialisation$^{[81]}$ of  Morse theory
 to exclude the possibility of critical points of $\varpi$. Ordinary Morse
theory establishes that the number of maxima plus the number of minima minus the
number of minimaxes is a topological invariant subject to fixed boundary
conditions and the assumption that no degenerate critical points occur. In the
harmonic case it is possible to make a stronger statement since maxima and
minima cannot occur while degenerate critical points do not need to be assumed
to be absent since their presence can easily be allowed for by labelling them
with an apropriate positive degeneracy index.  The resulting theorem$^{[81]}$
states that the index weughted sum over all critical points including possible
degenerate ones is a topological invariant which in the present application can
be seen (by considering any special case such as the Scchwarzschild solution) to
be zero.  Since the index is always of the same sign in the harmonic case no
cancellation is possible, so the fact that the total is zero in the case under
consideration makes it possible to deduce with certainty that there are no
critical points at all, degenerate or otherwise.

Having got to this point we introduce a further topological simplification
postulate to the effect that we are dealing with only a single topologically
spherical black hole. Our simple connectivity postulate has already excluded
conceivable toroidal black hole configurations but has left open the possibility
of having several topologically spherical black holes lined up on a common
rotation axis. A certain amount of work has been carried out, particularly by
Hawking$^{[47]}$ and Gibbons$^{[82]}$ towards showing that such configurations are
impossible except in the extreme Papapetrou Majumdar limit$^{[83][84]
[85][86][87][88]}$ of maximally charged non rotating configurations in which
electromagnetic repulsion balances gravitational attraction, but we shall not go
into the study of such exotic topological possibilities here. Assuming then that
the black hole topology is of simple spherical type we can fix the cylindrical
coordinate system symmetrically with respect to the black hole by  taking the
poles at which the horizon meets the rotation axis to be given by opposite
values of $z$ which, it can easily be seen, must be given explicitly by in terms
of the area and decay constant of the horizon by 
$$z=\pm c\ , \ \ \ \ \ \ c={\kappa{\cal A}\over 4\pi} \ .\eqno(5.8)$$

Leaving aside the awkward and still only partially understood
 special case for which the horizon is
degenerate in the sense of having $\kappa=0$ (corresponding to zero
temperature in the thermodynamic limit) we can conveniently proceed by
replacing the cylindrical coordinates $\varpi$,
$z$ by ellipsoidal type coordinates $\lambda$, $\mu$ according to the 
specifications
$$z=\lambda\mu \ , \ \  \ \ \varpi^2=(\lambda^2-c^2)(1-\mu^2) \eqno(5.9)$$
which are such as to arrange that the horizon is now given by the limit
$\lambda\rightarrow c$  while the two disconnected (``north" and ``south")
parts of the symmetry axis in the D.O.C. aregiven respectively by
$\mu\rightarrow \pm 1$, the whole D.O.C. being covered by the coordinate
range $c < \lambda< \infty$, $-1\leq\mu\leq\mu$.

In this system the metric takes the form
$$ds^2=\Xi \hat{ds}^2+Xd\phi^2+2Wd\phi dt-Vdt^2 \eqno(5.10)$$
for a conformally flat space metric given by
$$\hat {ds}^2={d\lambda^2\over\lambda^2-c^2}+{d\mu^2\over 1-\mu^2}
\eqno(5.11)$$
with
$$\Xi=(\lambda^2-c^2\mu^2)\Sigma \, \ \ \ W=X\omega \ ,\ \ \ \
V=X^{-1}(\varpi^2-W^2) \ . \eqno(5.12)$$
In terms of the two dimensional covariant differentiation operator
$\hat\nabla$ defined in terms of the known conformally flat metric
(5.11),    the system of independent source free Einstein Maxwell
equations reduces to a pair of Maxwell equations
$$\hat\nabla\left\{{X\over\varpi}(\hat\nabla\Phi-\omega\hat\nabla B)\right\}
=0\ , \eqno(5.13)$$
$$\hat\nabla\left\{ {\varpi\over X}\hat\nabla B+{\omega\over\varpi}
(\hat\nabla\Phi-\omega\hat\nabla B)  \right\}=0\ , \eqno(5.14)$$
together with a pair of Einstein equations
$$\hat\nabla\left\{{X^2\over\varpi}\hat\nabla\omega+
{4B\over\varpi}(\hat\nabla\Phi-\omega\hat\nabla B) \right\}=0\, \eqno(5.15)$$
$$\hat\nabla\left\{{\varpi\over X}\hat\nabla X \right\}
+{\vert\hat\nabla\omega\vert^2
\over\varpi}+{2X\over\varpi}\vert\hat\nabla\Phi-\omega\hat\nabla B\vert^2
+{2\varpi\over X}\vert\hat\nabla B\vert^2 =0 \ .\eqno(5.16)$$
Although this system is singular on the
axis where $X$ and $\varpi$ both vanish, and also on the
horizon where $\varpi$ also vanishes, it is garanteed to be regular
everywhere within the half plane under consideration where we have
$$\varpi>0 \ , \ \ \ \ \ X>0 \, \eqno(5.17)$$
the latter inequality being derived from the causality postulate. This is the
motivaion for having used a formulation giving the leading role to tha
axisymmetry Killing vector $m^\mu$ rather than using the more traditional
approach giving the leading role to the stationarity Killing vector $k^\mu$,
which would have given an analogous system but with $V$ turning up instead of
$X$ in the denominators, which would have the seriously inconvenient consequence
of making the system singular on the ergosurface that generically occurs within
the D.O.C.

The foregoing system can be made more tractable by performing the analogue
of the transformation introduced originally for the traditional formulation
based on $k^\mu$ rather than $m^\mu$ by Ernst$^{[80]}$. This is done by
first using the Maxwellian equation (5.13) to justify the introduction
of a stream function type electric potential $E$ given by
$$ X\left({\partial \Phi\over\partial\lambda}-\omega{\partial B\over
\partial\lambda}\right)=(1-\mu^2){\partial E\over\partial\mu}\ , \ \ \ \ \
X\left({\partial \Phi\over\partial\mu}-\omega{\partial B\over\partial\mu}
\right)=-(\lambda^2-c^2){\partial E\over\partial\lambda}\ , \eqno(5.18)$$
and by using the Einstein equation (5.15) to justify the introduction of an
analogous rotation potential $Y$ given by

$$X^2{\partial\omega\over\partial\lambda}=(1-\mu^2)\left\{{\partial Y\over
\partial\mu}+2E{\partial B\over\partial \mu}-2B{\partial E\over\partial\mu}
\right\}\ , $$       $$X^2{\partial\omega\over\partial\mu}=
-(\lambda^2-c^2)\left\{{\partial Y\over\partial\lambda}+
2E{\partial B\over\partial\lambda}-2B{\partial E\over\partial\lambda}
\right\}\ .\eqno(5.19) $$
Using the new potentals $E$ and $Y$ to replace $\Phi$ and $\omega$
one obtains the Maxwell equations in the form
$$\hat\nabla\left\{ {\varpi\over X}\hat\nabla B\right\}
+{\varpi\over X^2}\{(\hat\nabla Y+
2 E\hat\nabla B-2B\hat\nabla E\}\cdot\hat\nabla E=0\ , $$
$$\hat\nabla\left\{ {\varpi\over X}\hat\nabla E\right\}
-{\varpi\over X^2}\{(\hat\nabla Y+ 
2 E\hat\nabla B-2B\hat\nabla E\}\cdot\hat\nabla B=0\ , \eqno(5.20)$$
$$\hat\nabla\left\{ {\varpi\over X^2}(\hat\nabla Y+2E\hat\nabla B
-2B\hat\nabla E) \right\}  =0\ , $$
$$ \hat\nabla\left\{{\varpi\over X^2}\hat\nabla X\right\}+ {\varpi\over X^3}
\left\{\vert\hat\nabla X\vert^2+\vert\hat\nabla Y+2E\hat\nabla B-B\hat\nabla E
\vert^2\right\} +$$
$$\ \ \ \ \ \ \ \ \ \  {2\varpi\over X^2}\left\{\vert\nabla E\vert^2+\vert
\hat\nabla B\vert^2\right\}=0 \ . \eqno(5.21)$$

In terms of this new system the asymptotic boundary conditions for regularity at
large distance, i.e. as $\lambda\rightarrow \infty$, are more complicated than
in the traditional approach, being obtainable$^{[10]}$ as 
$$  \lambda^{-2}X=(1-\mu^2)+O(\lambda^{-1}) \ , \ \ \ \ \  \
E= -Q\mu +O(\lambda^{-1}) ,    $$
$$Y=2J\mu(3-\mu^2)+O(\lambda^{-1})\ ,\ \ \ \ \ \ B=O(\lambda^{-1})\ .
\eqno(5.22)$$
where $J$ is the asymptoticaly measured angular momentum about the rotation
axis while $Q$ is the total charge, and where the requirement that the total
magnetic monopole should vanish has been taken into account. The asymptotic mass
$M$ does not appear explicitly, but it is implicitly fixed by the overall scale
which is determined by the choice of the parameter $c$. In compensation for this
rather inhabitual degree of complication in the familiar large distance limit,
we get extremely simple boundary conditions in the less familiar limit at the
horizon as $\lambda \rightarrow c$, the only condition here being that the
unknowns $E$,\ $B$, $X$, $Y$ should be regular as differentiable functions of
the ellipsoidal type coordinates $\lambda$ and $\mu$. The most mathematically
delicate boundary conditions (for which however no physical considerations or
parameter values are involved) are those for geometrical regularity on the
rotation axis $\mu=\pm1$, i.e. for the limit $(1-mu^2) \rightarrow 0$,
which are given by 
$${\partial E\over \partial\lambda} =O(1-\mu^2) \ , \ \ \ \ \ \
{\partial B\over \partial\lambda} =O(1-\mu^2)\ , $$
$$ {\partial Y\over \partial\lambda} =O(1-\mu^2)\ , \ \ \ \ \
{\partial Y\over \partial\mu}+2E{\partial B\over \partial\mu}
-2B{\partial E\over \partial\mu} =O(1-\mu^2)   \ ,$$
$$X=0(1-\mu^2)\ , \ \ \ \  \ {(\mu^2-1)\over 2X}{\partial X\over \partial\mu}
=1+O(1-\mu^2)  \ . \eqno(5.23)$$

After I first obtained this system I succeeded deriving a pure vacuum ``no hair
theorem"$^{[911][10]}$ which Robinson was able soon after to generalise to the
full electromagnetic case$^{[90]}$, establishing that the solutions belong to
descrete families each depending continuously only on the three relevant
physical parameters involved in the boundary conditions namely $Q$, $J$, and
$C$, of which the latter fixes the overall scale and thus implicitly the mass
$M$.  Our method was to equate a certain divergence to a positive definite
function of the infinitesimal difference between nearby solutions for the same
parameter values and hence (using the boundary conditions) to show that the
latter must vanish. One such family consisted of course of the already known
Kerr solutions (subject to the condition $M^2>Q^2+J^2/M^2$) and in view of
various restrictions on special limits such as that of spherical symetry it
seemed unlikely from the outset that any others existed. Nevertheless it was
necessary to wait several years before it was established
beyond doubt that they do not.

The way that Robinson and I had constructed the divergence with the miraculously
positive definite form we needed was based on a purely trial and error approach
whose success in the electromagnetic case$^{[90]}$ required a veritable
algebraic tour de force. Robinson even succeeded in using the trial and error
metheod to construct a finite difference generalisation$^{[91]}$ that complelty
established the uniqueness of the original Kerr black hole solutions, with
$Q=0$, $M^2>J^2/M^2$, for the pure vacuuum case, but the trial and error method
was never able to cope with the finite difference case in its full
electromanetic generality, and so the complete solution had to await the
introduction of new and more sophisticated techniques by Bunting$^{[92][93]}$
and Mazur$^{[94]}$. The Bunting method is of great interest in its own right,
being potentially useful for much more general problems$^{[95]}$. I shall
however restrict myself here to the description of the Mazur method which is
more specialised but more explicit.

As soon as I obtained the system given above I noticed that the field equations
(5.21) and (5.22) have the striking feature$^{[10]}$ (which does not apply to
the traditional system defined in terms of $k^\rho$ rather than $m^\mu$) of
being derivable from a Lagrangian integral
$${\cal I}=\int{\cal L}d\lambda d\mu \eqno(5.24)$$ 
that is {\it positive definite} with the comparitively simple form
$${\cal L}={\vert\hat\nabla X\vert^2+\vert\hat\nabla Y+2E\hat\nabla B
-2B\hat\nabla E\vert^2\over 2X^2}  +2{\vert\hat\nabla E\vert^2
+\vert\hat\nabla B\vert^2\over X} \eqno(5.25)$$ 
but neither I nor Robinson had seen how to exploit this directly.  The
breakthrough by Mazur was based on work by Geroch$^{[96]}$ and
Kinnersley$^{[97]}$ who showed that Ernst type systems can be interpreted as
belonging to a class of non-linear $\sigma$ models whose field equations are
equivalent to a partially redundant set of ordinary divergence type conservation
laws of the form
$$\hat\nabla {\bf J}=0 \eqno(5.26)$$
where ${\bf J}$ is a matrix vector constructed according to the
prescription
$${\bf J}=\varpi^{-1}\bfPhi^{-1}\cdot\hat\nabla\bfPhi \eqno(5.27)$$
where $\varpi$ is the known positive weight function given by (5.9)
(which does not appear in traditional $\sigma$ models but whose appeareance
here adds no significant complication) and where $\bfPhi$ is a
hermitian matrix function with the important property of being
{\it positive definite} in the present case, its components being
given by
$$\Phi_{\dot a b}=\eta_{\dot a b}+2\bar v_{\dot a} v_b \eqno(5.28)$$
(using a bar to denote complex conjugation and placing a dot on
conjugately transforming indices)
where $\eta_{\dot a b}$ is just the the fixed Minkowski type hermitian
metric  in diagonam form with signature $(-1, 1, 1)$ for the space
of three dimensional complex vectors, in which the field vector
$v_a$ is given  in terms of the complex Ernst type variables
$$\varepsilon=-X+iY-\psi\bar\psi\ , \ \ \ \ \psi=E+i B \eqno(5.29)$$
by
$$(v_{_0}, v_{_1}, v_{_2}) = {_1\over^2}\vert X\vert^{-1/2}(\varepsilon-1,
\varepsilon+1, 2\psi) \eqno(5.30)$$
which is such as to make $v_a $ automatically ``timelike" in hermitian
space, with unit normalisation given by
$$\eta^{a\dot b}v_a \bar v_{\dot b} =-1 \ ,\eqno(5.31)$$
which is sufficient to guarantee the required positivity of the
hermitian matrix given by (5.28).

                 The preservation of the field equations by the
SU(2,1) action (leaving $\eta_{a\dot b}$ invariant) that Kinnersley
thus made manifest can also be seen to extend ``off shell" in the sense
that our Lagrangian (15.25) is also invariant, as can be seen by rewriting
it as
$${\cal L}=2\vert\eta^{a\dot b}v_a\hat\nabla\bar v_{\dot b}\vert^2
-\eta^{a\dot b}(\hat\nabla v_a)\hat\nabla v_{\dot b}
={_1\over^2}\hat g_{ij}{\rm tr}\{{\bf J}^i\cdot{\bf J}^j \} \eqno(5.32)$$
where $\hat g_{ij}$ ($i,j=1,2$) is the positive definite two dimensional metric
given by (5.11) that is used for the specification of $\hat\nabla$.

The Mazur method of establishing the uniqueness of the solutions of such
a system, subject to appropriate boundary conditions such as are given in
the present case, is essentally dependent on the positive definiteness
of both $\bfPhi$ and $\hat g_{ij}$. The objective is to prove the
vanishing of the difference
$$\upbPhi=\bfPhi_{[1]}-\bfPhi_{[0]}   \eqno(5.33)$$
between any pair of matrices representing conceivably distinct solutions.
Te vanishing of this difference is evidentlyequivalent to the vanishing of 
whatI refer to as the deviation matrix,
$$\bfDelta=\bfPhi_{[1]}\cdot \bfPhi_{[0]}^{\ -1}-{\bf 1}=
\upbPhi\cdot\bfPhi_{[0]}^{\ -1}  \eqno(5.34)$$
where ${\bf 1}$ is the unit matrix (in the complex 3 space).
The gradient of this deviation matrix will evidently be given by
$$\varpi\hat\nabla\bfDelta=\bfPhi_{[1]}\cdot \upbJ \cdot\bfPhi_{[0]}^{\ -1}
\ , \ \ \ \ \ \ \upbJ={\bf J}_{[1]}-{\bf J}_{[0]} \ . \eqno(5.35)$$
Taking the difference we obtain
$$\hat\nabla(\varpi\hat\nabla\bfDelta)=\bfPhi_{[1]}\cdot\left\{\hat\nabla\upbJ
+\varpi^{-1}\hat g_{ij}({\bf J}^i_{[1]}\cdot{\bf J}^j_{[1]} -2{\bf J}^i_{[1]}
\cdot{\bf J}^j_{[0]}+{\bf J}^i_{[0]}\cdot{\bf J}^j_{[0]} ) \right\}
\cdot \bfPhi_{[0]}^{\ -1}   \eqno(5.36)$$
The next step is to use the hermiticity property
$$\bfPhi=\bfPhi^\ast \ \ \ \Rightarrow \ \ \ {\bf J}^\ast=\bfPhi\cdot
{\bf J}\cdot \bfPhi^{-1}  \eqno(5.37)$$
(where the asterisk denotes the complex conjugate of the transpose)
to rewrite the quadratic terms in (5.36) as
$$\bfPhi_{[1]}\cdot({\bf J}^j_{[1]}\cdot\upbJ_j-\upbJ_j\cdot{\bf J}^j_{[0]})
\cdot\bfPhi_{[0]}^{\ -1}  ={\bf J}_{[1]}^{\ast\ j}\cdot\bfPhi_{[1]}\cdot
\upbJ_j\cdot\bfPhi_{[0]}^{\ -1} -\bfPhi_{[1]}\cdot\upbJ_j\cdot
\bfPhi_{[0]}^{\ -1}\cdot{\bf J}_{[0]}^{\ast \ j} \ . \eqno(5.38)$$
On taking the trace of (5.36) we thus obtain the scalar identity
$$\hat\nabla\left(\varpi\hat\nabla {\rm tr}\{\bfDelta \}\right)
- {\rm tr}\{\bfPhi_{[0]}^{\ -1}\cdot\bfPhi_{[1]}\cdot\upbJ\} =\varpi^{-1}
\hat g_{ij}{\ \rm tr}\{ \bfPhi_{[0]}^{\ -1}\cdot\upbJ{^{\ast\ j}}\cdot
\bfPhi_{[1]}\cdot \upbJ{^j} \} \ .\ \ \ \ \ \ \ \ \eqno(5.39)$$
This Mazur identity includes as special cases the identities
found for the linearised or uncharged limits by  Robinson and 
myself$^{[89][10][90][91]}$ using a less systematic approach.

It follows directly from the form (15.26) of the field equations 
that the current difference satisfies
$$\tilde\nabla\upbJ  =0 \eqno(5.40)$$
and hence that the left hand side of the identity (5.39) will reduce to
a divergennce whose integral can be converted using Green's theorem
to a surface contribution which will vanish,
$$\oint dS_i\ \varpi \hat g^{ij}\hat\nabla_j({\rm tr}\bfDelta)\rightarrow 0
 \eqno(5.41)$$
subject to the appropriate boundary conditions which can be verified
(using particular care for the axis where $X^{-1}$ diverges) to be in fact
satisfied in this case. Under these conditions one can deduce that the
right hand side of (5.39) vanishes since (by the positivity
of $\bfPhi$) it is clearly a positive definite function of
$\upbJ$ which must therefore vanish, i.e. using (5.35) we get
$$\upbJ=0 \ \ \ \ \rightarrow\ \ \ \ \ \bfDelta={\bf C}  \eqno(5.42)$$
where ${\bf C}$ is some constant matrix. Since the boundary conditions
as $\lambda\rightarrow \infty$ ensure that $\bfDelta\rightarrow 0$
there one ends  by getting
$${\bf C}=0 \ \ \ \rightarrow \bfPhi_{[1]}=\bfPhi_{[0]} \eqno(5.43)$$
which finally establishes the required uniqueness.

\bigskip

\bigskip\parindent=0 cm
{\bf 6. Special Properties of the Kerr Newman Vacuum Solutions.}
\medskip\parindent =1.2 cm

The theorem obtained at the end of Section 5 establishes conclusively that in
the source free electrovac case there are no (topologically simple) stationary
axisymmetric asymptotically flat black hole solutions with non degenerate
($\kappa>0$) horizon other than those of the Kerr Newman family$^{[102]}$ as
restricted by the condition
$$c^2>0 \ , \ \ \ \ c^2= M^2-a^2-Q^2\ , \ \ \ a =J/M \ , \eqno(6.1)$$
where this parameter $c$ represents the value for these solutions of the
quantity introduced more generally by (5.8)\ .
 This includes, for
$Q=0$, the pure vaccuum family of Kerr$^{[103]}$  solutions, whose black hole
nature, in the allowed parameter range $M^2>a^2$, was first clearly recognised
by Boyer$^{[71] [72]}$. (For the degenerate limit for which $c=0$ or
equivalently $\kappa=0$, corresponding to a horizon at zero temperature in the
thermodynamic analogy, the problem has still not been completely solved, but it
is known that in the non rotating case $J=0$ there is a
class$^{[83][84][85][86][87][88]}$ of
electrically balanced solutions with $M^2=Q^2$ that is much more general than
the corresponding Reisner Nordstrom subset within the Kerr Newman familly.)
Taking full advantage of the very special properties that will be briefly 
surveyed below, a long series of investigations, of which the first
was that of Vishveshwara$^{[104]}$ and the most recent that of Whiting$^{[13]}$
confirm that the Kerr solutions specified by (6.1) are effectively
 stable against the all the most obviously relevant kinds of perturbation.

In terms of
ordinary coordinates $r$, $\theta$ introduced by
$$\lambda = r-M\ ,\ \ \ \ \ \mu={\rm cos\ }\theta \eqno(6.2)$$
the explicit solutions for tha Ernst type variables are given by
$$X=\left\{ r^2+a^2 +{(2MR-Q^2) a^2 {\rm sin}^2\theta\over r^2+a^2{\rm 
cos}^2\theta } \right\}  {\rm sin}^2\theta $$
$$Y=\left\{\ M(2+{\rm sin}^2\theta) -{{\sin}^2\theta[Q^2 r-Ma^2{\rm sin}^2
\theta]\over r^2+a^2{\rm cos}^2\theta}  \right\} 2a{\ \rm cos\ }\theta $$
$$E={Q(r^2+a^2){\rm cos\ }\theta\over r^2+a^2{\rm cos}^2\theta} \ ,\ \ \ \ \ \
B={-Qra{\ \rm sin}^2\theta\over r^2+a^2{\rm cos}^2\theta} \ .\eqno(6.3)$$
Going back to ordinary metric and electromagnetic potential components
gives the rather simpler forms
$$V=1-{2Mr-Q^2\over r^2+a^2{\rm cos}^2\theta} \ , \ \ \ \ \ \ \Phi=
{Qr\over r^2+a^2{\rm cos}^2\theta } \ ,$$
$$W={-(2Mr-Q^2) a {\ \rm sin}^2\theta\over r^2+a^2{\rm cos}^2\theta }\ ,
\ \ \ \ \ \ \ \Xi=  r^2+a^2{\rm cos}^2\theta \ . \eqno(6.4)$$

Subject to (6.1) these solutions do in fact have turn 
out$^{[40][73][74][10]}$
to have the property (which was not assumed in advance in the approach
outlined above) of having a well behaved Kruskal type horizon crossover when
analytically extended towards the past. When analytically extended to the
interior they exhibit many amusing but one presumes physically irrelevant
features such as a time machine$^{[74][57]}$ in the region beyond the
Cauchy horizon, which (in this more general case, as in the
Reissner Nordstrom case discussed in Section 2 and in more detail in
the accompanuing lectures of Israel) is a sign of instability
occurring at $r=r_-$, using the standard abreviation
$$ r_\pm= M\pm c  \ . \eqno(6.5)$$

In these solutions the three quantities whose uniformity
over the black hole event horizon at $r=r_+$
  was guaranteed in advance by the results of
Section 4, namely the decay constant $\kappa$, the limiting value
$\Omega^{_H}$ of the zamo angular velocity $\omega$,
and the value $\Phi^{_H}$ of the comoving
potential $\Phi^{\dagger}=\Phi+\Omega^{_H} B$,  will be given
respectively  by
$$\kappa= {c\over 2Mr_+ -Q^2}  \ , \ \ \ \ \Omega^{_H}={a\over 2Mr_+ -Q^2}\ ,
\ \ \ \ \ \Phi^{_H}={Qr_+\over 2Mr_+-Q^2} \ . \eqno(6.6)$$

Neither the physically motivated global geometrical approach
that we have followed here,
 nor the even the analytical approach (based on the
assumption of a special form for the Weyl tensor) used originally by Kerr and
Newman makes it at all obvious in advance that the solutions should have such
remarkable special algebraic properties as they actually do. The special
simplicity of the metric that is finally obtained can be made most directly
manifest by expressing it in tetrad form
$$ds^2=dx^\mu dx^\nu(-e^{_0}_{\ \mu}e^{_0}_{\ \nu}+e^{_1}_{\ \mu}e^{_1}_{\ \nu}
+e^{_2}_{\ \mu}e^{_2}_{\ \nu}+e^{_3}_{\ \mu}e^{_3}_{\ \nu}) \eqno(6.7)$$
terms of a preferred {\it canonical separable tetrad} which I found very early
in the history of the study of the Kerr Newman solutions when the separablity of
the Hamilton jacobi equation for free particle trajectories and of the scalar
Klein Gordon  wave equation was first brought to light$^{[74][77]}$. However
although it is useful generally$^{[105][106][107][108]}$ 
and is even more strongly to be
preferred for obtaining separability of higher spin wave equations$^{[109]
[110][111][112]}$ , this canonical tetrad  has unfortunately tended to have been
neglected by workers on higher spin separability$^{[113][114][115]}$ in favor of
other tetrads (particularly that of Kinnersley$^{[116]}$ which has advantages in
other contexts but not in this one) thereby making many unavoidably heavy
calculations$^{[117]}$ even longer and more complicated than necessary. To get
maximum algebraical symmetry it is necessary to replace the geometrically
defined time and angle coordinates $t$ and $\phi$ as introduced in the previous
section (and which in the specific context of the Kerr solutions are commonly
referred to as the  coordinates of Boyer Linquist$^{[73]}$ in order to
distinguish them from the rather different time and angle coordinates used by
Kerr himself$^{[103]}$) by very closely related ignorable coordinates $\tilde t$
and $\tilde\phi$ say, and to introduce a recalibrated angle coordinate $q$ say
in place of $\mu$, according to the specifications
$$\tilde t=t-a\phi \ ,\ \ \ \ \tilde\phi=a^{-1}\phi\ , \ \ \ \ \
 q=a{\ \rm cos\ }\theta\ . \eqno(6.8)$$

The canonical maximally symmetric tetrad is then expressible as
$$e^{_0}_{\ \mu}dx^\mu=-\left({\Delta_r\over r^2+q^2}\right)^{1/2}
(q^2 d\tilde\phi+d\tilde t) \ ,\ \ \ \ \ \ e^{_1}_{\ \mu}dx^\mu=
\left({ r^2+q^2\over \Delta_r}\right)^{1/2} dr\ , \ \ \ \ \ \ \ $$
$$e^{_3}_{\ \mu}dx^\mu=-\left({\Delta_q\over r^2+q^2}\right)^{1/2}
(r^2 d\tilde \phi-d\tilde t) \ , \ \ \ \ \ \ e^{_2}_{\ \mu}dx^\mu=
\left({ r^2+q^2\over \Delta_q}\right)^{1/2} dq\ , \eqno(6.9)$$
where $\Delta_r$ and $\Delta_q$ are quadratic functions repectively of $r$
only and of $q$ only, the symmetry between these variables being broken
only by the presence of a linear mass term in the former but not the latter:
$$\Delta_r=r^2-2Mr+ a^2 \ , \ \ \ \ \ \Delta_q= a^2-q^2\ .  \eqno(6.10)$$
The electromagnetic field potential is expressible in the even simpler
form (proportional to the first of these tetrad forms as
$$A_\mu dx^\mu=\left({Qr\over r^2+q^2}\right)(q^2 d\tilde\phi+d\tilde t)
\eqno(6.11)$$

At the expense of violating the asymptotic boundary conditions the algebraic
symmetry between $r$ and $q$ could be made complete$^{[77]}$ by including
appropriate gravimagnetic and electromagnetic monopole terms. The full
symmetry is however already manifest in the corresponding expression$^{[12]}$
for the crucially important Killing Yano 2-form  $f_{\mu\nu}$ given by
$$f_{\mu\nu}dx^\mu dx^\nu=q dr\wedge(d\tilde t+q^2 d\tilde\phi)
+r dq\wedge(d\tilde t-r^2 d\tilde\phi) \ ,\eqno(6.12)$$
whose existence,  as a solution of the Killing Yano equations
$$ f_{\mu\nu}=0\ , \ \ \ \ \ \ \nabla_\rho 
f_{\mu\nu}=\nabla_{[\rho}f_{\mu\nu]}, \eqno(6.13)$$
underlies the remarkable hidden symmetries of the Kerr Newman family and is
by itself sufficient to characterise them completely among asymptotically
flat electrovac solutions. 

Why the purely local characterisation given by (6.13) should give the same
result as the global boundary conditions for a black hole equilibrium problem
remains mysterious, but once it is known that (as was first revealed by
the work of Penrose and his collaborators
$^{[118][119][120][121][122]}$ using a spinorial approach)
 there is a non zero solution of (6.13), most of the other
special properties of the Kerr solutions can be obtained by more or less
 straightforeward deduction without the intervention of any
further independent miracles, starting with the stationarity, whose generator is
given$^{[123]}$ directly by
$$ k^\mu ={1\over 3!}\varepsilon^{\mu\nu\rho\sigma}\nabla_\nu f_{\rho\sigma}
\ , \eqno(6.14) $$
which, by (6.13) will automatically satisfy the Killing equation (4.1).
In addition to this ``primary" Killing vector (6.11) also$^{[123][124][12]}$
ensures the existence of a ``secondary", generically independent, one given by
$$ h^\mu =a^\mu{_\nu}k^\nu\ ,\ \ \ \ \ \ \ a^\mu{_\nu}=f^\mu{_\rho}f^\rho{_\nu}
\ ,\eqno(6.15)$$
which again satisfies Killing's equation as a further automatic
consequence of (6.13) and which turns out when evaluated to
be given as a linear combination of the ``primary", stationarity generating
Killing vector $k^\mu$ and of the axisymetry generating Killing vector
$m^\mu$ (as distinguished by having closed trajectories) that is
interpretable$^{[124][12]]}$ as generating
rigid rotations with angular velocity $\Omega=a^{-1}=M/J$ about the axis:
explicitly it satisfies
$$\nabla_{(\mu}h_{\nu)}=0 \ , \ \ \ \ \ \ \ h^\mu =a^2 k^\mu+am^\mu \ .
\eqno(6.16)$$
The pair of independent Killing vectors thus obtained will of course give
rise to a corresponding pair of quantities $k^\nu u_\nu$ and $h^\nu u_\nu$
(or $k^\nu u_\nu$ and $m^\nu u_\nu$) that are conserved allong solutions
of the geodesic equations (4.8). 
The ``hidden symmetry" corresponding to the existence of the ``fourth"
constant of motion that is needed to provide a complete set of first 
integrals of the equations of motion (the third being given trivially just
bu $u^\mu u_\mu$ itself) is given by
the  tensor $a^{\mu\nu}$ as defined in (6.15) which is an ordinary
symmetric Stackel Killing tensor in the sense that, again as as
automatic consequence of (6.13), it satisfies the conditions
$$a_{[\mu\nu]}=0 \ ,\  \ \ \ \ \ \
\nabla_{(\rho}a_{\mu\nu)}=0 \eqno(6.17)$$
which evidently suffice to ensure that the quadratic combination
$a_{\mu\nu}u^\mu u^\nu$ will indeed be constant allong solutions of (4.8), thus
providing the  ``fourth" constant that is needed to make these equations
completely integrable$^{[74][10]}$. Moreover this tensor $a^{\mu\nu}$ is not
just a Killing Stackel tensor in the weak sense of satisfying (6.17) but
is automatically, by the integrability conditions$^{[12][124]}$ for (6.13),
is automatically a Killing tensor in the strong sense$^{[125]}$ that
it also satisfies
$$a^\rho{_{[\mu}}R_{\nu]\rho}=0 \eqno(6.18)$$
which is the suppmementary condition needed in conjunction with
(6.17) to ensure that the corresponding self adjoint differential operator
commutes with Dalembertian (scalar) wave operator, i.e.
$$[\nabla_\mu a^{\mu\nu}\nabla_\nu,\nabla^\rho\nabla_\rho]=0 \ ,\eqno(6.19)$$
Just as the scalar Dalembertian can be thought of as a sort of square
of the ordinary first order dirac operator $\gamma^\mu \nabla_\mu$ acting
on 4-spinors, so analogously$^{[126][30][12]}$
the operator $\nabla_\mu a^{\mu\nu}\nabla_\mu$
can be thought of as a sort of square of a first order generalised spinor
angular momentum operator $L$ that commutes with the Dirac operator
$$ [L, \gamma^\mu \nabla_\mu]=0 \ , \ \ \ \ \ \ L=i\gamma_\mu
(\gamma^{_5} f_\mu{^\nu}\nabla_\nu-k_\mu )\ , \eqno(6.20)$$
in the same way as does the ordinary Kosman$^{[127]}$ energy operator $K$ for
 any Killing vector,
$$[K, \gamma^\mu\nabla_\mu ]=0   \ , \ \ \ \ \ \ K=ik^\mu\nabla_\mu
+{_1\over^4}[\gamma^{[\mu} \gamma^{\nu]}k_\nu, i\nabla_\mu] \ .\eqno(6.21)$$

Just as the commutation law (6.19) is interpretable as resulting from the
separability of the scalar Klein Gordon equation$^{[77]}$ so analogously
the commutation law (6.20) is interpretable as resulting from the separability
of the Dirac equation$^{[108]}$. In addition to these cases of separability for massive
particle wave equations, the Kerre Newman solutions are also characterised by
analogous separability properties for massless higher spin wave
equations, including notably those for the separate electromagnetic
and electromagnetic perurbations that are relevant to
stability analysis$^{[114][115]}$, though so far not for the case when the
electromagnetic and gravitational perturbations are coupled as will be the
case for the generic perturbation in the charged Kerr Newman case.
The analysis of such wave equations is carried out most conveniently by
using not the orthonormal version but the corresponding
 null version of the canonical tetrad, the latter being given in terms
of the former by a transformation of the standard form
$$\ell_\mu={_1\over^{\sqrt 2}}(e^{_0}_{\ \mu}+e^{_1}_{\ \mu}  ) \ , \ \ \ \
\tilde\ell_\mu={_1\over^{\sqrt 2}}(e^{_0}_{\ \mu}-e^{_1}_{\ \mu}  ) \ , \ \ \ \
z_\mu={_1\over^{\sqrt 2}}(e^{_0}_{\ \mu}+ie^{_1}_{\ \mu}  ) \ ,\eqno(6.22)$$
where $\ell_\mu$ and $\tilde \ell_\mu$ are null and $z_\mu$ is complex.
Let us consider together the case
of an ordinary complex scalar field $\Psi_{_0}$ say, the case of an ordinary Maxwell 
field $F_{\mu\nu}$, for which we take a priviledged pair of complex
tetrad components
$$\Psi_{_1}=F_{\mu\nu}\ell^\mu z^\nu\ , \ \ \ \ \
\Psi_{_{-1}}=F_{\mu\nu}\bar z^\mu \tilde\ell^\nu\ , \eqno(6.23)$$
(where for reasons of notational convenience that will become obvious we do not
use the traditional counting system which would label these components as
$\Phi_{_0}$ and $\Phi_{_2}$ ), and finally the case of a gravitational
perturbations 
with Weyl tensor $C_{\mu\nu}{^\rho}{_\sigma}$ for which we similarly take
the priviledged pair of complex tetrad components
$$\Psi_{_2}=-C_{\mu\nu\rho\sigma}\ell^\mu z^\nu\ell^\rho z^\sigma\ ,\ \ \ \ \
\Psi_{_{-2}}=-C_{\mu\nu\rho\sigma}\tilde\ell^\mu \bar z^\nu
\tilde\ell^\rho \bar z^\sigma\ ,\eqno(6.24)$$
 (which in the traditional counting system would be labeled $\Psi_{_0}$
and $\Psi_{_1}$).  Then the upshot of the many studies referred
to above is that$^{[109]}$ corresponding field
equations are separable, in terms of our present notational system, by setting
$$(r-iq)^{\vert s\vert}\Psi_s= X_s(r)Y_s(q) e^{-i(E\tilde 
t-\tilde\Phi\tilde\phi)} \eqno(6.25)$$
(where the helicity index $s$ runs over the values $0, \ \pm1,\ \pm2$)
 with the resulting separated equations having the form
$$\left\{ {d\over dr}\Delta_r {d\over dr} +{(Er^2+ is(r-M)-\tilde\Phi)^2
\over \Delta_r } +4isEr \right\}X_s = \tilde K X_s $$
$$\left\{ {d\over dq}\Delta_q {d\over dq} +{(Eq^2+ sq+\tilde\Phi)^2
\over \Delta_q } +4sEq \right\}Y_s = -\tilde K Y_s \eqno(6.26)$$
where $\tilde K$ is the separation constant whose existence expresses the
hidden symmetry and where it can be seen from the relation
$$E\tilde t-\tilde\Phi\tilde\phi= Et-\Phi\phi\eqno(6.27) $$
that $E$ is interpretable as the ordinary energy associated with the
``primary" Killing vector $k^\mu$ while $\tilde\Phi$, which is analogously
associated with the secondary Killing vector $h^\mu$ is related to the
ordinary angular momentum constant $\Phi$ associated with the axial Killing
vector $m^\mu$ by the simple relation
$$\tilde \Phi=a\Phi-a^2 E\eqno(6.28) $$
In the case $s=0$ the above form agrees directly with what is obtained in
the limit of vanishing perticle charge and mass, $e=m=0$  from my original
separated form of the Klein Gordon equation, but in the higher spin cases it
differs from the forms originally obtained by Teukolsky due to his use of a
non canonical (Kinnersley$^{[116]}$ type) tetrad which lead to an
unnecessarily complicated form in which the symmetries manifest in the
version above ( the helicity symmetry between $s$ and $-s$, and the almost
perfect algebraic symmetry between $r$ and $iq$) are all spoiled by
subjecting (6.9) to a symmetry violating tetrad transformation consisting of 
a combined boost and rotation of the form
$$\ell^\mu\rightarrow \left({2(r^2+q^2)\over\Delta_r}\right)^{1/2}\ell^\mu\ ,
\ \ \ \ \ \ \ \tilde\ell^\mu\rightarrow
 \left({\Delta_r\over 2(r^2+q^2)}\right)^{1/2}\tilde\ell^\mu\ ,  $$
$$z^\mu\rightarrow {r-iq\over\sqrt {r^2+q^2}}z^\mu \ .  \eqno(6.29)$$
The dazzling prestige conferred on the Kinnersley tetrad by its successful
use in the original discovery of the higher spin separability has
unfortunately blinded many workers to the fact that the separation works
even more efficiently in terms of the original canonical tetrad, with the
result that many published calculations$^{117}$ are at least twice as long
as necessary The canonical tetrad has however tended to come back into use
in more recent work, whose achievements include separation of the equations
of parallel transport of a tetrad$^{[107][108][102]}$ and the equations for
the stationary equilibrium of  cosmic strings$^{[128][129][130]}$ of of
certain simple kinds (not just the ordinary Goto Nambu kind but also the
more general non dispersive model allowing for the averaged effect of noise
or wiggles) whose mechanics will be explained in the following sections.

\bigskip

\parindent=0 cm
{\bf 7. Basic Brane Mechanics.}
\rm\parindent=1.2 cm
\medskip

The purpose of the last part of this course is to give a brief introductory
overview (and some illustrative applications in the context of cosmic strings)
of  the general principles of brane dynamics using a recently developed fully
covariant approach$^{[14][15]}$ that avoids the use of excess mathematical
bagage (such as the use of distribution theory and specially adapted coordinates
for separate subsystems) that may be useful for detailed calculations in
specific applications, but that would obscure the simplicity and generality of
laws such as the general equation governing the extrinsic motion of any brane,
which is expressible in the formalism set up below (using underlining to
distinguish quantities defined with respect to a $p$-brane under consideration
from any higher dimensional analogue that may also be relevant) in terms of its
stress momentum energy tensor $\underline T^{\mu\nu}$, and its second
fundamental tensor $ K_{\mu\nu}{^\rho}$ in the form
$$\underline T^{\mu\nu}K{_{\mu\nu}}{^\rho}=
\af{^\rho}\ , \eqno(7.1)$$
where  $\af{^\rho}$ is the total orthogonally projected
force contribution (such as that of the wind on a sail, or of an external
electromagnetic field on the current in a cosmic string) from the various
external systems (if any) with which the brane may interact.

Following an increasingly popular usage$^{[131][132]}$, the term {\it brane} is used
here to designate a physical model of the category that includes continuous
media and point particles as extreme cases, with ordinary membranes (from which
the term is derived) and strings as the only other possibilities in a
4-dimensional background. Generally, a $(p-1)$ brane is to be understood to
be a dynamical system defined in terms of fields with support confined to a $p$
dimensional world sheet surface ${\cal S}$ in a background (flat or curved)
spacetime manifold of dimension $n\geq p$. The extreme case, with $n=p$ is that
of a continuous medium for which the confinement condition is redundant. The use
of this concept makes it possible to give a unified description of basic
properties that are common to a very wide range of physically diverse
phenomena. A simple and very important example is the universal rule that (as a
consequence of (7.1) and independently of the nature of any external forces so
long as their coupling does not involve gradients of internal field variables)
the condition for a (contravariant) vector $\eta^\mu$ say to be an  extrinsic
{\it  bicharacteristic} vector, i.e. to be tangent to the direction of ``group"
propagation of localised wave packets of small extrinsic displacements of the
localisation of the world sheet (which of course is meaningful only for $n<p$)
with a corresponding {\it characteristic} covector $\chi_\mu$ normal to the
direction of the associated ``brane wave" sheets, will be given$^{[123]}$ simply by
$$\eta^\mu=\underline T{^{\mu\nu}}\chi_\mu\ , \ \ \ \ \ \
\underline T{^{\mu\nu}}\chi_\mu \chi_\nu=0 \ . \eqno(7.2)$$
The hyperbolicity condition to the effect that the characteristic equation (7.2)
should define a real characteristic cone provides a restriction (trivial for a
point particle  and reducing just to a requirement of positivity of the
ordinary tension $T$ in the case of a string$^{[133]}$) that must always be
satisfied as a condition for local stability except of course in the case $p=n$
of a continuous medium for which there is no geometric possibility of
extrinsic perturbations, which is why an ordinary perfect fluid with postive
pressure $P$ can be stable after all, despite the fact that it is elliptic
(with no real roots) as far as the criterion (7.2) is concerned.

In an ordinary spacetime with $n=4$ a continuous medium (with $p=4$) counts as a
3-brane, the other possibilities being that of a membrane model (with $p=3$)
which counts as a 2-brane, a string model (with $p$=2) which counts as a
1-brane, and finally at the other extreme, a point particle model (with $p$=1)
which counts as a zero brane.   Employment of brane models of lower dimension,
$p<n$, for which the extrinsic confinement condition and the associated
hyperbolicity requirement derived fom (7.2) are essential, is often useful for
providing an approximate descriptions of higher dimensional case when the the
fields characterising the latter are highly concentrated in the neighbourhood of
a lower dimensional world sheet within a distance that is small compared with
the scales characteristic of dynamic variations in directions tangential to the
world sheet. Thus for example a point particle model might be useful for
describing the motion, with respect to a relatively slowly varying background,
of a small loop in a string model that might itself be just a opproximation for
decribing what at a more microscopically accurate level might need the use of a
continuum model.  The example that has been most important in motivating the
development of the relativistic formalism described here is that of the
representation ( as originally suggested by Kibble$^{[134]}$, Witten$^{[135]}$
and others) of vortex defects (due to spontaneous symmetry breaking) of the
vaccuum by (``cosmic") string models as a macroscopic approximation for use in
the (cosmologically important) cases in which the vortex thickness can be
treated as negligible compared with other relevant length scales. This lead to
the introduction of models of variational type in which the action was to be
thought of as being derived from the microscopic action of the relevant
underlying field theory by integral across the vortex in a local equilibrium
state.

Quite generally, in cases where a compound system has a variational formulation
in terms of a total action of the form $\sum  {\cal I}$, the action contribution
of an individual $p$ brane of the system will be given by a corresponding $p$
surface integral
$${\cal I}=\int \underline{\cal L} d\underline{\cal S}   \eqno(7.3)$$ 
where $d\underline{\cal S}$ denotes the induced surface measure
and $\underline{\cal L}$  is a Lagrangian scalar function of whatever internal
fields on the world sheet are involved and also of any relevant externally
induced fields such as those given by (7.2). In the simplest (non conducting)
cosmic string models originally envisaged by Kibble$^{[134]}$ it was sufficient to
use a Goto-Nambu$^{[136]}$ action in which (as in the analogous Dirac membrane
model$^{[137]}$) the scalar $L$ is specified trivially as a constant, which is
interpretable as the negative of the (spacially isotropic) tension $T$ which in
this case is not only uniform but (as an expression of the special property of
intrinsic Lorentz invariance which distinguishes these particular models) is
also equal in this case to the value of the energy density $U$ say. The quantity
$U$ can be defined, for a generic brane model, as the eigenvectvalue
corresonding to the timelike eigenvctor of its surface stress momentum energy
tensor $\underline T{^{\mu\nu}}$, while in a string model the tension $T$ is
unambiguously definable as the other eigenvalue, the case of a membrane being
more complicated in far as it admits the possibility of two possibly distinct
tension eigenvalues.

It typically occurs that the approximate macroscopic treatment of a system that
is conservative, with a variational formulation, at a microscopic level may
require the use of a non conservative macroscopic model involving averaging over
microscopic degrees of freedom that are taken into count as entropy.  Although
it may invalidate the conservative nature of the model as a whole, such an
averaging process does not invalidate the local conservation laws obeyed by
additive quantities such as energy momentum or electromagnetic charge: what
happens is that instead of having the status of {\it Noether identities}
expressing the invariance properties that hold for the underlying variational
model, such conservation laws are to be interpreted in the macroscopic model as
{\it constistency conditions} for the existence of a corresponding microscopic
variational model. The commonly but (not always) appropriate notion that a
macroscopic model under consideration is obtainable by integrating out the fine
details of a more complicated underlying model makes it seem physically natural
to try to preserve some of the spirit of the original finer model by using a
description in terms of Dirac distributions. However although very useful for
some purposes when used with discretion, use of Dirac distributions can easily
become addictive, and is often systematically abused in a manner that hinders
clear analysis and provides an archetypical example of the kind of excess
mathematical baggage that the present approach is designed to avoid.

The most important example of mathematical machinery that is very helpful for
many specific purpose but whose use needs to be avoided (as excess baggage) when
one wants to obtain a simple formulation of general principles such as that
embodied in the ``generalised sail equation"$^{[14]}$ (1.1), is that of the
introduction of a system of internal coordinates $\sigma^i$ say, $(i=0, ...,
p-1)$, on the $p$ dimensional world sheet of the $(p-1)$ brane under
consideration, whose imbedding is thereby describable as a mapping
$\sigma^i\mapsto x^\mu$ where the $x^\mu$, $(\mu=0,1, ...,n-1)$ are local
coordinates on the $n$ dimensional background spacetime. A possibility that is
of considerable practical utility in the intermediate stages of many calculation
is the use of what I call ``adapted coordinates" meaning a matched system of
internal and external coordinates in terms of which the imbedding mapping is
characterised by
$x^{_0}=\sigma^{_0},\ ...,\ x^{_{p-1}}=\sigma^{_{p-1}}\ \  $,
$x^{_p}=0,\ ...,\ x^{_{n-1}}=0\ $, but this obviously can not be done for the
simultaneous treatment of intersecting branes (as at the junctions in a cluster
of soap bubbles) and it is also obviously incompatible with the freedom to use
an objective characterisation of the background coordinates (e.g. in flat space
applications by the requirement that they be Minkowskian) which may be important
for the final presentation and utilisability of the results. 

One of the uses, as an intermediate step, of a coordinate mapping $\xi^i\mapsto
x^\mu$ is for the explicit construction of the corresponding intrinsic
components of the images induced in the imbedding of {\it covariant}
tensor fields on the background space, such as the electromagnetic
potential $A_\mu$ and most important of all the background space time
metric $g_{\mu\nu}$, whose respective images are given by
$$A_\mu\mapsto\alpha_i=A_\mu{\partial x^\mu\over\partial\sigma^i}\ , \ \ \ \
g_{\mu\nu}\mapsto h_{ij}=g_{\mu\nu}{\partial x^\mu\over\partial\sigma^i}
{\partial x^\nu\over\partial \sigma^j} \ .\eqno(7.4)$$
In cases where a compound system has a variational formulation as a sum in which
each distinct brane contributes a term of the form (7.3), the obvious analogue
of the traditional variational specification of the conserved current and stress
energy momentum tensor (whose local conservation equations are the Noether
identities expressing gauge invariance and general diffeomorphism covariance)
will take the form
$$j^i={\partial \underline{\cal L}\over\partial \alpha_i}\ ,
  \ \ \ \ \ t^{ij}= 2{\partial\underline{\cal L}\over\partial h_{ij}}
+\underline{\cal L}h^{ij} \ ,\eqno(7.5)$$
subject to the proviso (which is not necessary for the simple and conducting
cosmic string models models that will be considered below)that Eulerian
variational derivatives are to be used instead of simple
partial derivatives if derivatives of the potential and metric are involved.
The quantities $h^{ij}$ appearing in (7.5) are of course the components
of the {\it contravariant inverse} of the induced metric which is to be used
for raising and lowering internal indices.

Whether they are specified variationally, as in (7.5), or whether they are
specified in some more empirical way, as would be necessary in a general, non
conservative model, the internal current and stress energy momentum tensor will
determine corresponding background tensor fields by the natural pull back
mapping that is determined directly by the imbedding for any {\it contravariant}
vector fields, the corresponding coordinate expressions being given by
$$\underline J^\mu=j^i{\partial x^\mu\over\partial \sigma^i}\ ,\ \ \ \
\underline T^{\mu\nu}=t^{ij}{\partial x^\mu\over\partial\sigma^i}
{\partial x^\nu\over\partial \sigma^j} \ .\eqno(7.6)$$

The idea of the strategy developed here is that it is more efficient for general
theoretical (as opposed to specific computational) purposes not to work with
internal tensors such as $j^i$ and $t^{ij}$ but rather to work with the
corrsponding background spacetime tensors, which in this case  are $\underline
J^\mu$  (with the underlining as a reminder that it refersto a surface not
volume current) and $\underline T^{\mu\nu}$. When a variational specification is
available it is preferable (particularly for dealing with compound systems
involving several mutually interacting branes of diverse dimensions) to bypass
the passage via (7.4) and (7.6) through the internal coordinate versions by
replacing (7.5) by the equivalent but more direct background coordinate
specifications
$$\underline J^\mu={\partial \underline{\cal L}\over\partial A_\mu}\ ,
  \ \ \ \ \underline T^{\mu\nu}= 2{\partial\underline{\cal L}
\over\partial g_{\mu\nu}}
+\underline{\cal L}\olg{^{\mu\nu}} \ ,\eqno(7.7)$$
in which the only formal difference from the usual expression for a continuous
medium as opposed to lower dimensional brane model is the replacment in the last
term of the contravariant version $g^{\mu\nu}$ of the ordinary background metric
by what I call the (first) {\it fundamental tensor} $\olg{^{\mu\nu}}$ of the
brane world sheet, which is obtained here as the pull back of the contravariant
inverse of the induced metric, i.e.
$$\olg{^{\mu\nu}}=h^{ij}{\partial x^\mu\over\partial\sigma^i}
{\partial x^\nu\over\partial \sigma^j}\ ,\eqno(7.8)$$
where a double overline is introduced here to denote the {\it surface
tangential part} of any tensor as defined with respect to the background metric,
i.e. the result of contracting all its indices with respect to the mixed rank
$p$ projection tensor version $\olg{^\mu}{_\nu}$ of the
fundamental tensor itself. It is to be noted that any background tensor that is
obtained as the pullback of an intrinsic tensor within the imbedded surface will
automatically be equal to its own tangential part, so that in particular we
shall have $\ov{\ov{\underline J}}{^\mu} =\underline J{^\mu}$ and
$\ov{\ov{\underline T}}{^{\mu\nu}}=\underline T{^{\mu\nu}}$.

The ``fundamental" tensor of the imbedding that is thus specified in accordance
with (7.8) is of great (but still insufficiently widely recognised)  importance
as the starting point for the systematic tensorial analysis of imbedding
curvature as described in the next section, whose results are applicable not
just to a timelike brane world sheet but also to submanifolds that are spacelike
(though not to those that are null, i.e. metrically degenerate).

In the particular case of a Goto Nambu string model$^{[136]}$ or a Dirac
membrane model$^{[137]}$, as characterised by an action of the form (7.3) with
$\underline L=L_0$ for some {\it fixed} value $L_0$, which (see section 3) gives
a uniform isotropic tension $T$ that is equal to the corresponding energy
density $U$ and opposite to the Lagrangian itself, i.e. $U=T=-L_{_0}$,  the
introduction of the fundamental tensor $\ug{^{\mu\nu}}$ makes it easy to check
the well known property that the characteristic propagation speed of extrinsic
perturbations is, in this case, that of light (c=1 in the units used here) by
substituting in (7.1) the simple formula whereby the stress momentum energy
density for such a (Gotu Nambu or Dirac) model  is expressible directly as
$\underline T{^{\mu\nu}}=-U\ug{^{\mu\nu}}$.

As explained in the appendix, we shall adopt the systematic use of a
 convention using an overhead parallelism symbol, $^=$,
to indicate the effect of projection into the surface, and an
overhead perpendicularity symbol, $ ^{\underline{\ \vert\ }}$, to indicate
the effect of the complementary orthogonal projection operation, so that
the  surface tangentiality conditions that the surface current
and stress momentum energy tensors must satisfy by construction,
will simply take the form
$${{\alJ}}{^\mu}=0\ ,\ \ \ \ \ {{\alT}}{^{\mu\nu}}
= 0\ , \eqno(7.9)$$

 it can be seen from (7.7) that the variations in a brane
Lagrangian $\underline{\cal L}$ due to an infinitesimal electromagnetic gauge variation
$A_\mu\mapsto$ $A_\mu+\nabla_\mu \chi$ and an infinitesimal diffeorphism
variation $g_{\mu\nu}\mapsto$ $g_{\mu\nu}+\nabla_{(\mu}\xi_{\nu)} $ of the
metric will be expressible respectively as
$$\underline J{^\mu}\nabla_\mu\chi=\onab_\mu(\chi\underline J^\mu)-\chi
\onab_\mu\underline J^\mu \ , \eqno(7.10)$$
and
$$\underline  T{^{\mu\nu}}\nabla_\mu\xi_\nu =\onab_\mu (\xi_\nu \underline
T{^{\mu\nu}})-\xi_\nu\onab_\mu \underline T{^{\mu\nu}}\ .\eqno(7.11)$$
It can be seen that the first term on the right of each of these equations has
the form$^{[15]}$ that characterises a tangential current divergence within the
$p$ dimensional brane world sheet, and hence that by the appropriate $p$
dimensional version of Green's theorem the corresponding surface integral will
be expressible as the integral over the brane boundary (if any) of the
contraction of the tangential current with the unit world sheet tangent vector
normal to, and oriented towards, the boundary. 

Let us consider the very large class of situations$^{[14]}$ that can be
represented by a well behaved {\it brane complex} ( or ``rigging system") in
which direct action of a lower on a higher dimensional brane occurs only when
the former forms a smooth boundary segment of the latter (as when a monopole,
treated as a point particle, forms the termination of a string, or when a sail
forms the boundary between two external wind volumes), subject to dynamic
equations to the effect that the infinitesimal variation of the relevant fields
other than the externally determined background fields $g_{\mu\nu}$ and $A_\mu$,
gives no contribution to the variation of the combined action $\sum{\cal I}$
taken over the various brane constituents of the system, restricting ourselves
to cases in which derivatives of the external fields $g_{\mu\nu}$ and $A_\mu$
are not involved in the action. (The exclusion of more general derivative
couplings merely avoids the extra technical complications that are present in
more elaborate, e.g. polarised systems, but the exclusion of direct action
except on a smooth boundary is more essential, being needed to avoid the serious
divergence difficulties, exemplified by that of the radiation back reaction on a
point particle, which would otherwise be involved.) Then it can be seen (by
systematically using (7.10) to convert divergences to boundary contributions)
that the requirement that this combined action $\sum{\cal I}$ should  also be
identically {\it invariant under gauge transformations} generated by an
arbitrary field $\chi$ is equivalent to the condition that there should be a
total current conservation law expressed by the condition$^{[123]}$ that
for each $p$ brane of the system we should have
$$\onab_\mu\underline J{^\mu}=\sum\lambda{_\mu} J^\mu , \eqno(7.12)$$
where the summation is taken over the separate $(p+1)$ branes of which the $p$
brane under consideration forms part of the bondary, and where $J^\mu$ without
underline denotes the value on the boundary segment of the current vector in the
higher dimensional sheet while $\lambda{_\mu}$ denotes the unit normal
from the $p$ dimensional boundary into the relevant externally attatched brane
world sheet. Similarly (by analogous systematic use of (7.11) to convert
divergences to boundary contributions) it can be seen under the same conditions
that the general covariance requirement that the combined action be {\it
invariant under diffeomorphisms} generated by an arbitrary vector field
$\xi^\mu$ is equivalent to a local energy momentum conservtion law to the
effect$^{[1]}$ that for each brane of the system we should have
$$\onab_\nu{\underline T}{^{\mu\nu}}= f^\mu\ , \ \ \ \ \ f_\mu=
\sum\lambda{_\nu} T{^\nu}{_\mu}
+F_{\mu\nu}\underline J{^\nu}  \eqno(7.13)$$
in which the force density is obtained as the sum of  contact contributions from
the (non underlined) stress momentum energy density tensor $T{^{\mu\nu}}$ of
each of the attached $p$ branes (at most two if $p=n$, but arbitrarily many for
$p<n$) of which the $(p-1)$ brane under consideration is a boundary segment,
together with an external electromagnetic force contribution determined by the
Maxwellian field $F_{\mu\nu}=2\nabla_{[\mu}A_{\nu]}$.

Although the foregoing direct derivation starts from a variational postulate,
charge and energy momentum conservation laws of the form (7.12) and (7.13) can
still be expected to hold for more general dissipative models such as would be
obtained by macroscopic averaging over internal degrees of freedom whose net
effect would be taken into account in terms of entropy currents.  An alternative
(for some tastes more intuitive, but mathematically much more awkward) way of
deriving (3.4) and (3.5) in such cases would be to consider the brane system as
the infinitely thin limit of a continuous medium model where the current $J^\mu$
and stress energy momentum density $T^{\mu\nu}$  are no longer continuous
fields but have become Dirac distributions, whose coefficients are interpretable
as the corresponding smooth world sheet supported fields $\underline J{^{\mu}}$
and $\underline T{^{\mu\nu}}$. By whatever route they may have been obtained,
the ubiquitous generality of (7.12) and (7.13) - and of the extrinsic equation of
motion (7.1) that is obtainable via (7.9) as a direct consequence - cannot be
overemphasised.  In the particular case of a {\it free motion} for which
external electromagnetic and contact effects are absent we evidently get
$$ f_\mu=0 \ \ \ \Rightarrow\ \ \ \
\underline T{^{\mu\nu}}K_{\mu\nu}{^\rho}=0 \eqno(7.14)$$
In the case of a variational model with action simply proportional to the world
sheet measure, as in the case of a Dirac membrane, a Goto-Nambu string, or an
ordinary free point particle, the force free equation of extrinsic (``brane
wave") motion (7.14) obviously reduces  to the even simpler (``harmonic") form
$K^\mu=0$ which includes the equation for a geodesic in the one dimensional
case.

\bigskip

{\bf 8. Perfect Brane Models.}
\medskip\parindent =2 cm

For a general brane model, we can always define an energy density scalar, $U$
say, as the negative of the eigenvalue specified by
$$\underline T{^\mu}{_\nu}u^\nu=-U u^\nu \eqno (8.1) $$
where the corresponding eigenvector $u^\mu$ is distinguished by the requirement
that it be timelike or null. As a widely applicable special case (including the
Dirac membrane mentioned above, as well as {\it all} point particle and string
models) a $(p-1)$ brane may be described as ``perfect" if its surface stress
momentum energy tensor is isotropic with respect to the other orthogonal
directions, which in the generic case for which the eigenvector $u^\mu$ is
strictly timelike (not null) and hence normalisable to unity, one gets$^{[1]}$
the explicit form 
$$\underline T{^\mu}{_\nu}=(U-T)u^\mu u_\nu-T\ \olg{^\mu}{_\nu}
\ , \ \ \ \ \ \ u^\mu u_\mu=-1\ ,\eqno (8.2) $$
where $T$ (the negative of the other $(p-1)$ degenerate eigenvalues)
is what is interpretable as the {\it tension} of the $(p-1)$ brane.

The category of perfect branes includes, as the extreme case $p=n$, the example
of an ordinary ``perfect fluid" (with $U=\rho$, where $\rho$ is the ordinary
volume density of mass-energy, while $T=-P$ where $P$ is the ordinary, positive,
pressure). In the other cases, i.e.
 for a $(p-1)$ brane of lower dimension than the
background, i.e. $p<n$, for which extrinsic displacements are possible (so that
the tension must be non negative in order to avoid local 
instability$^{[14][133]}$)
the extrinsic motion will be governed by (7.1) or in the force free case
by (7.14) which, on substitution of (8.2) gives the dynamic equations for
a free perfect brane world sheet in the form
$$c_{_E}{^2}K^\mu=(1-c_{_E}{^2})\ag{^\mu}{_\nu}\dot u{^\mu} \ , \ \ \ \ \
\dot u^{\mu}=u^\nu\nabla_\nu u^\mu \ , \ \ \ \ c_{_E}=\sqrt{T\over 
U}\eqno(8.3)$$
where $\dot u^\mu$ is the acceleration vector of the unit eigenvector $u^\mu$
and $c_{_E}$ is interpretable as the speed of propagation - relative to the
preferred frame specified by $u^\mu$ - of extrinsic perturbations, as derived
from the general characteristic equation (7.2). It is to be noted that in the
ultra relativistic case of a Dirac membrane or Goto Nambu string one has
$c_{_E}=1$ which means that the right hand side of (8.3) will vanish. On the
other hand  the strings and membranes that are commonly used (in violins, drums,
etc.) by old fashionned non relativistic (i.e. non electronic) orchestras for
music generation, will also be describable to a very good approximation by this
{\it same} equation  but with $c_{_E}<<1$, which means that the coefficient
$c_{_E}{^2}$ will be able to be neglected on the right though not of course on
the left.

The extreme case of a ``zero brane" with $p=1$, i.e. that of an ordinary
(massive) point particle, can be considered as being automatically of the
perfect type characterised by (8.2) with $U=m$ where $m$ is its mass, and with
identically vanishing tension $T=0$ which is consistent with the obvious
necessity of having zero relative speed of propagation of any perturbation in
this one dimesional case. For a point particle trajectory the first and second
fundamental tensors will be given simply by
$$ \olg^{\mu}{_\nu}=-u^\mu u_\nu \ , \ \ \ \ K_{\mu\nu}{^\rho} =\olg{_{\mu\nu}}
K{^\rho}\ , \ \ \ \ \ K{^\mu}=-\dot u^{\mu}\ , \ \ \ \ \ \dot u^\mu=
u^\nu \nabla_\nu u^\mu \eqno(8.4)$$
while in terms of the particle mass $m$ and charge $e$ say substitution of the
appropriate expressions
$$ \underline T{^\mu}{_\nu}=-m\ \olg{^\mu}{_\nu} \ ,\ \ \ \ \
 \underline J^\mu=eu^\mu \ ,
\eqno(8.5)$$ 
into the general expressions (7.12) and (7.13) gives the dynamical equations in
the familiar form
$$u^\mu \nabla_\mu e=0 \ , \ \ \ \ u^\mu \nabla_\mu m=0 \ , \ \ \
\ \ -m K_\mu=eF_{\mu\nu}u^\mu \ , \eqno(8.6)$$
subject of course to the usual proviso (which in this context is to be taken
quite litterally!) that there are {\it no strings attached}, since otherwise
corresponding contact contributions on the right of (7.12) and (7.13) could cause
variations of the values of the charge and mass scalars, $e$ and $m$ as well as
modifying the acceleration equation in (7.14).

The case of a membrane in 4-dimensions (or more generally of an $(n-2)$ brane in
$n$ dimensions) shares with the opposite extreme case of a point particle the
property of having comparatively simple kinematic properties, since any timelike
hypersurface has first and second fundamental tensors that are expressible in
terms of its unit normal $\underline\lambda{_\mu}$ (as specified by an arbitrary
choice of orientation) in the form
$$\olg^{\mu}{_\nu}=g^\mu{_\nu}-\lambda{^\mu}\lambda{_\nu}\ ,
\ \ \ \ \ K_{\mu\nu}{^\rho}=K_{\mu\nu}\lambda{^\rho}\ , \ \ \ \ \
\lambda_\mu \lambda^\mu=1\ . \eqno(8.7)$$
Analogously to the way the first fundamental tensor $\olg{^{\mu\nu}}$ is
specifiable (by (7.8)) as the pull back of the contravariant version of the
induced metric, i.e. of what is commonly known as the first fundamental form of
the imbedding, so analogously the symmetric tensor $K^{\mu\nu}$ is the pull back
of the contravariant version of what is commonly known as the {\it second
fundamental form} on the hypersurface, a quantity whose specification, like that
of the unit normal $\lambda{_\mu}$ involves an arbitrary choice of sign.
(In addition to its principle advantage of being applicable to imbeddings of
arbitrary dimension, not just hypersurfaces, an advantage of our present
strategy of working with the three index second fundamental {\it tensor} rather
that the two index second fundamental {\it form} even in the hypersurface case
where the latter is available is that unlike that of $K_{\mu\nu}$ the
specification of $K_{\mu\nu}{^\rho}$ is quite unambiguous.) Whereas the
kinematic specifications (8.7) are simpler than their analogues for the
lower dimensional case of a string, on the other hand the dynamics of a
membrane are generally more complicated. Unlike the case of a string
model which must always, trivially, be perfect in the sense of (8.2)
(or of its null limit$^{[14]}$) the postulate of ``perfection" in this sense
is a serious restriction in the case of a membrane, being satisfied
for a Dirac membrane or an ordinary soap bubble type membrane,
(and even as a reasonable approximation to the way musical drum membranes
are most commonly tuned), but it will not be at all valid for
such applications as to a typical ship's sail.

Between the highpersurface supported case of a membrane and the curve
supported case of a point particle  the only intermediate kind of
brane that can exist in 4-dimensions is that of 1-brane, i.e.
a string model, which (for
any background dimension $n$) will have a first fundamental tensor
that is expressible as the square of the antisymmetric tangential tensor
 ${\cal E}{^{\mu\nu}}$ that is defineable$^{[138]}$ as the pullback of the
contravariant version of the induced measure tensor that is specified
modulo a choice of orientation by the imbedding, i.e. we shall have
$$\olg{^\mu}{_\nu}={\cal E}{^\mu}{_\rho}{\cal E}^{\rho}{_\nu} \ , \ \ \ \ \
{\cal E}{^{\mu\nu}}=\ov{\ov{\cal E}}{^{[\mu\nu]}}\ . \eqno(8.8)$$

A special feature distinguishing string models from point particle
models on one hand and from higher dimensional brane models on the other
is the dual symmetry$^{[139][14]}$ that exists at a formal level between the spacelike
and timelike eigenvectors $u^\mu$ (as already introduced) and $v^\mu$
that for a generic case (excluding the null state limit$^{[14]}$)
are characterised modulo a choice of orientation by the expression
$$\underline{T}^{\mu\nu}=Uu^\mu u^\nu-Tv^\mu v^\nu  \ ,
 v^\mu v_\mu=1=-u^\mu u_\mu   \eqno(8.9)$$
in which the tension $T$ appears as the dual analogue of the
``rest frame" energy per unit length $U$. This formal duality 
can also be made apparent in the expression for the extrinsic curvature
vector of the string, which can be expressed as
$$K^\mu=\ag{^\mu}{_\nu}(v^{\prime\nu}-\dot u^\nu) \ , \ \ \ \
v^{\prime\mu}=v^\nu\nabla_\nu v^\mu\ , \ \ \ \ \dot u^\mu=u^\nu\nabla_\nu
u^\mu  \eqno(8.10)$$
whose substitution in (8.3) enables the equation of extrinsic motion
of a free string to be expressed in the manifestly self dual form
$$U\ \ag{^\mu}{_\nu}\dot u{^\nu}=T\ \ag{^\mu}{_\nu}v^{\prime \nu}  \ .
\eqno(8.11)$$

Of course the extrinsic equation of motion, whether of the general form
(7.1) or the free string specialised form (8.11), cannot actually be used
to determine the evolution of the world sheet until the appropriate
prescription has been given for evaluating the necessary
stress momentum energy tensor components, which in the string case (8.11)
can be taken to be just $T$ and $U$. In the simple Goto-Nambu case, for
which these eigenvalues are specified in advance to have constant values,
$U=T=-L_{_0}$, no further preparation is needed for the integration
of (8.12) but in general, for a string model with non trivial intrinsic
structure the completion of the system of equations of motion will
involve the specification of other differential equations. The simplest
non trivial possibility, which is applicable to higher dimensionsional
perfect brane models as well as to strings, is what is known in the
specific context of perfect fluid theory as the ``barotropic" case,
meaning the case in which $T$ is specified (directly or parametrically)
 as a function only of
$U$ by a {\it single equation of state}. In this barytropic case
(which includes the Witten type conducting cosmic string models$^{[135]}$
whose investigation provided the original motivation for this work)
the only differential equations that are needed to supplement
the extrinsic equation of motion (8.3) or (8.11) are those that are
obtained from the projection into the world sheet of the full
local momentum energy conservation equation (7.13), which in the force
free case simply gives
$$\ov{\ov{\nabla_\mu\underline T{^{\mu\nu}}}} = 0   \eqno(8.12)$$
whose two independent components can be conveniently expressed
as a pair of mutually dual surface current conservation laws
given by
$$\onab_\mu(\nu u^\mu)=0 \ , \ \ \ \ \onab_\mu (\mu v^\mu)=0 \ ,\eqno(8.13)$$
in terms of an effective number density $\nu$ and an associated effective mass
density $\mu$ that obtained from the equation of state as functions of $U$ or
equivalently of $T$ by  a pair of (mutually dual) integral relations of the form
$${\rm ln\ }\nu=\int{dU\over U-T} \ ,\ \ \ \ \ \ {\rm ln\ }\mu
=\int{dT\over T-U}  , \eqno(8.14)$$
which fix them modulo a pair of constants of integration of which
one is conventionally fixed by imposing the (self dual) restraint condition
$$\mu\nu=U-T \ . \eqno(8.15)$$

Appart from the extrinsic perturbations of the world sheet location itself,
which propagate with the ``brane wave" speed $c_{E}$ (relative to the  frame
deterined by $u^\mu$) as already discused, the only other kind of perturbation
mode that can occur in a barytropic string are longitudinal modes specified by
the variation of $U$ or equivalently of $T$ within the world sheet. Such
longitudinal perturbations (the analogue of ordinary sound waves in a perfect
fluid) can easly be seen$^{[14][133]}$ to have a relative propagation velocity
given by
$$   c_{_L}=\sqrt{\nu d\mu\over\mu d\nu}=\sqrt{-dT\over dU}  \eqno(8.16)$$
which must be real in order for the string to be locally stable. Knowledge of
whether  the longitudinal perturbation speed $c_{_L}$ is greater or less than
the extrinsic speed $c_{_E}$ may be critically significant for questions such as
the stability of stationary rotating ring equilibrium
states$^{[14][140][141][142]}$ and
their deformed generalisations$^{[130]} $ which for Witten type cosmic strings
(as opposed to the ordinary Goto Nambu type for which no such states exist) may
be cosmologically important$^{[141][143]}$.  Most early, and many more recent
discussions$^{[144][145][146][147]}$ of Witten type strings  were implicitly based on
the use of an equation of state for which the sum $U+T$ remains constant, which
implies longitudinal propagation at a speed equal to that of light, $c_{_L}=1$
which thus necessarily exceeds $c_{_E}$ but more accurate
investigations$^{[148][149]}$
have recently been developed$^{[150]}$ to a stage at which it is becoming
increasingly clear that the opposite is usually the case, i.e. Witten type
models  would seem to be typified by $c_{_L} < c_{_E}$.

A very special interest attaches to the intermediate non-dispersive, case
characterised by $c_{_E}=c_{_L}$, which corresponds to an equation of state for
which the eingenvalue product $TU$ is constant, leading to dynamic
equations$^{[151]}$ that I
have shown to be {\it explicitly integrable} (like those of the
degenerate Goto Nambu case) in a flat spacetime background, the general form in
an arbitrary curved background being expressible as
$$L_\pm{^\nu}\nabla_\nu L_\mp{^\mu}=0\ ,\ \ \ \ \ \
 L_\pm{^\mu}={\sqrt U u^\mu\pm
\sqrt T v^\mu\over\sqrt{U-T}} \ ,\  \ \ \ T={U_{_0}{^2}\over U} \ ,\eqno(8.17)$$
where $U_{_0}$ is a constant and  the (timelike) unit vectors $L_\pm{^\mu}$ are
directed along the ``left" and ``right" moving unit characteristic directions,
the former being parallel propagated by the latter and vice versa. Another, more
recently established special property of the non-dispersive equation of state
$UT=U_{_0}^2$ is the remarkable solubility (by separation of the relevant
Hamilton Jacobi equation) of the corresponding equations of stationary
equilibrium not only in flat space but in a generalised Kerr black hole
background$^{[130]}$.  This special ``constant product" string model (which can be
recognised$^{[14]}$ as turning up spontaneously in Kaluza Klein
theory$^{[152][153][154][155]}$
 is not just of purely mathematical interest: my predction$^{[151]}$
that it should provide a good description of the  averaged effect of random
noise perturbations on an ``ordinary" Goto-Nambu type  cosmic string (on the
grounds that their presence should not introduce dispersion) has been confirmed
by Vilenkin's more detailed ``wiggly string" calculations$^{[156]}$.

\bigskip

\parindent = 0 cm \bf
Appendix: Background tensor analysis of the curvature of an imbedding.
\rm
\parindent=2 cm\medskip

In any $n$ dimensional manifold with a non degenerate (Riemannian or pseudo
Riemannian) metric tensor with components $g_{\mu\nu}$ (with respect to some
local coordinate patch) that is to be used for index lowering and raising, any
non-null (strictly spacelike or timelike) $p$-dimentional surface element at a
point determines a corresponding decomposition
$$g^\mu_{\ \nu}=\olg{^\mu}{_ \nu}+\ag{^\mu{_ \nu}} \eqno(A1)$$
where $\olg{^\mu}{_\nu}$ is the {\it fundamental} (rank $p$) projection
tensor of  the surface element, and $\ag{^\mu}{_ \nu}$ is the
complementary (rank $n-p$) tensor of projection orthogonal to the surface.
Consistently with (A1) we shall adopt the systematic use of a
 convention using an overhead parallelism symbol, $^=$,
to indicate the effect of projection into the surface, and an
overhead perpendicularity symbol, $ ^{\underline{\ \vert\ } }$, to indicate
the effect of the complementary orthogonal projection operation, so that
in particular for an arbitrary vector with components $\xi^\mu$, and for the 
standard operator $\nabla_\mu$  of Riemannian covariant differentiation (as
defined with respect to a symmetric connection such that $\nabla_\mu 
g_{\nu\rho}$ vanishes), we have
$$\ov{\ov\xi}{^\mu}\eqdef\olg{^\mu}{_\nu}\xi^\nu\ , \ \ \ \  \
\onab_\mu\eqdef\olg {^\nu}{_\mu}\nabla_\nu\ , \ \ \ \ \
{\axi}{^\mu}\eqdef\ag{^\mu}_\nu\xi^\nu\ , \ \ \ \ \
{\anab}_\mu\eqdef\ag{^\nu}_\mu\nabla_\nu\ . \eqno(A2)$$
In terms of this convention, the fundamental tangential and orthogonal
projection operators are thus characterised by the conditions that for any
vector $u^\mu$ that is tangent to the $p$-surface element, and any vector
$\lambda^\mu$ that is orthogonal to the $p$-surface element we must have
$$\ov{\ov u}{^\mu}=u^\mu\ , \ \ \ \ \ {\acu}{^\mu}=0 \ ,
 \ \ \ \ \ \ov{\ov\lambda}{^\mu}=0\ ,
\ \ \ \ \ {\alam}{^\mu}=\lambda^\mu \ . \eqno(A3)$$

Unlike the full covariant differentiation operator $\nabla_\mu$ and its
orthogonally projected part ${\anab}_\mu$, the {\it tangential covariant
differentiation operator}, $\onab_\mu$, has the property of being well defined
not only for (sufficiently smooth) fields defined on an open neighbourhood of
the background space but even for fields with support is confined to a
(sufficiently smooth) $p$-surface whose tangent surface element specifies the
projection. In particular, for any such $p$-surface there will be a well defined
{\it second} fundamental tensor, $K_{\mu\nu}{^\rho}$ defined$^{[133][14][15]}$
 in terms of its first fundamental tensor $\olg {^\mu}{_\nu}$ by
$$K_{\mu\nu}{^\rho}\eqdef \olg {^\sigma}{_\nu}\onab_\mu
\olg {^\rho}{_\sigma} \ , \eqno(A4)$$
which as a trivial algebraic identity is obviously tangential on the first
two indices and almost as obviously orthogonal on the last, i.e.
for an arbitrary vector $\xi^\mu$ it satisfies
$$K_{\mu\nu}{^\rho}\xi_\rho=\ov{\ov{K_{\mu\nu}{^\rho}\xi_\rho }}=
K_{\mu\nu}{^\rho}{\axi}{_\rho}\ . \eqno(A5)$$

Such a tensor $K_{\mu\nu}{^\rho}$ is of course definable  not only for the
fundamental projection tensor of a $p$-surface, but also for any (smooth) field
of rank $p$ projection operators $\olg {^\mu}{_\nu}$ as specified by a
field of arbitrarily orientated $p$-surface elements. What distinguishes the
integrable case, i.e. that in which the elements mesh together to form a well
defined $p$-surface through the point under consideration, is the condition that
the tensor defined by (A5) should also satisfy the {\it Weingarten identity}
$$K_{[\mu\nu]}{^\rho} =0 \eqno(A6)$$
(where the square brackets denote antisymmetrisation), this {\it non trivial}
symmetry property of the second fundamental tensor being derivable$^{[124]}$ as a
version of the well known Frobenius theorem.

The second fundamental tensor $K_{\mu\nu}{^\rho}$ has the property of fully
determining the tangential derivatives of the first fundamental tensor
$\olg{^\mu}{_\nu}$ by the formula
$$\onab_\mu\olg{_{\nu\rho}}=2K_{\mu(\nu\rho)} \eqno(A7)$$
(using round brackets to denote symmetrisation) and it can be seen to be
characterisable by the condition that the orthogonal projection of the
acceleration of any tangential vector field $u^\mu$ will be given by
$$\ag{^\rho}{_\mu}{u^\nu\nabla{_\nu} u^\mu} =u^\mu u^\nu K_{\mu\nu}{^\rho} \
,\eqno(A8)$$
as well as by the condition (in which the non-trivial role of the symmetry
property (A6) is more apparent) that the tangential projection of the derivative
of any field of surface normal vectors $\lambda^\mu$ should be given by
$$\ov{\ov{\nabla_\mu\lambda_\nu}}=-K_{\mu\nu}{^\rho}\lambda_\rho \ .
\eqno(A9)$$

Going on to higher order we can introduce the {\it third} fundamental 
tensor$^{[15]}$ in an analagous manner as
$$\Xi_{\lambda\mu\nu}{^\rho}=\olg{^\sigma}{_ \mu}\olg{^\tau}{_\nu}
\olg{^\tau}{_\nu}\ag{^\rho}{_\alpha}\onab_\lambda
K_{\sigma\tau}{^\alpha} \ , \eqno(A10)$$
which  by construction is obviously
symmetric between the second and third indices and tangential on
all  the first three indices while being,
 i.e. (for an arbitrary vector $\xi^\mu$)
it satisfies  the trivial identities
$$\Xi_{\lambda[\mu\nu]}{^\rho}=0\ , \ \ \ \ \
         \Xi_{\lambda\mu\nu}{^\rho}\xi_\rho=\ov{\ov
 {\Xi_{\lambda\mu\nu}{^\rho}\xi_\rho}}= \Xi_{\lambda\mu\nu}{^\rho}
{\axi}{_\rho}\ . \eqno(A11)$$
In a spacetime background that is flat (or of constant curvature as is the case
for the DeSitter universe model) this third fundamental tensor is fully
symmetric over all the first three indices by what is interpretable as the
generalised Codazzi identity which is expressible$^{[124]}$ in a background with
arbitrary Riemann curvature $R_{\lambda\mu}{^\rho}{_\sigma}$ as
$$\Xi_{\lambda\mu\nu}{^\rho}= \Xi_{(\lambda\mu\nu)}{^\rho} +{_2\over^3}
\olg{^\sigma}{_\lambda}\olg{^\tau}{_{(\mu}}  \olg{^\alpha}{_\nu)}
R_{\sigma\tau}{^\beta}{_\alpha}\ag{^\rho}{_\beta}
\ . \eqno(A12)$$

It is very useful for a great many purposes to introduce the {\it
extrinsic curvature vector} $K^\mu$, defined as the trace of the second
fundamental tensor, i.e.
$$ K^\mu\eqdef K^\nu_{\ \nu}{^\mu}\ .\ \ \ \ \ \ov{\ov K}{^\mu}=0
\eqno(A13)$$
The specification of this extrinsic curvature vector for a
timelike $p$-surface in a dynamic theory provides what can be taken as the
equations of extrinsic motion of the $p$-surface$^{[14]}$ (the simplest case
being the ``harmonic" condition $K^\mu=0$ obtained from a simple surface measure
variational principle such as that of the Goto-Nambu string model or the Dirac
membrane model). It is also useful for many purposes$^{[15]}$ to introduce the
{\it extrinsic conformation} tensor $C_{\mu\nu}{^\rho}$ defined as the trace
free part of the second fundamental tensor by $$C_{\mu\nu}{^\rho}\eqdef
K_{\mu\nu}{^\rho}-{_1\over ^p}\olg{_{\mu\nu}}K^\rho \ , \ \ \ \     C^\nu_{\
\nu}{^\mu}=0 \ .\eqno(A14)$$
which (like the Wey tensor of the background metric) has the noteworthy property
of being conformally invariant with respect to conformal modifications of
$g_{\mu\nu}\mapsto e^{2\sigma}g_{\mu\nu}$ of the background metric.

The condition of preserving the tangent element to an imbedded $p$-surface at a
point breaks down the full $n$ dimensional rotation group preserving the
background metric into the product of the restricted $p$ dimensional rotation
group preserving the induced metric in the imbedding with the restricted $(n-p)$
dimensional rotation group preserving the induced metric in the orthogonal
element. Associated with each of these subgroups there is a corresponding
naturally induced connection and covariant differentiation operator acting on
the corresponding bundles of tangent vectors $u^\mu$ and orthogonal vectors
$\lambda^\mu$ respectively, and for each there will be a corresponding,
respectively ``inner" and ``outer" bundle curvature, which will be represented
by a corresponding background tensor, the former ``inner" curvature tensor being
just the pull-back onto the background by the imbedding mapping of the ordinary
Riemann curvature of the intrinsic geometry induced by the imbedding. Explicitly
for any vector fields satisfying the appropriate tangentiallity and
orthogonality conditions (A2), the effects of the corresponding restricted
``inner" (tangentially projected) and ``outer" (orthogonally projected)
differentiation operations will be given respectively by
$$\ov{\ov{\nabla_\mu t^\nu}} \equiv
\olg{^\nu}{_\rho}\onab_\mu u^\rho=\onab_\mu u^\nu
-K_{\mu\rho}{^\nu}u^\rho\ , \ \ \ \ \ \ag{^\nu}{_\rho}
\nabla_\mu\lambda^\rho=\onab_\mu \lambda^\nu+K_\mu^{\ \nu}{_\rho}
\lambda^\rho \ . \eqno(A15)$$
Using the convention that an underline is inserted whenever necessary to
distinguish a quantity defined with respect to an imbedding from its higher
dimensional background analogue,  the corresponding {\it inner curvature}
tensor $\underline R{_{\kappa\lambda}}{^\mu}{_\nu}$ of the $p$-surface (as
distinct from the ordinary background Riemann tensor
$R_{\kappa\lambda}{^\mu}{_\nu}$)  and the corresponding
 {\it outer curvature} tensor
$\Omega_{\kappa\lambda}{^\mu}{_\nu}$  (for which no background analogue exists,
so that there is no need to underline it) are specifiable by the respective
conditions that for any tangential vector $u^\mu$ and any orthogonal vector
$\lambda^\mu$ to the surface, i.e. for any vectors satisfying (A3) we should
respectively have
$$2\ov{\ov{\nabla_{[\mu}\ov{\ov{\nabla_{\nu]} u^\rho}}}}\ \equiv\
 2\olg{^\rho}{_\lambda}\olg{^\sigma}{_{[\nu}}\onab{_{\mu]}}
(\olg{^\tau}{_\rho}\onab_\sigma u_\tau)       \
=\ \underline R{_{\mu\nu}}{^\rho}{_\sigma}u^\sigma  \ ,\eqno(A16)$$
and
$$2\ag{^\rho}{_\lambda}\olg{^\sigma}{_{ [\nu}}\onab{_{\mu]}}
(\ag{^\tau}{_\rho}\onab{_\sigma} \lambda_\tau)
\ =\ \Omega_{\mu\nu}{^\rho}{_\sigma}\lambda^\sigma  \ .\eqno(A17)$$
Then it can be verified that the inner curvature tensor is given in terms of
the tangential projection of its background analogue by the relation
$$\underline R{_{\mu\nu}}{^\rho}{_\sigma}= 2K^\rho{_{[\mu}}{^\tau}
K_{\nu]\sigma\tau}+ \ov{\ov R}_{\mu\nu}{^\rho}{_\sigma} \ , \eqno(A18)$$
which is the translation into the present scheme of what is well known
in other schemes as the generalised Gauss identity. The much less well
known analogue for the
(identically trace free and conformally invariant)
 outer curvature, for which the most
historically appropriate name is arguably that of Schouten, is given$^{[15]}$
by the expression 
$$\Omega_{\mu\nu}{^\rho}{_\sigma}= 2C_{[\mu}{^{\tau\rho}}
C_{\nu]\tau\sigma}+\olg{^\kappa}{_\mu} 
\olg{^\lambda}{_\nu}C_{\kappa\lambda}{^\alpha}{_\tau}
\ag{^\rho}{_\alpha}\ag{^\tau}{_\sigma} \ , \eqno(A19)$$
where $C_{\mu\nu}{^\rho}{_\sigma}$ is
(trace free conformally invariant)  background Wey tensor, which is definable
implicitly for a background of dimension $n>2$ by the decomposition of the
Riemann tensor into trace and trace free parts as
$$R_{\mu\nu}{^\rho}{_\sigma}=C_{\mu\nu}{^{\rho\sigma}} +{_4\over^{ n-2} }
g^{[\rho}_{\ [\mu}R^{\sigma]}_{\ \nu]}-{_2\over ^{(n-1)(n-2)} } R
g^{[\rho}_{\ [\mu}g^{\sigma]}_{\ \nu]} \ ,\eqno(A20)$$
where, as usual the background Ricci tensor and Ricci scalar are given by
$$ R_{\mu\nu}=
R_{\rho\mu}{^\rho}{_\nu}  \ , \ \ \ \ \ R=R^\nu_{\ \nu}\ . \eqno(A21)$$
It can be seen from the form of the identity (A19) that in a flat or
conformally flat background (for which it is necessary, and for $n\geq 4$
sufficient, that the Weyl tensor should vanish) the vanishing of the extrinsic
conformation tensor $C_{\mu\nu}{^\rho}$ will be sufficient (independently of the
behaviour of the extrinsic curvature vector $K^\mu$) for vanishing of the outer
curvature tensor $\Omega_{\mu\nu}{^\rho}{_\sigma}$, and hence (by (A17)) for
the possibility of constructing fields of orthogonal vectors $\lambda^\mu$ that
satisfy the generalised Fermi-Walker propagation  condition to the effect that
$\ag{^\rho}{_\mu}\onab_\nu\lambda_\rho$ should vanish. It can
also be shown (taking special trouble for the case $p=3$ )  that in a
conformally flat background (of arbitrary dimension $n$) the vanishing of the
conformation tensor $C_{\mu\nu}{^\rho}$ is always sufficient (though by no means
necessary) for conformal flatness of the induced geometry in the imbedding.

  \vfill\eject

\bigskip\parindent=0 cm
{\bf Acknowledgements}
\medskip \parindent=2cm

I wish to thank Claude Barrabes, Jean-Alain Marck,
Patrick Peter, Tsvi Piran, and David Polarski, for many conversations
that have helped to clarify the ideas summarised in the final sections.

\medskip
\bigskip \parindent=0 cm
\bf  References.\rm   {\bf }
\medskip

[1] C.W. Misner, K.S. Thorne, J.A. Wheeler, {\it Gravitation},
(Freeman, San Francisco, 1973). \smallskip
[2] R.M. Wald, {\it General Relativity}, (University of Chicago Press, 1984).
\smallskip

[3] Ya. B. Zel'dovich, I.D. Novikov, {\it Relativistic Astrophysics I},
(University of Chicago Press, 1971).
\smallskip

[4] M. Demianski {\it Relativistic Astrophysics}, (Pergamon, Oxford, 1985).
\smallskip

[5] S.L. Shapiro, S.A. Teukolsky, {\it Black Holes, White Dwarves, and
Neutron Stars; the Physics of Compact Objects}, (Wiley, New York, 1983).
\smallskip

[6] K.S. Thorne, R.H. Price, D.A. Macdonald, {\it Black Holes: the Membrane
Paradigm}, (Yale University Press, New Haven, 1986).
\smallskip

[7] S.W. Hawking, G.F.R. Ellis, {\it The Large Scale Structure of Space Time},
(Cambridge University Press, 1973).
\smallskip

[8] I.D. Novikov, V. Frolov, {\it Physics of Black Holes} (Kluwer
Academic Publishers, Dordrecht, 1989).
\smallskip

[9] B. Carter, Journal de Physique C7, {\bf 34 }, 7 (1973).
\smallskip

[10] B. Carter, in {\it Black Holes} (1972 Les Houches Summer School),
 ed. C. and B. DeWitt, (Gordon and Breach, New York, 1973).
\smallskip

[11] B. Carter in {\it General Relativity: an Einstein Centenary Survey},
ed S.W. Hawking, W. Israel, (Cambridge University Press, 1979).
\smallskip

[12] B. Carter, in {\it Gravitation in Astrophysics} (Carg\`ese 1986),ed. B. Carter, 
J.B. Hartle (Plenum Press, New York, 1986).
\smallskip

[13] B. Whiting, J. Math. Phys. {\bf 30}, 1301 (1989).
\smallskip

[14] B. Carter , ``Covariant Mechanics of Simple and Conducting
Strings and Membranes", in {\it The Formation and Evolution of Cosmic Strings},
ed G. Gibbons, S. Hawking, T. Vachaspati, pp 143-178 (Cambridge U.P., 1990).
\smallskip

[15] B. Carter, ``Outer Curvature and Conformal Geometry of an Imbedding",
in the Penrose Festchrift Volume {\it Complex Geometry and Mathematical
Physics}, to be published in Journal of Geometry and Physics (1991).
\smallskip

[16] B. Carter in {\it Active Galactic Nuclei} ed. C. Hazard, S. Mitton,
(Cambridge University Press, 1979).
\smallskip

[17] S. Chandrasekhar, Astroph. J. {\bf 74 }, 81 (1931).
\smallskip

[18] J.R. Oppenheimer, G. Volkhoff, Phys. Rev. {\bf 55 }, 374 (1939).
\smallskip

[19] B.K. Harrison, K.S. Thorne, M. Wakano, J.A. Wheeler {\it Gravitation
Theory and Gravitational Collapse}, (University of Chicago Press, 1965).
\smallskip

[20] S.W. Hawking, Commun. Math. Phys. {\bf 43}, 199 (1975).
\smallskip

[21] A.S. Eddington, {\it The Internal Constitution of the Stars},
(Cambridge University Press, 1926).
\smallskip

[22] J. Michell, Phil. Trans. Roy. Soc. Lond. {\bf LXXIV}, 35 (1784).
\smallskip

[23] P.S. Laplace, {\it Expos. du Syst\`eme du Monde }, 305, (Paris, 1796).
\smallskip

[24] D. Lynden-Bell, Nature {\bf 233}, 690 (1969).
\smallskip

[25][ J.G. Hills, Nature {\bf 254}, 295 (1975).
\smallskip

[26] B. Carter, J-P. Luminet, Astron., Astroph. {\bf 121}, 97 (1983).
\smallskip

[27] D. Christodoulou, Commun. Math. Phys. {\bf 93}, 171 (1984);
   Commun. Math. Phys. {\bf 109}, 613 (1987).
\smallskip

[28] D. Christodoulou, Commun. Pure, Appl. Math. {\bf XLIV}, 339 (1991).
\smallskip

[29] C.W. Misner, D.H. Sharp, Phys. Rev. {\bf B136}, 571 (1964).
\smallskip

[30] B. Carter in {\it Recent Developments in Gravitation} (1978 Carg\`ese Summer
School), ed. M. Levy, S. Deser, (Plenum, New York, 1979).
\smallskip

[31] A. Lichnerowicz, {\it Relativistic Hydrodynamics and
Magnetohydrodynamics} (Benjamin, New York, 1967).
\smallskip

[32] B. Carter, B. Gaffet, J. Fluid. Mech. {\bf 186}, 1 (1988).
\smallskip

[33] M. May, R.H. White, Phys. Rev. {\bf 141}, 1232 (1966).
\smallskip

[34] D.M. Eardley, L. Smarr, Phys. Rev. {\bf D19}, 2239 (1979).
\smallskip

[35] R. Oppenheimer, H. Snyder, Phys. Rev. {\bf 56}, 455 (1939).
\smallskip

[36] B. Carter, J.Math. Phys. {\bf 10}, 70 (1969).
\smallskip

[37] M.D. Kruskal, Phys. Rev. {\bf 119}, 1743 (1960);
G. Szekeres, Publ. Mat. Debrecen {\bf 7}, 285 (1960).
\smallskip

[38] J.C. Graves, D.R. Brill, Phys. Rev. {\bf 120}, 1507 (1960).
\smallskip

[39] B. Carter, Physics Letters {\bf 21}, 423 (1966).
\smallskip

[40] B. Carter, Phys. Rev. {\bf 141}, 1242 (1966).
\smallskip

[41] R. Penrose  in {\it Relativity, Groups and Topology} (1963 Les Houches Summer School),
 ed. C. and B. DeWitt, (Gordon and Breach, New York, 1964).
\smallskip

[42] R. Penrose, Phys. Rev. Lett. {\bf 14 }, 57 (1965).
\smallskip

[43] S.W. Hawking, R. Penrose, Proc. Roy. Soc. Lond. {\bf A324}, 529 (1970).
\smallskip

[44] R. Penrose, Riv. Nuovo Cimento I {\bf 1}, 252 (1969).
\smallskip

[45] D.M. Eardley, in {\it Gravitation in Astrophysics} (Carg\`ese 1986), 
ed. B. Carter, J.B. Hartle (Plenum Press, New York, 1986).
\smallskip

[46] A. Ori, T. Piran, Phys. Rev. Lett. {\bf 59}, 2137 (1987); J. Gen. Rel.
Grav. {\bf 20}, 7 (1988).
\smallskip

[47] S.W. Hawking, Commun. Math. Phys. {\bf 25}, 152 (1972).
\smallskip

[48] A. Raychaudhuri, Phys. Rev. {\bf 98}, 1123 (1955); Phys. Rev. {\bf 106},
172 (1957).
\smallskip

[49]  D.H. Hartley and R.W. Tucker, in {\it Geometry of Low Dimensional
Manifolds, I} (Lond. Math. Soc. Lecture Notes {\bf 150}),
ed. S.K. Donaldson, C.B. Thomas, (Cambridge University Press, 1990).
\smallskip

[50] S.W. Hawking in {\it Black Holes} (1972 Les Houches Summer School),
 ed. C. and B. DeWitt, (Gordon and Breach, New York, 1973).
\smallskip

[51] J.B. Hartle, S.W. Hawking, Commun. Math. Phys. {\bf 27}, 283 (1972).
\smallskip

[52] J. Bardeen, B. Carter, S.W. Hawking, Commun. Math. Phys. {\bf 31}, 181 (1973).
\smallskip

[53] J. Beckenstein, Phys. Rev. {\bf D7}, 949 (1973).
\smallskip

[54] B. Carter, Commmun. Math. Phys. {\bf 17}, 233 (1970).
\smallskip

[55]  S. Chandrasekhar, {\it Ellipsoidal Figures of Equilibrium},
(Yale University Press, New Haven, 1969).

[56] J.M. Bardeen in {\it Black Holes} (1972 Les Houches Summer School),
 ed. C. and B. DeWitt, (Gordon and Breach, New York, 1973).
\smallskip

[57] B. Carter, J. Gen. Rel. Grav.  {\bf 9}, 437 (1968).
\smallskip

[58] B. Carter, in {\it General Relativity and Relativistic Astrophysics,
Proc. 2nd Canad. Conf}, (Toronto 1987), ed. A. Coley, C. Dyer, B. Tupper,
(World Scientific, Singapore, 1988).
\smallskip

[59] S.M. Wagh, N. Dadhich, Physics Reports {\bf 183}, 137 (1989).
\smallskip

[60] B. Carter, J. Math. Phys. {\bf 29}, 224 (1988).
\smallskip

[61] A. Lichnerowicz, {\it Theories Relativistes de la Gravitation et de
l'Electromagnetism}, (Masson, Paris, 1955).
\smallskip

[62] W. Israel, Phys. Rev. {\bf 164}, 1776 (1967).
\smallskip

[63] W. Israel, Commun. Math. Phys. {\bf 8}, 245 (1968).
\smallskip

[64] H. M$\ddot u$ller zum Hagen, D.C. Robinson, H.J. Seifert, J. Gen. Rel.
Grav. {\bf 4}, 53 (1973).\smallskip

[65] H. M$\ddot u$ller zum Hagen, D.C. Robinson, H.J. Seifert, J. Gen. Rel.
Grav. {\bf 5}, 59 (1974). \smallskip

[66] D.C. Robinson, J. Gen. Rel. Grav. {\bf 8}, 659 (1977).
\smallskip

[67] G. Bunting, A.K.M. Massood-ul-Alam, J. Gen. Rel. Grav, ...to appear.
\smallskip

[68] A. Papapetrou, Ann. Inst. H. Poincar\'e, {\bf A4}, 83 (1966).
\smallskip

[69] W. Kundt, M. Trumper, Ann. Physik {\bf 192}, 414 (1966).
\smallskip

[70] B. Carter, Nature (Phys. Sci.) {\bf 238}, 71 (1973).
\smallskip

[71] R.H. Boyer, Proc. Roy. Soc. Lond. {\bf A311}, 245 (1969).
\smallskip

[72] R.H. Boyer, T.G. Price, Proc. Camb. Phil. Soc. {\bf 62}, 531 (1965).
\smallskip

[73] R.H. Boyer, R.W. Lindquist, J. Math. Phys. {\bf 8}, 265 (1967).
\smallskip

[74] B. Carter, Phys. Rev. {\bf 174}, 1559 (1968).
\smallskip

[75] W. Boucher, G. Gibbons, G.T. Horowicz, Phys. Rev.{\bf D30},
2447 (1984). \smallskip

[76] B. Carter Phys. Letters {\bf A26}, (1968).
\smallskip

[77] B. Carter, Commun. Math. Phys. {\bf 10}, 280 (1968).
\smallskip

[78] G.W. Gibbons, S.W. Hawking, Phys. Rev. {\bf D15}, 2738 (1976).
\smallskip

[79] A. Papapetrou, Ann. Physik {\bf 12}, 309 (1953).
\smallskip

[80] F.J. Ernst, Phys. Rev. {\bf 167}, 1175 (1968).
\smallskip

[81] M. Morse, M. Heins, Ann. of Math. {\bf46}, 625 (1945).
\smallskip

[82] G.W. Gibbons, Commun. Math. Phys. {\bf 27}, 87 (1972).
\smallskip

[83] S.J. Majumdar, Phys. Rev. {\bf 72}, 930 (1947); A. Papapetrou, Proc.
R. Irish. Acad. {\bf A51}, 191 (1947).
\smallskip

[84] J.B. Hartle, S.W. Hawking, Commun. Math. Phys. {\bf 26}, 37 (1972).
\smallskip

[85] W. Israel, G.A. Wilson, J. Math. Phys. {\bf 13}, 865 (1972).
\smallskip

[86] W. Israel, J.T.J. Spanos, Nuovo Cimento Lett. {\bf 7}, 245 (1973).
\smallskip

[87] J.P. Ward, Int. J. Th. Phys. {\bf 15}, 293 (1976).
\smallskip

[88] G.W. Gibbons, C.M. Hull, Phys. Letters {\bf 109B}, 190 (1982).
\smallskip

[89] B. Carter, Phys. Rev. Lett. {\bf 26}, 331 (1971).
\smallskip

[90] D.C. Robinson, Phys. Rev. {\bf D10}, 458 (1974).
\smallskip

[91] D.C. Robinson, Phys. Rev. Lett. {\bf 34}, 908 (1875).
\smallskip

[92] G. Bunting, ``Proof of the Uniqueness Conjecture for Black Holes",
(Thesis, University of New England, Armidale, N.S.W., 1983).
\smallskip

[93] B. Carter, Commun. Math. Phys. {\bf 99}, 563 (1985).
\smallskip

[94] P.O. Mazur, J. Phys. {\bf A15}, 3173 (1982).
\smallskip

[95] B. Carter, Phys. Rev. {\bf D33}, 991 (1986).
\smallskip

[96] R. Geroch, J. Math. Phys. {\bf 12}, 918 (1971).
\smallskip

[97] W. Kinnersley, J. Math. Phys. {\bf 14}, 651 (1973);
\smallskip

[98] D. Christodoulou, Phys. Rev. Lett. {\bf 25 }, 1596 (1970).
\smallskip

[99] B. Carter, Commun. Math. Phys. {\bf 30}, 261 (1973).
\smallskip

[100] R.L. Znajek, Mon. Not. R. Astr. Soc. {\bf 182}, 639 (1978).
\smallskip

[101] T. Damour, Phys. Rev. {\bf D18}, 3598 (1978).
\smallskip
*
[102] E.T. Newman {\it et al}, J. Math. Phys. {\bf 6}, 918 (1965).
\smallskip

[103] R.P. Kerr, Phys. Rev. Letters, {\bf 11}, 238 (1963).
\smallskip

[104] C.V. Vishveshwara, J. Math. Phys. {\bf 9}, 1319 (1968).
\smallskip

[105] R. Debever. Bull. Soc. Math. Belgique {\bf XXIII}, 360 (1971).
\smallskip

[106] R.L. Znajek, Mon. Not. R. Astr. Soc. {\bf 179}, 457 (1977).
\smallskip

[107] J.A. Marck, Proc. Roy. Soc. Lond. {\bf A385}, 431 (1983).
\smallskip
[
108] J.A. Marck, Phys. Lett. {\bf A97}, 140 (1983).
\smallskip

[109] B. Carter, R.G. M$^c$Lenaghan, in {\it Recent Developments in General
Relativity}, ed. R. Ruffini (North Holland, Amsterdam, 1982).
\smallskip

[110] R. Debever, N. Kamran, R.G. M$^c$Lenaghan, J. Phys. {\bf 25}, 1955
(1984). \smallskip

[111] N. Kamran, R.G. M$^c$Lenaghan, J. Math. Phys. {\bf 25}, 1019 (1984).
\smallskip

[112] N. Kamran, J.A. Marck., J. Math. Phys. {\bf 27}, 1589 (1986).
\smallskip

[113] S. Chandrasekhar, Proc. Roy. Soc. Lond. {\bf A349}, 571 (1976).
\smallskip

[114] S.A. Teukolsky, Astroph. J. {\bf 185}, 283 (1973).
\smallskip

[115] W. Press, S.A. Teukolsky, Astroph. J. {\bf 185}, 649 (1973).
\smallskip

[116] W. Kinnersley, J. Math. Phys. {\bf 10}, 1195 (1969).
\smallskip

[117] S. Chandrasekhar, {\it The Mathematical Theory of Black Holes},
(Clarendon Press, Oxford, 1983).

[118] M. Walker, R. Penrose, Commun. Math. Phys. {\bf 18}, 265 (1970).
\smallskip

[119] L.P. Hughston, R. Penrose, P. Sommers, M. Walker,
Commun. Math. Phys. {\bf 27}, 303 (1972).
\smallskip

[120] L.P. Hughston, P. Sommers, Commun. Math. Phys. {\bf 32}, 147 (1973).
\smallskip

[121] L.P. Hughston, P. Sommers, Commun. Math. Phys. {\bf 33}, 129 (1973).
\smallskip

[122] R. Penrose, Ann. N.Y. Acad. Sci {\bf 224}, 125 (1973).
\smallskip

[123] W. Dietz, R. Rudiger, Proc. Roy. Soc. Lond. {\bf A375}, 361 (1981).
\smallskip
[
124] B. Carter, J. Math. Phys. {\bf 28}, 1535 (1987).
\smallskip

[125] B. Carter, Phys. Rev. {\bf D16}, 3414 (1977).
\smallskip

[126] B. Carter, R.G. M$^c$Lenaghan, Phys. Rev. {\bf CD19}, 1093, 1979).
\smallskip

[127] Y. Kosman, Ann; di Mat. Pura ed Appl. {\bf IV}, 317 (1972).
\smallskip

[128]   V.P. Frolov, V.D. Skarzhinsky, A.I. Zelnikov, O. Heinrich,
Phys. Lett. {\bf 224}, 255 (1989).
\smallskip

[129]  B. Carter, V.P. Frolov, Class. and Quantum. Grav. {\bf 6}, 569 (1989).
\smallskip

[130] B. Carter, V.P. Frolov, O. Heinrich, Class. and Quantum Grav.,
{\bf 8},135 (1991).
\smallskip

[131] K.S. Stelle; P.K. Townsend, ``Are 2-branes better than 1?",
Imperial College (London) T.P./87-88/5 (in proc. C.A.P. Summer Institute,
Edmonton, Alberta, 1987).
\smallskip

[132] G.T. Horowicz, A. Strominger, ``Black Strings and p-Branes",
preprint UCSBRH-91-06 (Dept. of Physics, U.C. Santa Barbara, 1991).
\smallskip

[133] B. Carter, Phys. Lett. {\bf B228 }, 446 (1989).
\smallskip

[134] T.W.B. Kibble, Physics Reports {\bf 67 }, 183 (1980)
\smallskip

[135] E. Witten, Nucl. Phys. {\bf B249}, 557 (1985).
 \smallskip

[136] Y. Nambu, in proc. 1969 Detroit conf. on Symmetries and Quark Modes
(Gordon and Breach, New York, 1970).
\smallskip

[137] P.A.M. Dirac, Proc. Roy. Soc. London, {\bf A268}, 57 (1962).
\smallskip

[138] J. Stachel, Phys. Rev. {\bf D21}, 2171 (1980).
\smallskip
[
139] B. Carter, Phys. Lett. {\bf 224}, 61 (1989).
\smallskip

[140] R.L. Davis, E.P.S. Shellard, Phys. Lett. {\bf B209 }, 485 (1988).
\smallskip

[141] R.L. Davis, E.P.S. Shellard, Nucl. Phys. {\bf 323}, 209 (1989).
\smallskip

[142] B. Carter, Phys. Lett. {\bf B238 }, 166 (1990).
\smallskip

[143] B. Carter, in proc. XXVth Rencontre de Moriand,
 {\it Particle Astrophysics: The Early Universe and Cosmic Structures}, ed.
J.M. Alimi, A. Blanchard, A. Bouquet,F. Martin de Volnay, J. Tran Thanh Van, 
pp 213-221 (Editions Fronti\`res, Gif-sur-Yvette, 1990).
\smallskip

[144] D.N. Spergel, T. Piran, J. Goodman, Nucl. Phys. {\bf B291}, 847 (1987).
\smallskip

[145] A. Vilenkin, T. Vachaspati, Phys. Rev. Lett {\bf 58}, 1041 (1987).
\smallskip

[146] D.N. Spergel, W.H. Press, R.J. Scherrer, Phys. Rev. {\bf D39},
379 (1989).
\smallskip

[147] P. Amsterdamski, Phys. Rev. {\bf D39}, 1534 (1989).
\smallskip

[148] C.T. Hill, H.M. Hodges, M.S. Turner, Phys Rev. {\bf D37}, 263 (1988).
\smallskip

[149] A. Babul, T. Piran, D.N. Spergal, Phys. Lett. {\bf 202B}, 307 (1988).
\smallskip
[
150] P. Peter, {\it Superconducting cosmic string: Equation of state for
spacelike and timelike current in the Neutral limit}, D.A.R.C. preprint
 (Observatoire de Paris-Meudon, 1991).
\smallskip

[151] B. Carter, Phys. Rev. {\bf D41}, 3886 (1990).
\smallskip

[152] N.K. Nielsen, Nucl. Phys. {\bf B167}, 248 (1980).
\smallskip

[153] N.K. Nielsen, P. Olesen, Bucl. Phys. {\bf B291}, 829 (1987).
\smallskip

[154] A. Davidson, K.C. Wali, Phys. Lett. {\bf 213B}, 439 (1988).
\smallskip

[155] A. Davidson, K.C. Wali, Phys. Rev. Lett., {\bf 61}, 1450 (1988).
\smallskip

[156] A. Vilenkin, Phys. Rev. {\bf D41}, 3038 (1990).
\bigskip

 \end{document}